\documentclass[12pt]{article}

\usepackage{amsfonts}
\usepackage{dsfont}
\usepackage{amsmath}
\usepackage{amssymb}
\usepackage{amsthm}
\usepackage[title]{appendix}
\usepackage{algorithm}
\usepackage{algpseudocode}
\usepackage{graphicx}
\usepackage{enumerate}
\usepackage{natbib}
\usepackage{url}
\usepackage{geometry}
\usepackage{hyperref}
\usepackage{pgfplots}
\usepackage{pdfsync}
\usepackage{epstopdf}
\usepackage{caption}
\usepackage{subcaption}
\usepackage{multirow}
\usepackage[affil-it]{authblk}

\newtheorem{condition}{Condition}
\newtheorem{lemma}{Lemma}
\newtheorem{corollary}{Corollary}
\newtheorem{example}{Example}
\newtheorem{proposition}{Proposition}
\newtheorem{remark}{Remark}

\def\1{\mathds{1}}

\def\ACC{\mathsf{ACC}}

\def\bbeta{\bar{\beta}}

\def\CPU{\mathsf{CPU}}
\def\Cov{\text{Cov}}
\def\comp{\mathcal{K}}
\def\DM{\text{D}_{\text{M}}}
\def\tD{\tilde{D}}

\def\event{\mathcal{E}}

\newcommand{\eg}{\textit{e.g.}\;}
\def\eps{\epsilon}
\newcommand{\esp}{\mathbb{E}}

\def\event{\mathcal{E}}
\def\ESS{\mathsf{ESS}}
\def\EFF{\mathsf{EFF}}

\newcommand{\iid}{\textit{i.i.d.\,}}
\newcommand{\ie}{\textit{i.e.}\;}

\def\JMP{\mathsf{JMP}}
\def\KL{\mathsf{KL}}

\def\Law{\mathcal{L}}

\def\Mmax{M_{\text{max}}}
\newcommand{\nset}{\mathbb{N}}
\newcommand{\norm}{\mathcal{N}}
\newcommand{\neigh}{\mathfrak{N}}

\newcommand{\proba}{\mathbb{P}}

\def\oP{\overline{P}}
\def\oQ{\overline{Q}}

\newcommand{\rset}{\mathbb{R}}
\newcommand{\rmd}{\mathrm{d}}

\def\tQ{\tilde{Q}}

\newcommand{\targ}{\pi}

\newcommand{\tX}{\tilde{X}}

\def\tY{\tilde{Y}}
\def\uW{\underline{W}}

\newcommand{\Wst}{\overline{W}}
\newcommand{\X}{X}
\newcommand{\Xset}{\mathsf{X}}
\newcommand{\Xalg}{\mathcal{X}}
\newcommand{\Ycal}{\mathcal{Y}}
\def\1{\mathds{1}}

\def\fevent{\mathcal{F}}

\begin{document}
  \title{Adaptive Incremental Mixture\\ Markov chain Monte Carlo}

\author[1,2]{Florian Maire}
\author[1,2]{Nial Friel}
\author[3,4]{Antonietta Mira}
\author[5]{Adrian E. Raftery}
\affil[1]{School of Mathematics and Statistics, University College Dublin}
\affil[2]{The Insight Centre for Data Analytics}
\affil[3]{Data Science Center, ICS, Universit\`{a} della Svizzera italiana}
\affil[4]{DISAT, Universit\`{a} dell'Insubria}
\affil[5]{Department of Statistics, University of Washington}
\maketitle

\begin{abstract}
We propose Adaptive Incremental Mixture Markov chain Monte Carlo (AIMM), a novel approach to sample from challenging probability distributions defined on a general state-space. While adaptive MCMC methods usually update a parametric proposal kernel with a global rule, AIMM locally adapts a semiparametric kernel. AIMM is based on an independent Metropolis-Hastings proposal distribution which takes the form of a finite mixture of Gaussian distributions. Central to this approach is the idea that the proposal distribution adapts to the target by locally adding a mixture component when the discrepancy between the proposal mixture and the target is deemed to be too large. As a result, the number of components in the mixture proposal is not fixed in advance. Theoretically, we prove that there exists a process that can be made arbitrarily close to AIMM and that converges to the correct target distribution. We also illustrate that it performs well in practice in a variety of challenging situations, including high-dimensional and multimodal target distributions. \\
\textbf{keywords}: Adaptive MCMC, Bayesian inference, Independence Sampler, Importance weight, Local adaptation.
\end{abstract}

\baselineskip=18pt
\section{Introduction}
\label{sec:intro}
We consider the problem of sampling from a target distribution defined on a general state space. While standard simulation methods such as the Metropolis--Hastings algorithm \citep{metropolis1953equation,hastings1970monte} and its many variants have been extensively studied, they can be inefficient in sampling from complex distributions such as those that arise in modern applications. For example, the practitioner is often faced with the issue of sampling from distributions which contain some or all of the following: multimodality, very non-elliptical high density regions, heavy tails, and high-dimensional support. In these cases, standard Markov chain Monte Carlo (MCMC) methods often have difficulties, leading to long mixing times and potentially to Monte Carlo estimators with large asymptotic variance. The adaptive MCMC framework originally developed by \cite{gilks1992adaptive}, \cite{gilks1998adaptive} and \cite{haario1999adaptive} can help overcome these problems. The theoretical properties of these algorithms have been extensively analyzed by \cite{andrieu2006ergodicity} and \cite{roberts2007}.  Adaptive MCMC methods improve the convergence of the chain by tuning its transition kernel on the fly using knowledge of the past trajectory of the process. This learning process causes a loss of the Markovian property and the resulting stochastic process is therefore no longer a Markov chain.

Most of the adaptive MCMC literature to date has focused on updating an initial parametric proposal distribution. For example, the Adaptive Metropolis--Hastings algorithm \citep{haario1999adaptive,haario2001adaptive}, hereafter referred to as AMH, adapts the covariance matrix of a Gaussian proposal kernel, used in a Random Walk Metropolis--Hastings algorithm. The Adaptive Gaussian Mixtures algorithm \citep{giordani2010adaptive,luengo2013fully}, hereafter referred to as AGM, adapts a mixture of Gaussian distributions, used as the proposal in an Independent Metropolis--Hastings algorithm.

When knowledge on the target distribution is limited, the assumption that a \textit{good} proposal kernel can be found in a specific parametric family may lead to suboptimal performance. Indeed, a practitioner using these methods must choose, sometimes arbitrarily, (i) a parametric family and (ii) an initial set of parameters to start the sampler. However, poor choices of (i) and (ii) may hamper the adaptation and result in slow convergence.

In this paper, we introduce a novel adaptive MCMC method, called Adaptive Incremental Mixture Markov chain Monte Carlo (AIMM). This algorithm belongs to the general Adaptive Independent Metropolis class of methods, developed in \cite{holden2009adaptive}, in that AIMM adapts an independence proposal density. However, the adaptation process here is quite different from others previously explored in the literature. Although our objective remains to reduce the discrepancy between the proposal and the target distribution along the chain, AIMM proceeds without any global parameter updating scheme, as is the case with adaptive MCMC methods belonging to the framework developed in \cite{andrieu2006ergodicity}.

Our idea is instead to add a \textit{local} probability mass to the current proposal kernel, when a large discrepancy area between the target and the proposal is encountered. The local probability mass is added through a Gaussian kernel located in the region that is not sufficiently supported by the current proposal. The decision to increment the proposal kernel is based on the importance weight function that is implicitly computed in the Metropolis acceptance ratio. We stress that, although seemingly similar to \cite{giordani2010adaptive} and \cite{luengo2013fully}, our adaptation scheme is local and semiparametric, a subtle difference that has important theoretical and practical consequences. In particular, in contrast to AGM, the approach which we develop does not assume a fixed number of mixture components. The AIMM adaptation strategy is motivated by the quantitative bounds achieved when approximating a density using the recursive mixture density estimation algorithm proposed in \cite{li2000mixture}. The main difference is that while in AIMM the sequence of proposals is driven by the importance weight function at locations visited by the AIMM Markov chain, it is constructed in \cite{li2000mixture} by successively minimizing the entropy with respect to the target distribution, an idea which was also put forward in \cite{cappe2008adaptive}. As in most adaptive MCMC, AIMM requires the specification of some tuning parameters and we provide a default setting that works well in the examples we have considered. We also present a faster version of AIMM that guarantees that the proposal is not too costly to evaluate, a situation which occurs when the number of components in the incremental mixture gets large, especially if the state space is high dimensional.

Proving the ergodicity of adaptive MCMC methods is often made easier by expressing the adaptation as a stochastic approximation of the proposal kernel parameters \citep{andrieu2006ergodicity}. AIMM cannot be analysed in this framework as the adaptation does not proceed with a global parameter update step. We do not study the ergodicity of the process in the framework developed by \cite{holden2009adaptive}, and also used in \cite{giordani2010adaptive}, as the Doeblin condition required there is essentially equivalent to assuming that the importance function is bounded above. This condition is generally hard to meet in most practical applications, unless the state space is finite or compact. Instead, our ergodicity proof relies on the seminal work by \cite{roberts2007}, which shows that (i) if the process adapts less and less (\textit{diminishing adaptation}) and (ii) if any Markov kernel used by the process to transition reaches stationarity in a bounded time (\textit{containment}), then the adaptive process is ergodic. We show that AIMM can be implemented in a way that guarantees diminishing adaptation, while the containment condition remains to be proven on a case by case basis, depending on the target distribution. Moreover, we show using the recent theory developed in \cite{craiu2015stability} and \cite{rosenthal2017ergodicity} that there exists a process, which can be made arbitrarily close to the AIMM process, which is ergodic for the target distribution at hand.

The paper is organized as follows. We start in Section \ref{sec:ex} with a pedagogical example which shows how AIMM succeeds in addressing the pitfalls of some adaptive MCMC methods. In Section \ref{sec:aimm}, we formally present AIMM, and study its theoretical properties in Section \ref{sec:conv}. Section \ref{sec:sim} illustrates the performance of AIMM on three different challenging target distributions in high dimensions, involving two heavy-tailed distributions and a bimodal distribution. AIMM is also compared with other adaptive and non-adaptive MCMC methods. In Section \ref{sec:disc} we discuss the connections with importance sampling methods, particularly Incremental Mixture of Importance Sampling \citep{raftery2010}, from which AIMM takes inspiration.

\section{Introductory example}
\label{sec:ex}
We first illustrate some of the potential shortcomings of the adaptive methods mentioned in the introduction and outline how AIMM addresses them. We consider the following pedagogical example where the objective is to sample efficiently from a one-dimensional target distribution. Appendix \ref{app5} gives more details of the various algorithms which we compare with AIMM and specifies the parameters of these algorithms.

\begin{example}
\label{example_pi1}
Consider the target distribution
$$\targ_1=(1/4)\,\norm(-10,1)+(1/2)\,\norm(0,.1)+(1/4)\,\norm(10,1)\,.$$
For this type of target distribution it is known that Adaptive Metropolis--Hastings (AMH) \citep{haario2001adaptive} mixes poorly since the three modes of $\targ_1$ are far apart, a problem faced by many non-independent Metropolis algorithms. Thus an adaptive independence sampler such as the Adaptive Gaussian Mixture (AGM) \citep{luengo2013fully} is expected to be more efficient. AGM uses the history of the chain to adapt the parameters of a mixture of Gaussians on the fly to match the target distribution. For AGM, we consider here two families of proposals: the set of mixtures of two and three Gaussians, referred to as AGM--2 and AGM--3, respectively. By contrast, our method, Adaptive Incremental Mixture MCMC (AIMM) offers more flexibility in terms of the specification of the proposal distributions and in particular does not set a fixed number of mixture components.
\end{example}

We are particularly interested in studying the tradeoff between the convergence rate of the Markov chain and the asymptotic variance of the MCMC estimators, which are known to be difficult to control simultaneously \citep{mira2001ordering,rosenthal2003asymptotic}. Table \ref{tab1} gives information about the asymptotic variance, through the Effective Sample Size (ESS), and about the convergence of the chain, through the mean squared error (MSE) of the tail-event probability $\targ_1(X>5)$. Further details of these performance measures are given in Appendix \ref{app5} (supplementary material).

\begin{table}
\centering
\caption{Example \ref{example_pi1}, results for $\targ_1$: Effective Sample Size (ESS) (the larger the better), Mean Squared Error of $\pi_1(X>5)$ (MSE) (the smaller the better). Estimates obtained through 100 independent runs of the four methods, each of 20,000 iterations, discarding the first 10,000 iterations for burn in. For the ESS statistic, the mean and variance are provided. \label{tab1}}
\begin{tabular}{c|c|c}
 & ESS & MSE ($\times 10^4$)\\
\hline
AMH & $(.08,.004)$ & $6030$\\
AGM--2 & $(.12,.001)$ & $120$\\
AGM--3 & $(.51,.091)$ & $76$\\
AIMM & $(.47,.004)$ & $7$
\end{tabular}
\end{table}

The ability of AMH to explore the state space appears limited as it regularly fails to visit the three modes in their correct proportions after 20,000 iterations (see the MSE of $\pi_1(X>5)$ in Table \ref{tab1}). The efficiency of AGM depends strongly on the number of mixture components of the proposal family. In fact AGM--2 is comparable to AMH in terms of the average ESS,
indicating that the adaptation failed in most runs. AIMM and AGM--3 both achieve a similar exploration of the state space and AGM--3 offers a slightly better ESS. An animation corresponding to this toy example
can be found at \href{http://mathsci.ucd.ie/~fmaire/AIMM/toy_example.html}{this link}.

From this example we conclude that AGM can be extremely efficient provided
%AM that
some initial knowledge on the target distribution is available, for example, the number of modes, the location of the large density regions and so on.
If a mismatch between the family of proposal distributions and the target occurs, inference can be jeopardized.
%AM Otherwise, the inference can be jeopardized if a mismatch between the family of proposal distributions and the target occurs.
Since
%AM this
misspecifications are typical in real world models where one encounters high-dimensional, multi-modal distributions and other challenging situations, it leads one to question the efficiency of
%AM these types of
AGM samplers.
On the other hand, AMH seems more robust to \textit{a priori} lack of knowledge of the target, but the quality of the mixing of the chain remains a potential issue, especially when high density regions are disjoint.

Initiated with an naive independence proposal, AIMM adaptively builds a proposal that approximates the target iteratively by adding probability mass to locations where the proposal is not well supported relative to the target; see Section \ref{sec:aimm:ex}
for details.
%AM there was a comma here with no end. Maybe somebody delated something accidentally?
As a result, very little, if any, information regarding the target distribution is needed, hence making AIMM robust to realistic situations. Extensive experimentation in Section~\ref{sec:sim} confirms this point.

\section{Adaptive Incremental Mixture MCMC}
\label{sec:aimm}

We consider target distributions $\targ$ defined on a measurable space $(\Xset,\Xalg)$ where $\Xset$ is an open subset of $\rset^d$ ($d>0$) and $\Xalg$ is the Borel $\sigma$-algebra of $\Xset$. Unless otherwise stated, the distributions we consider are dominated by the Lebesgue measure and we therefore denote the distribution and the density function (w.r.t the Lebesgue measure) by the same symbol. In this section, we introduce the family of adaptive algorithms referred to AIMM, the acronym for Adaptive Incremental Mixture MCMC.

\subsection{Transition kernel}
AIMM belongs to the general class of Adaptive Independent Metropolis algorithms originally introduced by \cite{gasemyr2003} and studied by \cite{holden2009adaptive}. AIMM generates a stochastic process $\{X_n,\,n\in\nset\}$ that induces a collection of independence proposals $\{Q_n,\,n\in\nset\}$. At iteration $n$, the process is at $X_n$ and attempts a move to $\tX_{n+1}\sim Q_n$ which is accepted with probability $\alpha_n$. In what follows, $\{\tX_n,\,n\in\nset^\ast\}$ denotes the sequence of proposed states that are either accepted or rejected.

More formally, AIMM produces a time inhomogeneous process whose transition kernel $K_n$ (at iteration $n$) is the standard Metropolis--Hastings (M--H) kernel with independence proposal $Q_n$ and target $\targ$. For any $(x,A)\in(\Xset,\Xalg)$, $K_n$ is defined by
\begin{equation}
\label{eq:transition}
K_n(x;A):=\int_A Q_n(\rmd x')\alpha_n(x,x')+\delta_{x}(A)\int_{\Xset}Q_n(\rmd x')\left(1-\alpha_n(x,x')\right)\,.
\end{equation}
In \eqref{eq:transition}, $\alpha_n$ denotes the usual M--H acceptance probability for independence samplers, namely
\begin{equation}
\label{eq:alpha}
\alpha_n(x,x'):=1\wedge \frac{W_n(x')}{W_n(x)}\,,
\end{equation}
where $W_n:=\targ\slash Q_n$ is the importance weight function defined on $\Xset$. Central to our approach is the idea that the discrepancy between $\targ$ and $Q_n$, as measured by $W_n$, can be exploited in order to
adaptively
improve the independence proposal. This has the advantage that $W_n$ is computed as a matter of course in the M--H acceptance probability \eqref{eq:alpha}.

\subsection{Incremental process}
We assume that available knowledge about $\targ$ allows one to construct an initial proposal kernel $Q_0$, from which it is straightforward to sample. When $\targ$ is a posterior distribution, a default choice for $Q_0$ could be the prior distribution. The initial proposal $Q_0$ is assumed to be relatively flat, in the spirit of a defensive distribution \citep{hesterberg1995weighted}.  Initiated at $Q_0$, a sequence of proposals $\{Q_n,\,n\in\nset\}$ is produced by our algorithm. In particular, the proposal kernel adapts by adding probability mass where the discrepancy between $Q_n$ and $\targ$ is deemed too large.

The adaptation mechanism is driven by the random sequence $\{W_n(\tX_{n+1}),\,n\in\nset\}$, which monitors the discrepancy between $\targ$ and the proposals $\{Q_n,\,n\in\nset\}$ at the proposed states $\{\tX_{n},\,n\in\nset^\ast\}$. Let $\Wst>0$ be a user-defined parameter such that the proposal kernel $Q_n$ is incremented upon the event $\event_n$ defined as:
\begin{equation}
\label{eq:increment}
\event_n:=\left\{W_n(\tX_{n+1})>\Wst\,,\quad\tX_{n+1}\sim Q_n\right\}\,.
\end{equation}
The event $\event_n$ exposes subsets of $\Xset$ where the proposal $Q_n$ does not support $\pi$ \textit{well enough}. The parameter $\Wst$ controls the tolerated discrepancy between the proposal and the target distribution.

\subsection{Proposal kernel}
The proposal
%AM proposition
kernel $Q_n$ takes the form of a mixture
\begin{equation}
\label{eq:prop_mixture}
Q_n=\omega_n Q_0+(1-\omega_n)\sum_{\ell=1}^{M_n}\beta_\ell \phi_\ell\bigg\slash \sum_{\ell=1}^{M_n}\beta_\ell\,,
\end{equation}
where $\{\omega_n\}_n$ is a sequence of nonincreasing weights such that $\omega_0=1$ and $\omega_n>0$ (see Section \ref{sec:conv}), $M_n$ is the number of components added to the mixture up to iteration $n$ and $\{\phi_1,\phi_2,\ldots,\phi_{M_n}\}$ are the incremental mixture components created up to iteration $n$. To each incremental component $\phi_\ell$ is attached a weight proportional to $\beta_\ell>0$ (see Section \ref{sec:aimm:inc}).

\subsection{Increment design}
\label{sec:aimm:inc}
Upon the occurrence of $\event_n$, a new component $\phi_{M_n+1}$ is created.
Specifically, $\phi_{M_n+1}$ is a multivariate Gaussian distribution with mean $\mu_{M_n+1}:=\tX_{n+1}$ and with covariance matrix $\Sigma_{M_n+1}$ defined by
\begin{equation}
\label{eq:cov}
\Sigma_{M_n+1}:=\Cov\left\{\neigh(\tX_{n+1}\,|\,X_1,\ldots,X_{n})\right\}\,,
\end{equation}
where for a set of vectors $V$, $\Cov(V)$ denotes the empirical covariance matrix of $V$ and $\neigh(\tX_{n+1}\,|\,V)$ is a subset of $V$ defined as a  neighborhood of $\tX_{n+1}$. In \eqref{eq:cov}, $\neigh(\tX_{n+1}\,|\,X_1,\ldots,X_{n})$ is the neighborhood of $\tX_{n+1}$ defined as
\begin{equation}
\label{eq:neigh}
\neigh(\tX_{n+1}\,|\,X_1,\ldots,X_{n}):=\left\{X_i\in(X_1,\ldots,X_{n})\,:\,\DM(X_i,\tX_{n+1})\leq
%AM I would put a : instead of a , to define the set of neighbors
%fm: makes sense
\tau\rho_n\targ(\tX_{n+1})\right\}\,,
\end{equation}
where $\tau\in(0,1)$ is a user-defined parameter controlling the neighborhood range and $\rho_n$ the number of accepted proposals up to iteration $n$. In \eqref{eq:neigh}, $\DM(X_i,\tX_{n+1})$ denotes the Mahalanobis distance between $X_i$ and $\tX_{n+1}$ which relies on a covariance matrix $\Sigma_0$. When $\targ$ is a posterior distribution, $\Sigma_0$ could be, for example, the prior covariance matrix. Finally, a weight is attached to the new component $\phi_{M_n+1}$ proportional to
\begin{equation}
\label{eq:weight}
\beta_{M_n+1}:=\targ(\tX_{n+1})^{\gamma}\,,\quad \gamma\in(0,1)\,,
\end{equation}
where $\gamma\in(0,1)$ is another user-defined parameter.
%AM I would add here that $\gamma$ is another user defined parameter

%The rationale behind the upper bound in \eqref{eq:neigh} is that the neighborhood of a state $\tX_{n+1}$ should be made larger if the target has been reasonably well explored by the chain in the vicinity of $\tX_{n+1}$.
%This
%AM not clear to me what THIS refers to. The subject of the previous sentence is THE RATIONALE
%is expected to be positively correlated with $\targ(\tX_{n+1})$ and $\rho_n$.
%FM: I agree these two sentences (and the following) are better removed from the paper.

%Components $\phi_1,\phi_2,\ldots$ located in high target probability areas are given higher weights if $\gamma$ is close to $1$. The AIMM acceptance rate is therefore expected to be higher when moves to lower density regions are proposed. In contrast, components will have uniform weights when $\gamma$ is close to $0$, a setup that will foster a faster exploration of the state space but may reduce the AIMM acceptance rate.

Note that in Eqs \eqref{eq:neigh} and \eqref{eq:weight}, one can possibly plug the unnormalized probability density function instead of $\pi$. Indeed, in Eq. \eqref{eq:neigh} the upper bound is proportional to $\tau$ which may be set so as to account for the normalizing constant.  In Eq. \eqref{eq:weight}, the weights are eventually self-normalized (see Eq. \eqref{eq:prop_mixture}), hence unknown normalizing constants are not problematic here neither.

\subsection{AIMM}
Algorithm \ref{alg:aimm} summarizes AIMM. Note that during an initial phase consisting of $N_0$ iterations, no adaptation is made. This serves the purpose of gathering sample points required to produce the first increment. Also, we have denoted by $N$ the Markov chain length which may depend on the dimension of $\Xset$ and the computational resources available. In any case, we recommend setting $N\gg N_0$.

\begin{algorithm}
\caption{Adaptive Incremental Mixture MCMC}
\label{alg:aimm}
\begin{algorithmic}[1]
 \State {\bf{Input:}} $
        \begin{aligned}[t]
				& \text{user-defined parameters:}~\Wst, \gamma, \tau, \{\omega_n\}_n, N_0, N\\
				& \text{a proposal distribution:}~Q_0
		\end{aligned}$
\vspace{.2cm}
\State {\bf{Initialize:}} $X_0\sim Q_0$, $W_0=\targ\slash Q_0$,  $M_0=0$ and $\omega_0=1$
\For{${n}~=~0,1,\ldots,N$}
  	\State Propose $\tX_{n+1}\sim Q_n=\omega_n Q_0+(1-\omega_n)\sum_{\ell=1}^{M_n}\beta_\ell \phi_\ell\big\slash \sum_{\ell=1}^{M_n}\beta_\ell $
	\State Draw $U_n\sim\text{Unif}(0,1)$ and set
\begin{equation*}
\left\{
\begin{array}{lll}
\X_{n+1}=\tX_{n+1} & \text{if} & U_n\leq \left(1\wedge W_n(\tX_{n+1})\slash W_n(X_{n})\right)\\
\X_{n+1}=\X_n & \text{otherwise}
\end{array}
\right.
\end{equation*}
	
    \If{$n>N_0$ and $\event_n:=\{W_n(\tX_{n+1})>\Wst\}$ is true}
    %AM we should say that we require the event $\event_n$ to be TRUE!
    \State Update $M_{n+1}=M_n+1$
    \State Increment the proposal kernel with the new component
    \begin{equation*}
       \phi_{M_n+1}=  \norm\left(\tX_{n+1},\Cov(\neigh(\tX_{n+1}\,|\,X_1,\ldots,X_{n}))\right)
    \end{equation*}
   %% \begin{equation*}
%%    \phi_{M_n+1}=
%%    \left\{
%%    \begin{array}{cc}
%%    \norm\left(\tX_{n+1},\Cov(\neigh(\tX_{n+1}\,|\,X_1,\ldots,X_{n}))\right) & \text{if } \sevent_n\\
%%    Q_0& \text{if } \uevent_n
%%    \end{array}
%%     \right.
%%    \end{equation*}
    \hspace{2cm}weighted by $\beta_{M_n+1}=\targ(\tX_{n+1})^\gamma$
    \State Update $W_{n+1}=\targ\slash Q_{n+1}$
    \Else
    \State Update $W_{n+1}=W_n$ and $M_{n+1}=M_n$
	\EndIf
\EndFor
\vspace{.2cm}
\State {\bf{Output:}} $\{X_n,\,n\leq N\}$, $\{Q_n,\,n\leq N\}$	
\end{algorithmic}
\end{algorithm}

\subsection{Example~\ref{example_pi1} (ctd.)}
\label{sec:aimm:ex}

\begin{figure}
\centering
\includegraphics[scale=1]{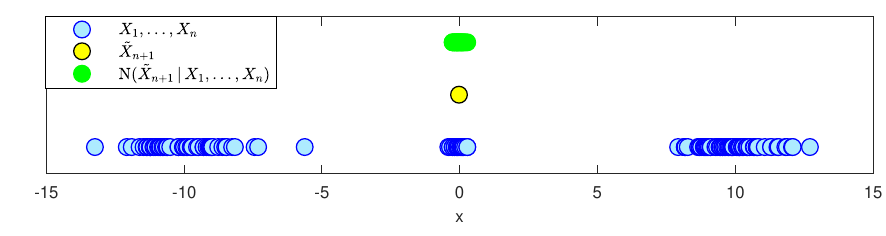}
\includegraphics[scale=1]{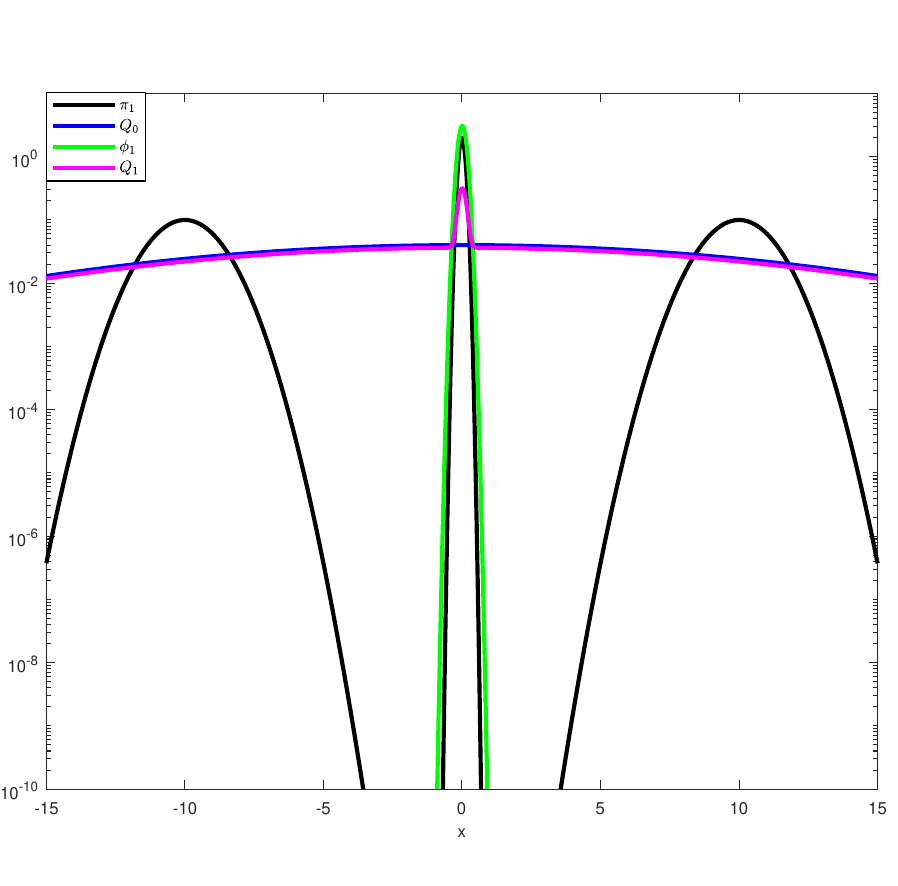}
\caption{Example \ref{example_pi1}: Illustration of one AIMM increment for the target $\pi_1$. Top: states $X_1,\ldots,X_n$ of the AIMM chain, proposed new state $\tilde{X}_{n+1}$ activating the increment process (\ie satisfying $W_n(\tilde{X}_{n+1})\geq \Wst$) and neighborhood of $\tilde{X}_{n+1}$, $\neigh(\tX_{n+1}\,|\,X_1,\ldots,X_n)$. Bottom: target $\pi_1$, defensive kernel $Q_0$, first increment $\phi_1$ and updated kernel $Q_1$ plotted on a logarithmic scale.}
\label{fig:ill1}

\end{figure}

For the toy example in Section~\ref{sec:ex} we used the following parameters:
$$
\Wst=1,\;\gamma=0.5,\;\tau=0.5,\;N_0=1,000,\;\omega_n=(1+M_n/10)^{-1},\;
Q_0=\norm(0,10).$$
%AM moved this here
Figure \ref{fig:ill1} shows the different states of the chain that,
%AM using $Q_0=\norm(0,10)$ IT IS NOT TRUE THAT THEY ARE ALL SAMPLED FROM THIS Q_0
% in the initial phase, or am I wrong? They are sampled from Q_n. But is Q_n the same as Q_0 in the initial phase? If this is the case then I think step 4 is a little confusing.
%FM I fixed the figure with the correct Q_0, and yes, before the first increment we always have Q_n=Q_0 (because \omega_n=1, since M_n=0)...
before the first increment takes place, are all sampled from $Q_0$.
The proposed state $\tX_{n+1}$ activates the increment process when the condition $\{W_n(\tX_{n+1})>\Wst\}$ is satisfied for the first time after the initial phase ($N_0=1,000$) is completed. At $\tX_{n+1}$ there is a large discrepancy between the current proposal,
$Q_n$,
%AM here I would write $Q_n$ and not $Q_0$
%FM OK
and the target, $\targ_1$. The neighborhood of $\tX_{n+1}$ is identified and defines $\phi_1$, the first component of the incremental mixture.

\begin{figure}
\centering
\includegraphics[scale=.65]{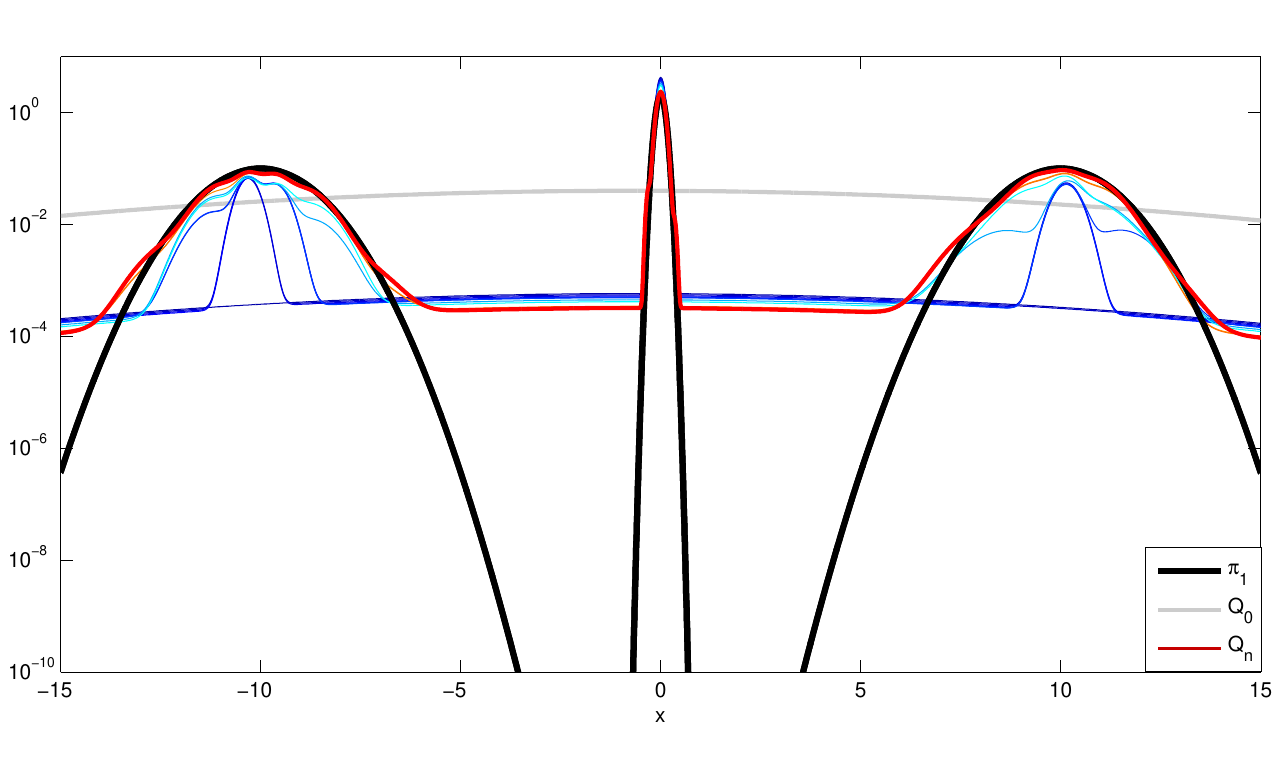}
\includegraphics[scale=.65]{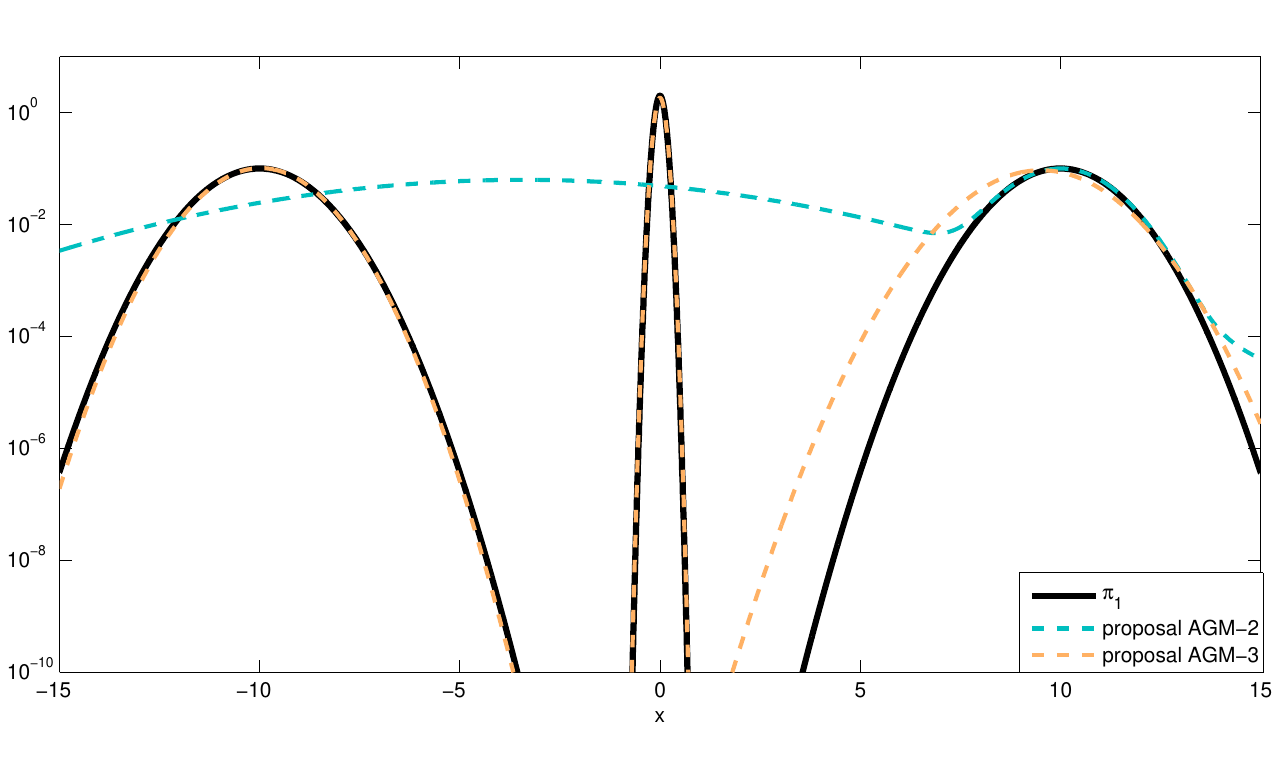}
\caption{Example \ref{example_pi1}: Illustration of AIMM and AGM sampling from the target $\pi_1$ for $n=20,000$ iterations. Top: sequence of proposals from $Q_0$ to $Q_n$ produced by AIMM. Bottom: proposals produced by AGM-2 and AGM-3. Both plots are in log-lin scale.}
\label{fig:ill2}
\end{figure}

\begin{figure}
\centering
\includegraphics[scale=.77]{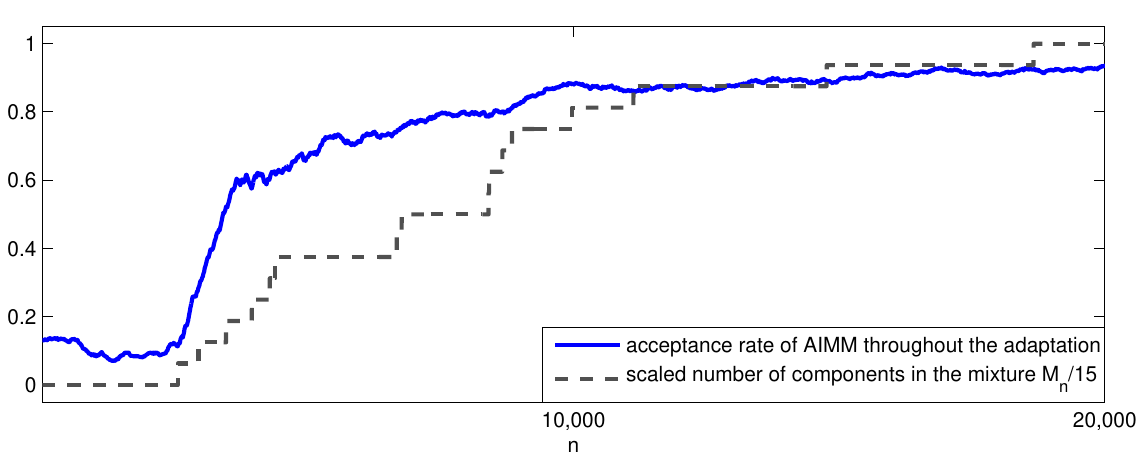}
\caption{Example \ref{example_pi1}: Effect of AIMM increments on the sampling efficiency for the target $\pi_1$ - acceptance rate and number of increments throughout $n=20,000$ iterations of AIMM.}
\label{fig:ill3}
\end{figure}

The top panel of Figure \ref{fig:ill2} shows the sequence of proposal distributions up to $n=20,000$ iterations. AIMM has incremented the proposal 15 times. The first proposals are slightly bumpy and as more samples become available, the proposal distribution gets closer to the target. Thus, even without any information about $\pi_1$, AIMM is able to increment its proposal so that the discrepancy between $Q_n$ and $\targ_1$ vanishes. This is confirmed by Figure \ref{fig:ill3} which reports the acceptance rate (averaged on a moving window) and the number of components in the mixture throughout the algorithm.
Finally, the proposal kernels obtained after 20,000 iterations of AGM-2 and AGM-3 are reported at the
bottom panel of Figure \ref{fig:ill2}
%AM of which figure? Figure 2? We should write this. Plus I think it is the bottom panel and not the right panel
%FM OK
and are in line with their performances as given in Table \ref{tab1}.

The AGM-3 proposal density declines to zero in $\pi_1$'s low density regions,  which is an
%apparently
%AM why apparently ?
appealing feature in this simple example. However, AGM shrinks its proposal density in locations that have not been visited by the chain.
This can be problematic if the sample space is large and it takes
%AM some
time to  reasonably explore it (see Section \ref{sec:sim}).

\section{Ergodicity of AIMM}
\label{sec:conv}

\subsection{Notation}
Let $(\Omega,\mathcal{A},\proba)$ be the probability space generated by AIMM. For any event $A\in\mathcal{A}$, $\proba_A$ denotes the probability measure $\proba$ conditionally on $A$. With some abuse of notations, for any $x\in\Xset$, $\proba_x$ refers to $\proba$ conditionally on the event $\{X_0=x\}$.

Let $\nu$ and $\mu$ be two probability measures defined on $(\Xset,\Xalg)$. Recall that the Kullback--Leibler (KL) divergence between $\nu$ and $\mu$ is defined as
$$
\KL(\nu,\mu)=\int_{\Xset}\log\frac{\nu(x)}{\mu(x)}\rmd\nu(x)\,.
$$
Moreover, the total variation distance between $\nu$ and $\mu$ can be written as
$$
\|\nu-\mu\|=\sup_{A\in\Xalg}|\nu(A)-\mu(A)|=\frac{1}{2}\int_{\Xset}|\mu(x)-\nu(x)|\rmd\rho(x)\,,
$$
where the latter equality holds if $\mu$ and $\nu$ are both dominated by a common measure $\rho$. %Note also that Pinsker's inequality allows one to bound the total variation in terms of KL divergence:
%\begin{equation}
%\label{eq:pinsker}
%\|\nu-\mu\|\leq \sqrt{\frac{1}{2}\KL(\nu,\mu)}\,.
%\end{equation}
Finally, for a sequence of random variables $\{X_n\}_{n>0}$, the notations $X_n=o_{p}(1)$ stands for the convergence of $\{X_n\}_{n>0}$ to zero in probability  $p$ (i.e. with respect to the probability measure $p$),
and $X_n=O_p(1)$ means that $\{X_n\}_{n>0}$ is bounded in probability $p$.

%Let $(\Omega,\mathcal{A},\proba)$ be the probability space generated by AIMM. For any event $A\in\mathcal{A}$, $\proba_A$ denotes the probability measure $\proba$ conditionally on $A$ and with some abuse of notations, for any $x\in\Xset$, $\proba_x$ refers to $\proba$ conditionally on the event $\{X_0=x\}$.
%\begin{itemize}
%\item $(\Omega,\mathcal{A},\proba)$ is the probability space generated by AIMM
%\item $\tQ_n$ is the incremental part of the proposal $Q_n$ at iteration $n$, \ie $\tQ_n\propto\sum_{k=1}^{n-1}\beta_k\phi_k$
%\item at iteration $n$, only $M_n\leq n-1$ components among $\phi_1,\ldots,\phi_{n-1}$ have a non-zero weight
%\item $\event_n$ is the event that the proposal increments at iteration $n$ of AIMM, \ie $\beta_n>0$ almost surely
%\item $\event$ is the event that the proposal increments infinitely often \ie $\event=\cap_n\cup_{k>n}\event_n$
%\item For any event $A\in\Xalg$, $\proba_A$ denotes the probability measure $\proba$ conditionally on $A$ and with some abuse of notations, for any $x\in\Xset$, $\proba_x$ refers to $\proba$ conditionally on the event $\{X_0=x\}$.
%\end{itemize}

\subsection{Ergodicity of adaptive Markov chains}
In this section, we establish assumptions under which the AIMM process is ergodic, \ie
\begin{equation}
\label{eq:ergodicity}
\forall\,x\in\Xset,\quad\lim_{n\to\infty}\|\proba_x(X_n\in \,\cdot\,)-\pi\|=0\,.
\end{equation}

We use the theoretical framework developed in \cite{roberts2007}. In particular, Theorem 2 in \cite{roberts2007} states that if the AIMM process satisfies the \textit{Diminishing adaptation} and \textit{Containment} conditions (see below) then the process is ergodic and Eq. \eqref{eq:ergodicity} holds.

\begin{condition}\textbf{Diminishing adaptation.}

For all $x\in\Xset$, the stochastic process $\{\Delta_n,\,n\in\nset\}$, defined as
\begin{equation}
\label{eq:da}
\Delta_n:=\sup_{x\in\Xset}\|K_{n+1}(x,\,\cdot\,)-K_{n}(x,\,\cdot\,)\|
\end{equation}
converges to zero in $\proba_x$-probability, \ie $\Delta_n=o_{\proba_x}(1)$.
\end{condition}

\begin{condition}\textbf{Containment.}

For all $\eps>0$ and $x\in\Xset$, the stochastic process $\{M_n(\eps),\,n\in\nset\}$, defined as
\begin{equation}
\label{eq:cont}
C_n(\eps):=\inf\{N\in\nset,\;\|K_n^N(X_n,\,\cdot\,)-\targ\|<\eps\}\
\end{equation}
%AM we have used M_n before with a different meaning. I think it is a little confusing. It is true that this is
% M_n(\eps), still, maybe we should change name
%FM agree
is bounded in $\proba_x$-probability, \ie $C_n(\eps)=O_{\proba_x}(1)$.
\end{condition}

Even though containment is not a necessary condition for ergodicity (see \cite{fort2011convergence}) and in fact seems to hold in most practical situations (see \eg \cite{rosenthal2011optimal}), it remains a challenging assumption to establish rigorously for the setup considered in this paper, \ie a broad class of target distributions defined on a not necessarily finite or compact state space. In the next section we show that diminishing adaptation holds for the AIMM process described at Section \ref{sec:aimm}, up to minor implementation adjustments, while containment needs to be proven on a case by case basis. We also show that there exists
a
%AM another
version of AIMM, conceptually different
from
%AM to
the process described at Section \ref{sec:aimm}
%AM but
%FM ok
that can be made arbitrarily close to it, which is ergodic for any absolutely continuous target distribution $\pi$.

\subsection{Main ergodicity results}
We study two variants of AIMM.

\subsubsection{Proposal with an unlimited number of increments}
\label{sec:conc:aimm1}
The first version
of AIMM
is similar to the algorithm presented in Section \ref{sec:aimm}. For this
version of the
%AM
algorithm, we prove only that the diminishing adaptation assumption holds under some minor assumptions. Indeed, proving  the containment condition
%AM holds for this algorithm
is challenging without further assumptions on $\pi$, \eg compact support \citep{craiu2009learn} or tail properties (see \citet{bai2009containment} in the context of adaptive random walk Metropolis samplers). Moreover, most proof techniques establishing containment require the space of the adapting parameter to be compact, something which does not hold in this version of AIMM as the number of incremental components can, theoretically, grow to infinity.

\begin{proposition}
\label{prop1}
Assume that there exist three positive constants $(\delta,\eta,\lambda)$ such that
\begin{enumerate}[\textbf{A}1.]
\setcounter{enumi}{0}
\item The covariance matrix of any component of the mixture satisfies $\det\Sigma_k>\delta>0$.
\item The (unormalized) incremental mixture weights are defined as:
\begin{equation}
\label{eq:new_weights}
\beta_{M_n+1}=\frac{\eta+\pi(\tX_{n+1})^\gamma}{(1+\eta)^{M_{n+1}}}
\qquad\text{and}\qquad\omega_n=\frac{1}{1+\sum_{k=1}^{M_n}\beta_k}\vee \lambda\,.
\end{equation}
\item The initial kernel $Q_0$ is subexponential or satisfies $Q_0(x)\propto \exp\{\psi(x)\}$ where $\psi=o(x^2)$ $(x\to\infty)$, \ie $Q_0$ has heavier tail than a Gaussian, \eg multivariate Laplace or t-distribution.
\item There is a parameter $0<\uW<\Wst$ such that the mixture increments upon the event $\event_n\cup\fevent_n$,
$$
\fevent_n:=\{W_n(\tX_{n+1})<\uW\}\,
$$
and when it increments upon $\fevent_n$, the new component is
equal to
%AM set to
$Q_0$ and the corresponding weight is defined as in Eq. \eqref{eq:weight}.
\end{enumerate}
Then, the AIMM process sampled using conditions \textbf{A}1--\textbf{A}4 satisfies diminishing adaptation, \ie Eq. \eqref{eq:da} holds.
\end{proposition}
The proof is given in Appendix \ref{app1}.

\begin{remark}
Proposition \ref{prop1} holds for parameters $(\delta,\eta,\lambda)$ arbitrarily close to zero. Also, Assumption \textbf{A}1 can be enforced simply by
expanding
%AM expending
the neighborhood of $\tX_{n+1}$ such that when $\det\Cov\{\neigh(\tX_{n+1})\}<\delta$,
$$
\Sigma_{M_{n}+1}:=\Cov\{X^\ast_1,\ldots,X^\ast_k\}
$$
where $\{X^\ast_1,\ldots, X^\ast_n\}$ is a permutation of $\{X_1,\ldots,X_n\}$ such that $\DM(\tX_{n+1},X^\ast_i)\leq \DM(\tX_{n+1},X^\ast_{i+1})$ and $k:=\inf\{i\leq n,\, \Cov(X_1^\ast,\ldots,X_i^\ast)\geq \delta\}$.
\end{remark}

\begin{remark}
The event $\fevent_n$ is the counterpart of $\event_n$ and exposes subsets of $\Xset$ where the proposal
puts
%AM put
\textit{too much} probability mass at locations of low $\pi$-probability.
In this case, the rationale for setting $\phi_{M_n+1}=Q_0$ is
%AM here
to reduce the probability mass of the proposal locally by increasing the weight associated to the initial proposal $Q_0$, assumed to be vague.
\end{remark}

\subsubsection{Proposal with adaptation on a compact set}
\label{sec:conv:aimm2}

The second version of AIMM that we study here represents somewhat a slight conceptual departure from the original algorithm presented in Section \ref{sec:aimm}. We stress that the assumptions \textbf{A}1--\textbf{A}4 from Section \ref{sec:conc:aimm1} are not required here.
\begin{proposition}
\label{prop2}
Assume that:
\begin{enumerate}[\textbf{B}1.]
\setcounter{enumi}{0}
\item The AIMM process $\{X_n,\,n\in\nset\}$ has bounded jumps, \ie there exists $D>0$ such that for all $n\in\nset$
\begin{equation}
\label{eq:bounded}
 \proba\{\|X_n-X_{n+1}\|\leq D\}=1\,.
\end{equation}
\item There is a compact set $\comp\subset\Xset$ such that the weight $\{\omega_n\}_n$ in the proposal \eqref{eq:prop_mixture} is replaced by $\{\overline{\omega}_n^\comp\}_n$
\begin{equation}
\label{eq:compact}
\overline{\omega}_n^\comp=\omega_n\vee \1_{X_n\not\in\comp}\,,
\end{equation}
\ie if the process is such that $X_n\not\in\comp$ then the proposed state is generated from $Q_0$ with probability one. Conversely, if the process is such that $X_n\in\comp$, then the proposed state is generated as explained in Section \ref{sec:aimm}, \ie from $Q_n$ \eqref{eq:prop_mixture}. We denote this proposal by $\overline{Q}_n^\comp$.
\item The number of incremental components in the adaptive proposal is capped, \ie there is a finite $M\in\nset$ such that $\proba(M_n\leq M)=1$ and the mean $\mu_n$ and covariance $\Sigma_n$ of each component are defined on a compact space.
\end{enumerate}
If the AIMM process presented in Section \ref{sec:aimm} satisfies \textbf{B}1--\textbf{B}3, then it is ergodic.
\end{proposition}

This result is a consequence of Theorem 2 of \cite{roberts2007} combined with the recent developments in \cite{craiu2015stability} and \cite{rosenthal2017ergodicity}. The proof is given in Appendix \ref{app2}.

We now explain how, in practice, a version of AIMM compatible with the assumptions of Proposition \ref{prop2} can be constructed and made arbitrarily close to the version of AIMM presented in Section \ref{sec:aimm}. First, fix an arbitratrily large constant $D<\infty$. Assumption \textbf{B}1 holds by construction if $\tX_{n+1}\sim \overline{Q}_n^\comp$ is automatically rejected when $\|\tX_{n+1}-X_n\|>D$. Assumption \textbf{B}2 holds by construction if $\overline{Q}_n^\comp$ is used instead of $Q_n$ in the AIMM algorithm presented in Section \ref{sec:aimm}. Finally, Assumption \textbf{B}3 is satisfied by slightly modifying the adaptation mechanism proposed in Section \ref{sec:aimm}. Define two arbitrarily large constants $L>0$ and $M\in\nset$ and increment the proposal upon the event $\overline{\event}_n:=\event_n\cap \{M_n<M\}$ and in the following way:
\begin{equation*}
\overline{\mu}_{M_n+1}:=\tX_{n+1}|_{L}\,,\qquad
\overline{\Sigma}_{M_n+1}:=\Cov\{\overline\neigh_L(\tX_{n+1}\,|\,X_1,\ldots,X_n)\}\,,
\end{equation*}
where for any vector $x\in\rset^d$ and any $L>0$
\begin{equation*}
x|_L=\{\{x_1\wedge L\}\vee-L,\ldots,\{x_d\wedge L\}\vee-L\}\quad\text{and}\quad
\overline{\neigh}_L=\left\{X_i|_L,\;X_i\in\neigh(\tX_{n+1}\,|\,X_1,\ldots,X_n)\right\}\,.
\end{equation*}
The definition of the unormalized weight $\beta_{M_n+1}$ (Eq. \eqref{eq:weight}) is unchanged.

\section{Simulations}
\label{sec:sim}
In this Section, we consider three target distributions:
\begin{itemize}
\item $\targ_2$, the banana shape distribution used in \cite{haario2001adaptive};
\item $\targ_3$, the ridge like distribution used in \cite{raftery2010};
\item $\targ_4$, the bimodal distribution used in \cite{raftery2010}.
\end{itemize}
Each of these distributions has a specific feature, resulting in a diverse set of challenging targets:
 $\targ_2$ is heavy tailed, $\targ_3$ has a narrow, nonlinear  and ridge-like support and $\targ_4$ has two distant modes.
We consider $\targ_2$ and $\targ_4$ in different dimensions: $d\in\{2, 10, 40\}$ for $\targ_2$ and $d\in\{4,10\}$ for $\targ_4$. We compare AIMM (Algorithm \ref{alg:aimm}) with several other algorithms that are briefly described in Appendix \ref{app5} using some performance indicators that are also defined, explained and justified in Appendix \ref{app5}. We used the following default parameters for AIMM:
$$
\Wst=d\,,\quad\gamma=0.5\,,\quad \tau=0.5\,,\quad \kappa=0.1\,,\quad N_0=1,000\sqrt{d}\,.
$$

For each scenario, we implemented AIMM with different thresholds $\Wst$ valued around $d$ to monitor the tradeoff between adaptation and computational efficiency of the algorithm. In our experience, the choice $\Wst=d$ has worked reasonably well in a wide range of examples. As the dimension of the state space increases a higher threshold is required, too many kernels being created otherwise. However, a satisfactory choice of $\Wst$ may vary depending on the target distribution and the
available computational budget.

\subsection{Banana-shaped target distribution}

\begin{example}
\label{example_pi2}
Let $\Xset=\rset^d$, $\targ_2(x)=\Psi_{d}(f_b(x);m,S)$ where $\Psi_{d}(\,\cdot\,;\,m,S)$ is the $d$-dimensional Gaussian density with mean $m$ and  covariance matrix $S$, and let
$f_b:\rset^d\to\rset^d$ be the mapping defined, for any $b\in\rset$, by
\begin{equation*}
\label{eq:banana_twist}
f_b:\left(
\begin{array}{c}
x_1\\
x_2\\
x_3\\
\vdots\\
x_d
\end{array}\right)
\to
\left(
\begin{array}{c}
x_1\\
x_2+b\,x_1^2-100b\\
x_3\\
\vdots\\
x_d
\end{array}\right)\,.
\end{equation*}
\end{example}

We consider $\targ_2$ in dimensions $d=2$, $d=10$ and $d=40$ and refer to the marginal of $\targ_2$ in the $i$-th dimension as $\targ_2^{(i)}$. The parameters $m=\boldsymbol{0}_d$ and $S=\text{diag}([100,\boldsymbol{1}_{d-1}])$ are held constant. The target is banana-shaped along the first two dimensions, and is challenging since the high density region has a thin and wide ridge-like shape with narrow but heavy tails. Unless otherwise stated, we use $b=0.1$ which accentuates these challenging features of $\targ_2$. We first use the banana-shaped target $\targ_2$ in dimension $d=2$ to study the influence of AIMM's user-defined parameters on the sampling efficiency.

\subsubsection{Influence of the defensive distribution}
\begin{figure}
\centering
\includegraphics[scale=.68]{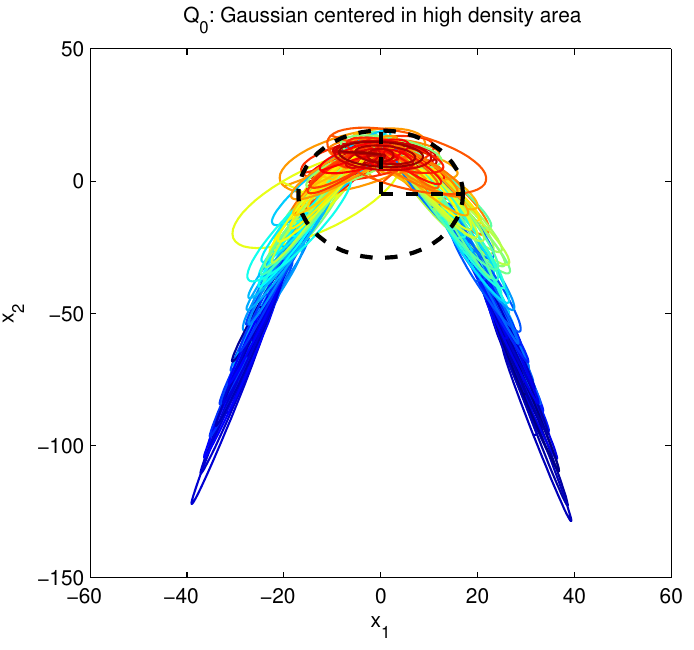}\includegraphics[scale=.68]{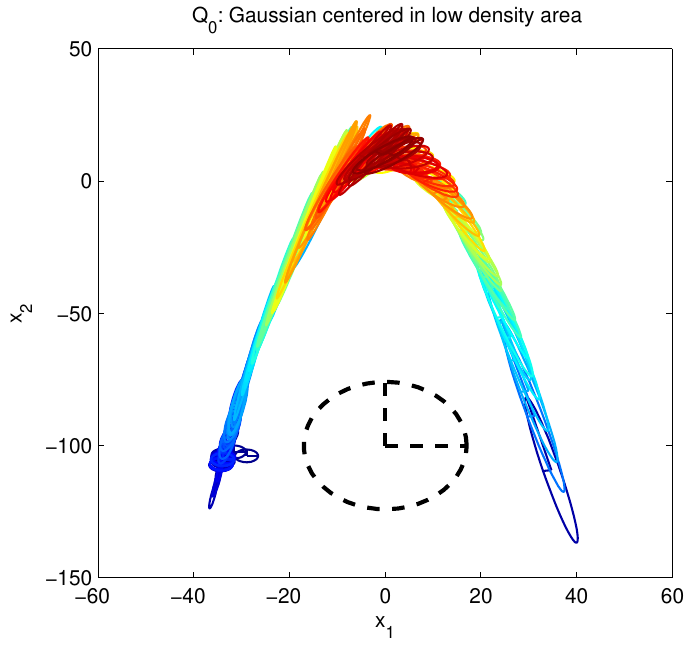}

\includegraphics[scale=.68]{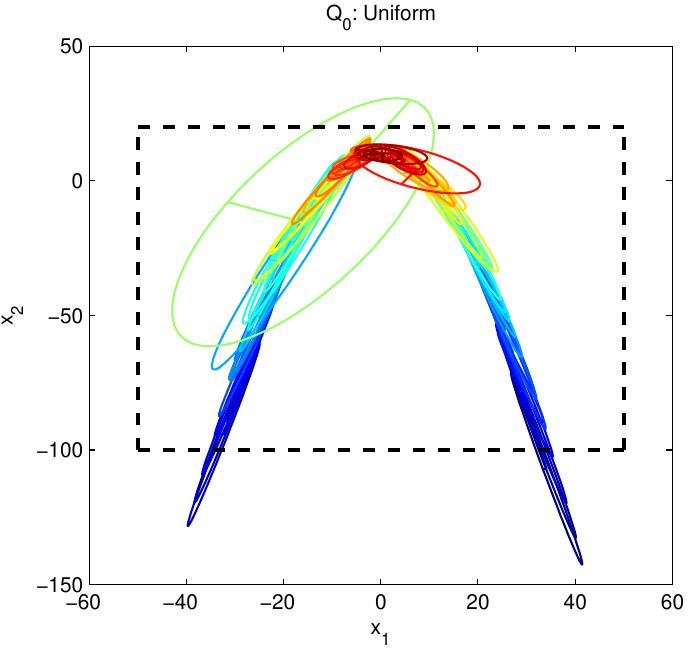}
\caption{Example \ref{example_pi2}: $\pi_2$ target, $d=2$ - Incremental mixture created by AIMM for three different initial proposals. Top row: $Q_0$ is a Gaussian density
(the region inside the black dashed ellipses contains $75\%$ of the Gaussian mass) centred on a high density region (left) and a low density region (right). Bottom row: $Q_0$ is Uniform (with support corresponding to the black dashed rectangle). The components $\phi_1,\ldots,\phi_{M_n}$ of the incremental mixture obtained after $n=100,000$ MCMC iterations are represented through (the region inside the ellipses contains $75\%$ of each Gaussian mass). The color of each ellipse illustrates the corresponding component's relative weight $\beta_\ell$ (from dark blue for lower weights to red).\label{fig:banana_defensive}}
\end{figure}

With limited prior knowledge of $\targ$, choosing $Q_0$ can be challenging. Here we consider three defensive distributions $Q_0$, represented by the black dashed contours in Figure \ref{fig:banana_defensive}. The first two are Gaussian, respectively, located close to the mode and in a nearly zero probability area and the last one is the uniform distribution on the set $\mathfrak{S}=\{x\in\Xset\,,x_1\in(-50,50),x_2\in(-100,20)\}$. Table~\ref{tab:banana_defensive} and Figure~\ref{fig:banana_defensive} show that AIMM is robust with respect to the choice of $Q_0$. Even when $Q_0$ yields a significant mismatch with $\targ_2$ (second case), the incremental mixture reaches high density areas, progressively uncovering higher density regions. The price to pay for a poorly chosen $Q_0$ is that more components are needed in the incremental mixture to match $\targ_2$, as shown by Table \ref{tab:banana_defensive}. The other statistics reported in Table \ref{tab:banana_defensive} are reasonably similar for the three choices of $Q_0$.

\begin{table}[h!]
\centering
\caption{Example \ref{example_pi2}: $\pi_2$ target, $d=2$ - Influence of the initial proposal $Q_0$ on AIMM outcome after $n=100,000$ MCMC iterations (replicated 20 times). $M_n$ is the number of components created by AIMM, $\ESS$ is the effective sample size, $\ACC$ is the acceptance rate, $\KL$ is the KL divergence between $\targ_2$ and the chain distribution, $\JMP$ is the average distance between two consecutive states of the chain and $\EFF$ is a time-normalized ESS. \label{tab:banana_defensive}}
\begin{tabular}{c||c|c|c|c|c|c}
$Q_0$ & $M_n$ & $\ESS$ & $\ACC$ & $\KL$ & $\JMP$ & $\EFF$ ($\times 10^{-4}$)\\
\hline
Gaussian on a high density region & 41 & .19 & .35 & .59  & 197 & 11\\
Gaussian on a low density region & 157 & .23 & .37 & .71  & 245 & 4\\
Uniform on $\mathfrak{S}$ & 39 & .24 & .38 & .62 & 320 & 11
\end{tabular}
\end{table}

\subsubsection{Influence of the threshold}
%The threshold $\Wst$ controls the number of kernels $M_n$ created by AIMM, and hence its computational efficiency.

\begin{figure}
\centering

\begin{subfigure}{\textwidth}
\centering
\includegraphics[scale=.75]{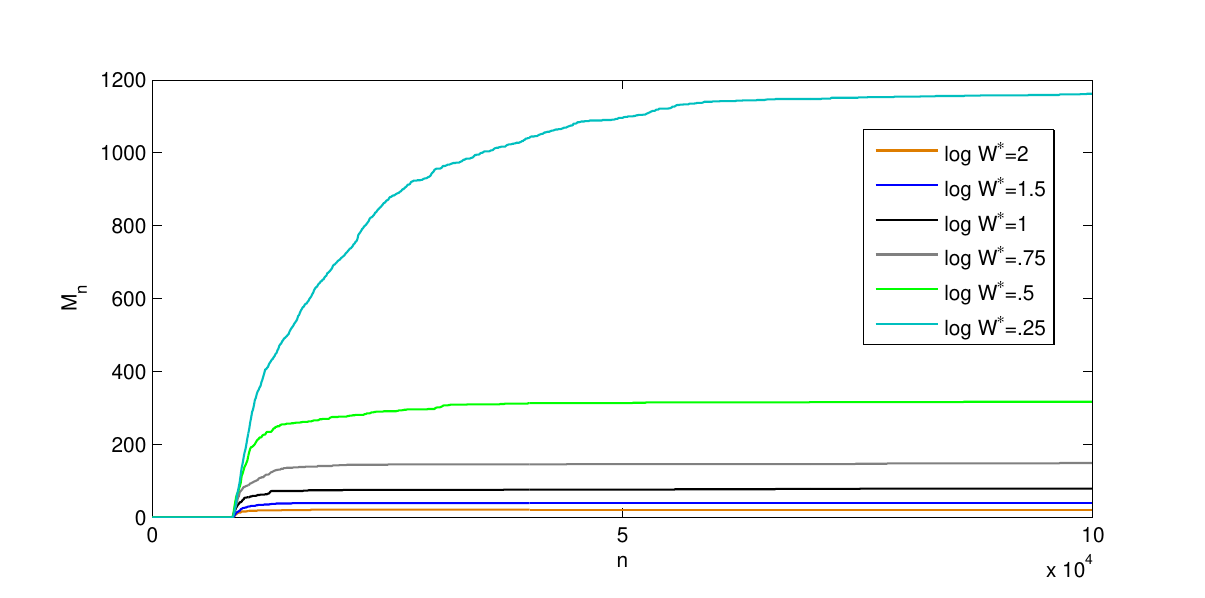}
\caption{Evolution of the number of kernels $M_n$ created by AIMM for different thresholds.\label{fig:mixture_devlopment_a}}
\end{subfigure}

\begin{subfigure}{\textwidth}
\centering
\includegraphics[scale=.75]{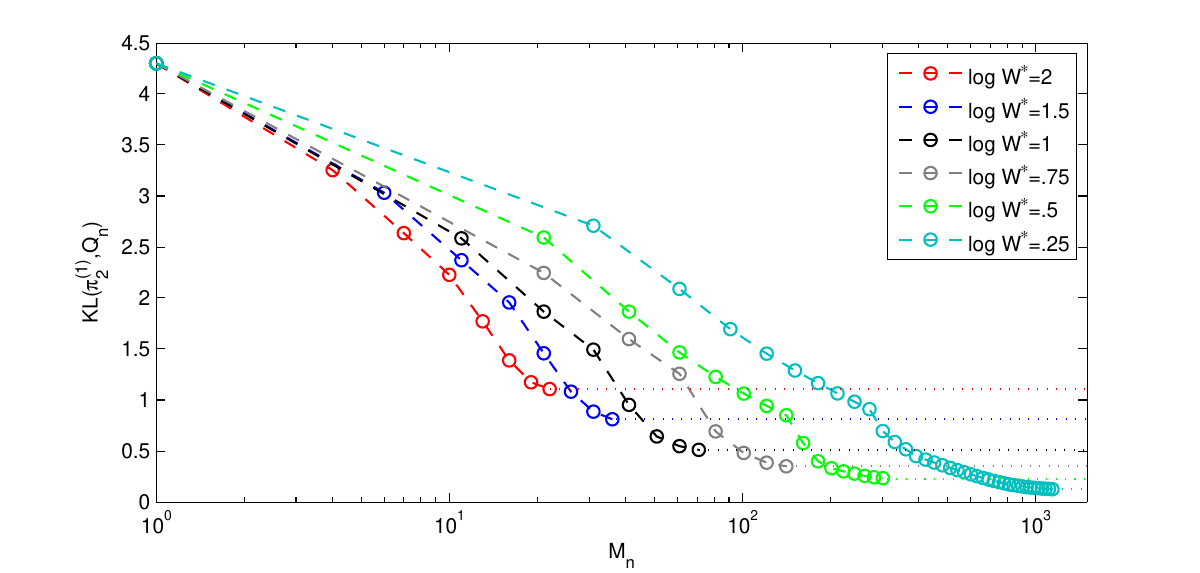}
\caption{Evolution of the KL divergence between $\targ_2$ and the incremental proposal $Q_n$, plotted in lin-log scale. \label{fig:mixture_devlopment_b}}
\end{subfigure}

\caption{Example \ref{example_pi2}:  $\pi_2$ target, $d=2$ - AIMM's incremental mixture design after $n=100,000$ MCMC iterations.}
\label{fig:mixture_devlopment}
\end{figure}

Even though we recommend the default setting $\Wst= d$, we analyse the performances of AIMM for different values of $\Wst$. We consider the same setup as in the previous subsection with the uniform defensive distribution $Q_0$ and report the outcome of the AIMM algorithm in Table~\ref{tab:banana_threshold_a}. As expected, the lower the threshold $\Wst$, the larger the number of kernels and the better is the mixing. In all cases, the adaptive transition kernel eventually stabilizes (faster if $\Wst$ is large), as illustrated by Figure~\ref{fig:mixture_devlopment_a}, resulting from the fact that the event $\event_n=\{W_n(\tX_{n+1})>\Wst\,,\tX_{n+1}\sim Q_n \}$ occurs less often as $n$ increases. %Figure~\ref{fig:mixture_devlopment_b} shows that the event $\{\KL(\targ,Q_n)>\KL(\targ,Q_{n+1})\}$ holds with a probability that increases with $n$, implying that the increment process progressively improves the proposal distribution.
As $\Wst$ decreases, the KL divergence between $\targ_2$ and the chain reduces while the CPU cost increases since more components are created. Therefore, the sampling efficiency is best for an intermediate threshold such as $\log\Wst=.75$. Finally, when the threshold is too low, the distribution of the chain converges more slowly to $\targ_2$ ; see e.g.~Table~\ref{tab:banana_threshold_a} where the $\KL$ indicator (defined as the K--L divergence between $\targ_2$ and the sample path of the AIMM chain; see Appendix \ref{app5}) is larger for $\log\Wst=0.25$ than for $\log\Wst=0.5$. Indeed when $\Wst\ll 1$, too many kernels are created to support high density areas and this slows down the exploration process of the lower density regions.%\footnote{The beginning of this section says that the default
%choice is $\bar{W} = d$, but this subsubsection describes and explores a
%completely different choice. This contradiction needs to be resolved,
%or at least explained. - AER, March 10, 2018}
%AM this note should be removed!

\begin{table}
\centering
\caption{Example \ref{example_pi2}: $\pi_2$ target, $d=2$ - Influence of the threshold $\Wst$ on AIMM and f-AIMM outcomes after $n=100,000$ MCMC iterations (replicated 20 times).\label{tab:banana_threshold}}
\begin{subtable}{\textwidth}
\centering
\caption{AIMM\label{tab:banana_threshold_a}}
\begin{tabular}{c||c|c|c|c|c|c|c}
$\log\Wst$ & $M_n$ & $\ESS$ & $\ACC$ & $\KL$ & $\JMP$ & $\CPU$ & $\EFF$ $(\times 10^{-4})$\\
\hline
10  &  0    &  .03  & .01   &   14.06     & 10    &   45  &  2\\
%AM why do we have an \hline here?
%FM OK
2   & 21    &  .16  & .27   &   .72        & 169   & 154   &  10\\
1.5 & 39    &  .24  & .38   &   .62         & 235   & 245   &  10\\
1   & 79    &  .37  & .53   &   .54         & 331   & 330   &  12\\
.75 & 149   &  .48  & .64   &   .53         & 286   & 658   &  7\\
.5  & 317   &  .64  & .75   &   .51         & 451   & 2,199  &  3\\
.25 & 1162  &  .71  & .87   &   .54         & 505   & 6,201  &  1
\end{tabular}
\end{subtable}

\vspace{.5cm}

\begin{subtable}{\textwidth}
\centering
\caption{f--AIMM\label{tab:banana_threshold_b}}
\begin{tabular}{c|c||c|c|c|c|c|c}
$\log\Wst$ & $\Mmax$ & $\ESS$ & $\ACC$ & $\KL$ &  $\JMP$ & $\CPU$ & $\EFF$ $(\times 10^{-4})$\\
\hline
 1.5 & 25 &   .29 & .51 & .59 & 268 & 255 &  $11$\\
 .75 & 100 & .53 & .69 & .53 & 409 & 421 & $13$\\
 .5 & 200 & .67 & .80 & .51 & 474 & 1,151 & $6$ \\
\end{tabular}
\end{subtable}

\end{table}

\subsubsection{Speeding up AIMM}
The computational efficiency of AIMM depends, to a large extent, on the number of components added in the proposal distribution, $M_n$. For this reason we consider a slight modification of the original AIMM algorithm to limit the number of possible components, thereby improving the computational speed of the algorithm. This variant of the algorithm is outlined below and will be referred to as \textit{fast} AIMM and denoted by f--AIMM:
\begin{itemize}
\item Let $\Mmax$ be the maximal number of components allowed in the incremental mixture proposal. If $M_n>\Mmax$, only the last $\Mmax$ added components are retained, in a moving window fashion. This \textit{truncation} has two main advantages, (i) approximately linearizing the computational burden once $\Mmax$ is reached, and (ii) forgetting the first, often transient components used to jump to high density areas (see \eg the loose ellipses in Figure \ref{fig:banana_defensive}, especially the very visible ones in the bottom panel).
\item The threshold $\Wst$ is adapted at
the beginning
%AM the start
of the algorithm in order to get the incremental process started. If the initial proposal $Q_0$ misses $\targ$, then $\esp_0\{W_0(\tX)\}\ll 1$ and the initial threshold $\Wst_0$ should be set as low as possible, to start the adaptation. However, as soon as the proposal is incremented, $\esp_n\{W_n(\tX)\}$ increases and leaving the threshold at its initial level will result in $\proba_n\{W_n(\tX)>\Wst\}\approx 1$, i.e.~in adding too many irrelevant components. The threshold adaptation produces a sequence of thresholds $\{\Wst_n\}_n$ such that $Q_{n}\{W_n(X)>\Wst_n\}\approx 10^{-3}$. Since sampling from $Q_n$ (a mixture of Gaussian distributions), can be performed routinely, $\Wst_n$ is derived from a Monte Carlo estimation.
As
little
%AM no
precision is required, the Monte Carlo approximation should be rough in order to limit the computational burden generated by the threshold adaptation. Also, those samples used to set $\Wst_n$ can be recycled and serve as proposal states for the AIMM chain so that this automated threshold mechanism comes for free. The threshold adaptation stops when the adapted threshold $\Wst_n$ reaches a neighborhood of the prescribed one $\Wst$, here defined as $|\Wst_n-\Wst|<1$.
 \end{itemize}

Table \ref{tab:banana_threshold} shows that f--AIMM outperforms the original AIMM, with some setups being nearly twice as efficient; compare, e.g.~AIMM and f--AIMM with $\log \Wst=.5$ in terms of efficiency (last column). Beyond efficiency, comparing $\KL$ for a given number of mixture components ($M_n$), shows that $\targ_2$ is more quickly explored by f--AIMM than by AIMM.

\subsubsection{Comparison with other samplers}

We now compare f--AIMM with four other samplers of interest: the adaptive algorithms AMH and AGM and their non-adaptive counterparts, random walk Metropolis-Hastings (RWMH) and Independent Metropolis (IM). We consider $\targ_2$ in dimensions $d=2, 10, 40$. Table~\ref{tab:banana_comp} summarizes the simulation results and emphasizes the need to compare the asymptotic variance related statistics ($\ESS$, $\ACC$, $\JMP$) jointly with the convergence related statistic ($\KL$). Indeed, from the target in dimension $d=2$, one can see that AGM yields the best ESS but is very slow to converge to $\pi_2$ as it still misses a significant portion of the state space after $n=200,000$ iterations; see the $\KL$ column in Table~\ref{tab:banana_comp} and Appendix \ref{app5} (especially Figure~1 therein) where we shed light on AGM's performance in this situation.

AIMM seems to yield the best tradeoff between convergence and variance in all dimensions. Inevitably, as $d$ increases, ESS shrinks and KL increases for all algorithms but AIMM still maintains a competitive advantage over the other approaches. It is worth noting that, if computation is not an issue, AIMM is able to reach a very large ESS (almost one), while having converged to the target. In such a situation, AIMM is essentially simulating \iid draws from $\pi_2$.

\begin{figure}
\centering
\includegraphics[scale=.9]{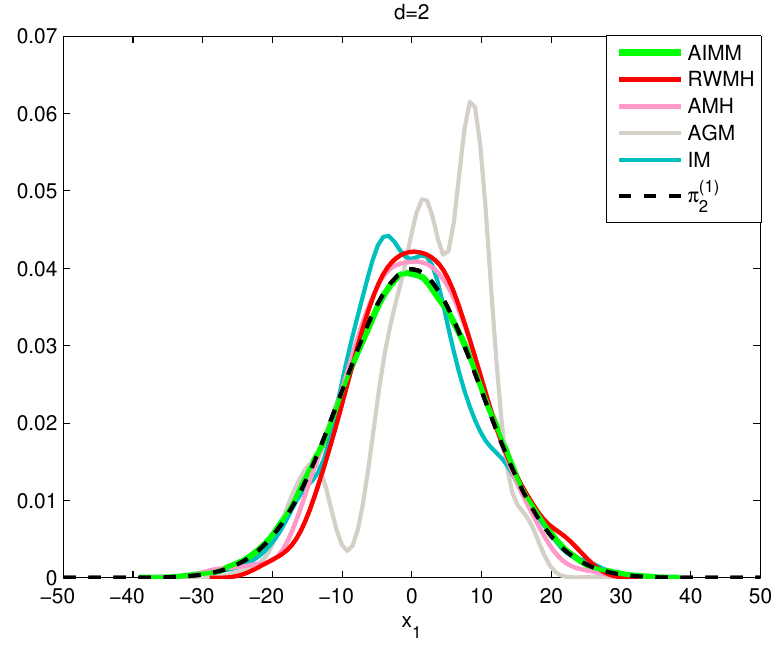}
\includegraphics[scale=.9]{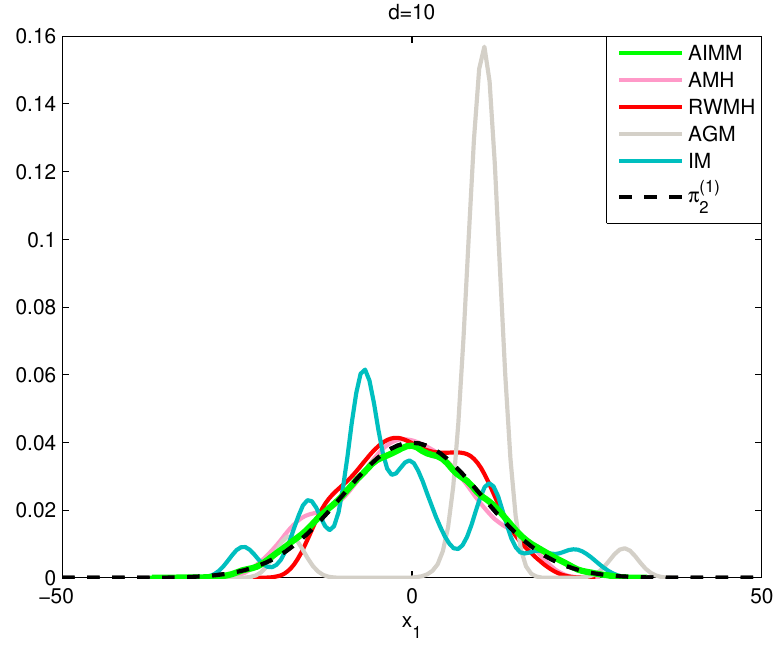}
\caption{Example \ref{example_pi2}: $\pi_2$ target, $d\in\{2,10\}$ - Kernel approximation of the marginal density $\targ_2^{(1)}$ provided by $n=200,000$ samples of the five (Markov) chains AIMM, AMH, RWMH, AGM and IM for the banana shape target example in dimensions $d=2$ (top) and $d=10$ (bottom). \label{fig:banana_marg}}
\end{figure}

Figure~\ref{fig:banana_marg} compares the kernel density approximation of the (Markov) chain marginal distribution of the first component for the five algorithms, with the true marginal $\targ_2^{(1)}=\norm(0,10)$. AIMM converges to $\pi_2$ faster than the other samplers, which need much more than $n=200,000$ iterations to converge. The bumpy shape of the AGM and IM marginal densities is caused by the accumulation of samples in some locations and reflects the slow convergence of these algorithms.

The other four algorithms struggle to visit the tail of the distribution,
see Table~\ref{tab:banana_tail}. Thus AIMM is doing better than (i) the non-independent samplers (RWMH and AMH) that manage to explore the tail of a target distribution but need a large number of transitions to return there, and (ii) the independence samplers (IM and AGM) that can quickly explore different areas but fail to venture into the lower density regions. An animation corresponding to the exploration of $\pi_2$ in dimension 2 by AIMM, AGM and AMH can be found at \href{http://mathsci.ucd.ie/~fmaire/AIMM/banana.html}{this link}.

\begin{table}
\centering

\caption{Example \ref{example_pi2}: $\pi_2$ target, $d\in\{2,10,40\}$ - Comparison of f--AIMM with RWMH, AMH, AGM and IM. Statistics were estimated from 20 replications of the five corresponding (Markov) chains after $n=200,000$ iterations. AGM fails to give sensible results for $d=40$. $\dag$: the ESS column are $\times 10^{-3}$ ; $\ddag$: the EFF column are $\times 10^{-6}$.
 \label{tab:banana_comp}}

 \vspace{-.5cm}
\begin{subtable}{\textwidth}
\centering
\vspace{.5cm}
\caption{$\pi_2$ in dimension $d=2$\label{tab:banana_comp_dim2}}
\begin{tabular}{c|c|c||c|c|c|c|c|c}
&$\log\Wst$ & $\Mmax$ & $\ESS$ $^\dag$ & $\ACC$ & $\KL$ & $\JMP$ & $\CPU$ & $\EFF$ $^\ddag$ \\
\hline
f-AIMM & 1.5 & 25 &   290 & .51 & .59 & 268 & 355 &  $810$\\
f-AIMM & .5 & 200 & 670 & .80 & .51 & 474 & 1,751 & $380$ \\
\hline
\hline

\multicolumn{2}{c|}{RWMH}& -- &  1 & .30 & 3.75 & 2 & 48 & $17$ \\
\multicolumn{2}{c|}{AMH} & -- & 4 & .23 & 2.69 & 7 & 60 & $63$ \\
\multicolumn{2}{c|}{AGM}
& 100 & 800 & .78 & 29.2 & 155 & 2,670 & $230$\\
\multicolumn{2}{c|}{IM}& -- & 6 & .01 & 10.1 & 8 & 92 & $63$
\end{tabular}
\end{subtable}

\begin{subtable}{\textwidth}
\centering
\vspace{.5cm}
\caption{$\pi_2$ in dimension $d=10$\label{tab:banana_comp_dim10}}
\begin{tabular}{c|c|c||c|c|c|c|c|c}
&$\log\Wst$ & $\Mmax$ & $\ESS^\dag$  & $\ACC$ & $\KL$ & $\JMP$ & $\CPU$ & $\EFF^\ddag$ \\
\hline
f-AIMM & 3 &   50 & 110 & .27 & 1.46 & 135 & 1,160 &  $94$\\
f-AIMM & 2.5 &  150 & 170 & .37  & .84 & 205 & 1,638 & $100$ \\
\hline
\hline
\multicolumn{2}{c|}{RWMH} & -- & $1$ & .22 & 9.63 & 1.1 & 61 & $15$\\
\multicolumn{2}{c|}{AMH} & -- & $1.4$ & .17 & 5.29 & 2.5 & 143 & $10$ \\
\multicolumn{2}{c|}{AGM}
& 100 & 600 & .67 &  119 & 153 & 4,909 &  $120$\\
\multicolumn{2}{c|}{IM} & -- & $.7$ & $10^{-3}$ & 97  & .9 & 316 & $2$
\end{tabular}
\end{subtable}

\begin{subtable}{\textwidth}
\centering
\vspace{.5cm}
\caption{$\pi_2$ in dimension $d=40$\label{tab:banana_comp_dim40}}
\begin{tabular}{c|c|c||c|c|c|c|c|c}
&$\log\Wst$ & $\Mmax$ & $\ESS^\dag$  & $\ACC$ & $\KL$ &  $\JMP$ & $\CPU$ & $\EFF^\ddag$\\
\hline
f-AIMM & 5 & 200 & 4 & .04 & 12.1 & 24 & 4,250 & $.7$\\
f-AIMM & 4 & 1,000 & 17 &.08 & 5.9 & 61 & 10,012 & $1.8$\\
\hline
\hline
\multicolumn{2}{c|}{RWMH} & -- & .8  & .24 & 11.6 & .8 & 71 & $11$\\
\multicolumn{2}{c|}{AMH} & -- &  .8 & .13 & 6.7 & 1.5 & 1,534& $.7$\\
\multicolumn{2}{c|}{IM} & -- & .5 & $10^{-4}$ & 147.1 & .03 & 1,270 & $.6$
\end{tabular}
\end{subtable}
\end{table}

\begin{table}
\caption{Example \ref{example_pi2}: $\pi_2$ target, $d\in\{2,10,40\}$ - Tail exploration achieved by the five different samplers. Tail events are defined as $\{\text{TAI}_1\}=\{X^{(2)}<-28.6\}$ and $\{\text{TAI}_2\}=\{X^{(2)}<-68.5\}$ so that $\pi_2\{\text{TAI}_1\}=0.05$ and $\pi_2\{\text{TAI}_2\}=0.005$. $\{\text{RET}_1\}$ and $\{\text{RET}_2\}$ corresponds to the expected returning time to the tail events $\{\text{TAI}_1\}$ and $\{\text{TAI}_2\}$, respectively. Estimation of these statistics was achieved with $n=200,000$ iterations of the corresponding samplers, repeated $20$ times. \label{tab:banana_tail}}

\centering
\begin{tabular}{ccc||c|c|c|c}
& & $\Mmax $ & $\text{TAI}_1$ & $\text{RET}_1$ & $\text{TAI}_2$ & $\text{RET}_2$ \\
\hline
\hline
\multirow{7}{*}{$d=2$} &  f--AIMM & 25 &   .05 & 43 &  .003 &  557 \\
& f--AIMM & 200 &  .05 & 23 &  .005 & 281  \\
& RWMH & -- &.04 & 2,945 & .002 & 69,730 \\
& AMH & -- & .04 & 3,676 & .004 & 74,292\\
& AGM & 100 &  .01 & 1,023 & $5.0\,10^{-5}$ & $155,230$\\
& IM & -- & .05 & 1,148 & .004 & 8,885\\
\hline
\hline
\multirow{4}{*}{$d=10$} & f--AIMM & 50 &  .04 & 98  & .001 & 5,463 \\
& f--AIMM & 150 &  .05 & 57  & .002 & 1,708 \\
& RWMH & -- & .03 & 30,004 & 0 & $\infty$ \\
& AMH & -- &  .04 &  4,081   &  .001 & 166,103\\
& AGM & 100 & .04 & 120,022 &.001  & 180,119 \\
& IM & -- & .05 & 13,140 & .003  & 103,454\\
\hline
\hline
\multirow{4}{*}{$d=40$} & f--AIMM & 200 & .01 & 1,350 & 0 & $\infty$\\
& f--AIMM & 1000 & .05 & 175 & 2,0 $10^{-4}$   & 16,230 \\
& RWMH & -- &.002 & 158,000 & 0 & $\infty$\\
& AMH &--& 0 & $\infty$ & 0 & $\infty$\\
& IM & -- & .006 & 28,552 & 0 & $\infty$
\end{tabular}
\end{table}

\subsection{Ridge like example}

\begin{example}
\label{example_pi3}
Let $\Xset=\rset^6$ and $\targ_3(x)\propto \Psi_6(x;\mu_i,\Gamma_i)\Psi_4(g(x);\mu_o,\Gamma_o)$, where $\Psi_{d}(\,\cdot\,;\,m,S)$ is the $d$-dimensional Gaussian density with mean $m$ and  covariance matrix $S$. The target parameters $(\mu_i,\mu_o)\in\rset^6\times\rset^4$ and $(\Gamma_i,\Gamma_o)\in\mathcal{M}_6(\rset)\times\mathcal{M}_4(\rset)$ are, respectively, known means and covariance matrices and $g$ is a nonlinear deterministic mapping $\rset^6\to\rset^4$ defined as:
\begin{equation*}
g(x_1,\ldots,x_6)=\left\{
\begin{array}{l}
\prod_{i=1}^6 x_i\,,\\
x_2x_4\,,\\
x_1\slash x_5\,,\\
x_3x_6\,.
\end{array}
\right.
\end{equation*}
In this context, $\Psi_6(\,\cdot\,;\mu_i,\Gamma_i)$ can be regarded as a prior distribution and $\Psi_4(\,\cdot\,;\mu_o,\Gamma_o)$ as a likelihood, the observations being some functional of the hidden parameter $x\in\rset^6$.
\end{example}

\begin{figure}
\centering
\includegraphics[scale=.45]{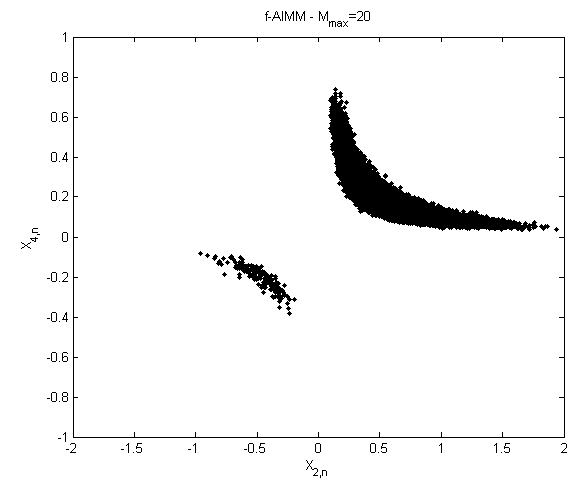}
\includegraphics[scale=.45]{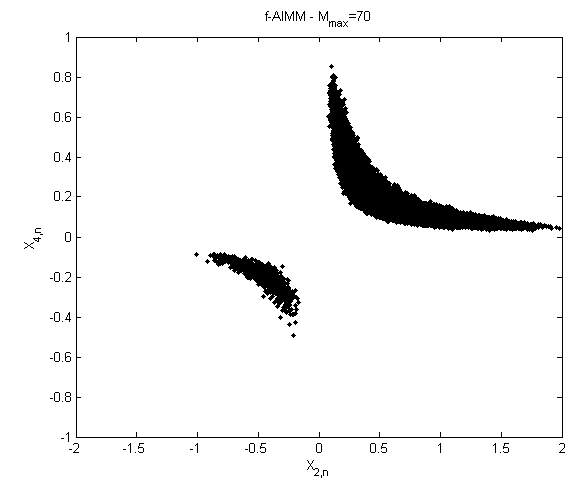}

\includegraphics[scale=.45]{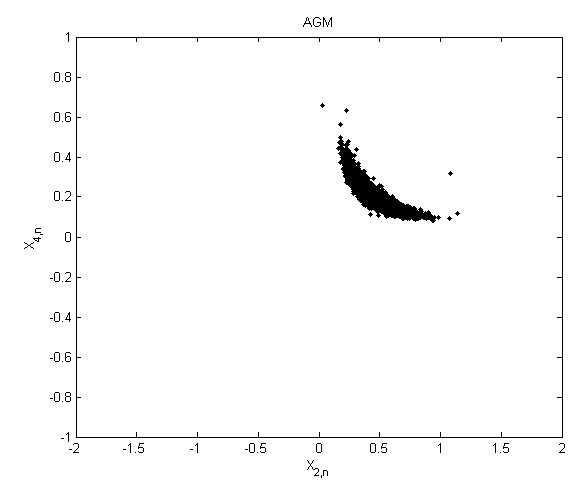}
\includegraphics[scale=.45]{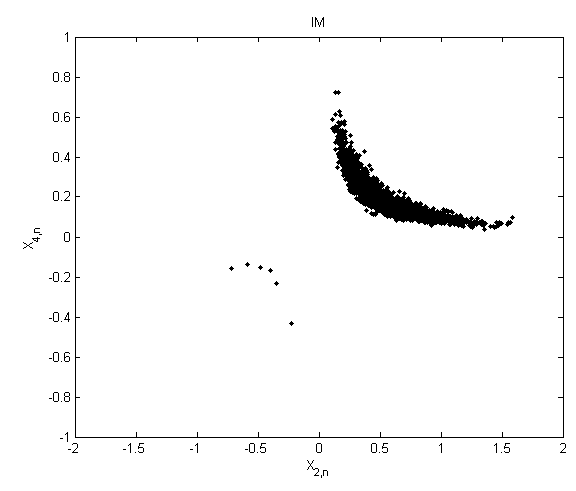}

\includegraphics[scale=.45]{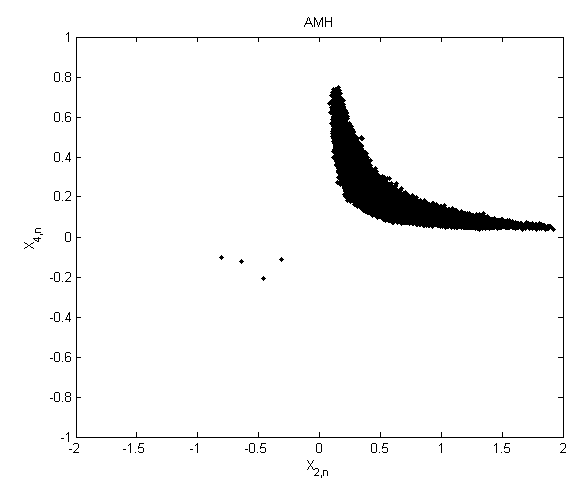}
\includegraphics[scale=.45]{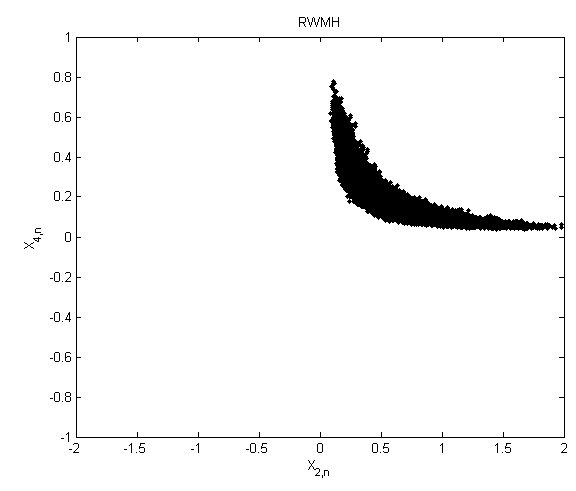}

\caption{Example \ref{example_pi3}: $\pi_3$ target, $d=6$ - Samples $\{(X_{2,n},X_{4,n}),\,n\leq 200,000\}$ from the five different algorithms (from left to right, top to bottom): AIMM in two settings $\Mmax=20$ and $\Mmax=70$, AGM, IM, AMH and RWMH. \label{fig:ridge_samples}}
\end{figure}

Such a target distribution often arises in physics. Similar target distributions often arise in Bayesian inference for deterministic mechanistic models and in other areas of science, engineering and environmental sciences; see \eg \cite{poole2000inference,bates2003bayesian}.
%AM which models are these?
They are hard to sample from because the probability mass is concentrated around thin curved manifolds. We compare f--AIMM with RWMH, AMH, AGM and IM first in terms of their mixing properties; see Table \ref{tab:ridge_comp}. We also ensure that the different methods agree on the mass localisation by plotting a pairwise marginal; see Figure \ref{fig:ridge_samples}. In this example, f--AIMM clearly outperforms the four other methods in terms of both convergence and variance statistics. We observe in Figure \ref{fig:ridge_samples}, that f--AIMM is the only sampler able to discover a secondary mode in the marginal distribution of the second and fourth target components $(X_2,X_4)$.

\begin{table}
\centering
\caption{Example \ref{example_pi3}: $\pi_3$ target, $d=6$ - Comparison of f--AIMM with the four other samplers after $n=200,000$ iterations (replicated $10$ times).\label{tab:ridge_comp}}
\begin{tabular}{c|c|c||c|c|c|c|c}
&$\Wst$ & $\Mmax$ & $\ACC$ & $\ESS$ & $\CPU$ & $\EFF$ & $\JMP$\\
\hline
f--AIMM & 10 & 20 &  .049 & .015  & 274 & $5.2\,10^{-5}$ & .08\\
f--AIMM & 1 & 70 & .169 & .089 & 517 & $1.7\,10^{-4} $& .27 \\
f--AIMM & .1 & 110 & .251 & .156 & 818 & $1.9\,10^{-4} $ & .38 \\
\hline
\hline
\multicolumn{2}{c|}{RWMH} & -- & .23 & .001 & 109 & $9.2\,10^{-6}$ & .007\\
\multicolumn{2}{c|}{AMH} & -- & .42 & .002 & 2,105 & $9.5\,10^{-7}$ & .004\\
\multicolumn{2}{c|}{AGM--MH} & 100 & .009 & .002 & 2,660 & $3.4\,10^{-6}$ & .003\\
\multicolumn{2}{c|}{IM} & -- & .003 & .001 & 199 & $5.5\,10^{-6}$ & .003
\end{tabular}
\end{table}

\subsection{Bimodal distribution}

\begin{example}
\label{example_pi4}
In this last example, $\targ_4$ is a posterior distribution defined on the state space $\Xset=\rset^d$, where $d\in\{4,10\}$. The likelihood is a mixture of two $d$-dimensional Gaussian distributions with weight $\lambda=0.5$, mean and covariance matrix as follows
\begin{eqnarray*}
&\mu_1=\textbf{0}_\text{d}\,,\qquad \Sigma_1= \text{AR}_\text{d}(-.95)\,,\\
&\mu_2=\textbf{9}_\text{d}\,,\qquad \Sigma_2= \text{AR}_\text{d}(.95)\,,
\end{eqnarray*}
where for all $\rho>0$, $\text{AR}_\text{d}(\rho)$ is the $d$-dimensional first order autoregressive matrix whose coefficients are:
$$
\text{for}\;1\leq (i,j)\leq d\,,\qquad m_{i,j}(\rho)=\rho^{\max(i,j)-1}\,.
$$
The prior is the uniform distribution $\mathcal{U}([-3,12]^d)$. For f--AIMM and IM, $Q_0$ is set as this prior distribution, while for AGM, the centers of the
Gaussian components in the initial proposal kernel are drawn according to the same uniform distribution. For AMH, the initial covariance matrix $\Sigma_0$ is set to be
the identity matrix.
\end{example}

\begin{figure}
\centering
\includegraphics[scale=.5]{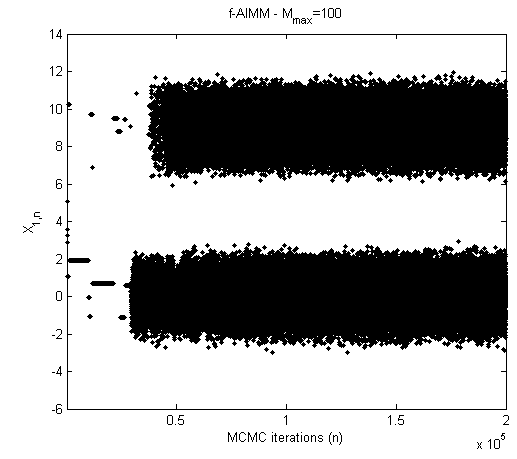}
\includegraphics[scale=.5]{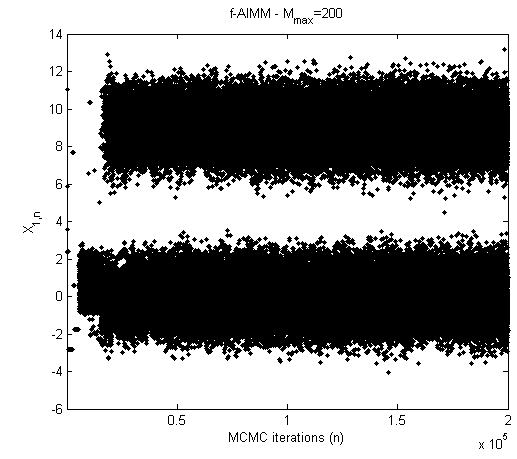}

\includegraphics[scale=.5]{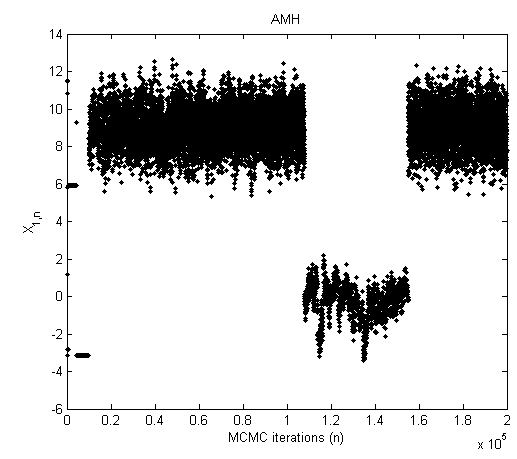}
\includegraphics[scale=.5]{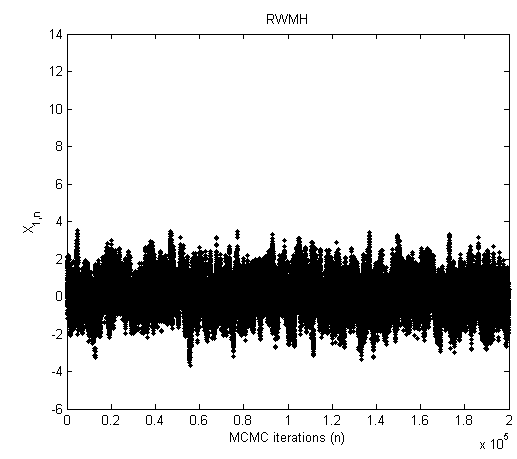}

\caption{Example \ref{example_pi4}: $\pi_4$ target, $d=4$ - Samples $\{X_{1,n},\,n\leq 200,000\}$ from four different algorithms (from left to right, top to bottom): f--AIMM in two settings $\Mmax=100$ and $\Mmax=200$, AMH and RWMH.
\label{fig:bimodal4_samples}}
\end{figure}

\begin{figure}
\centering
\includegraphics[scale=.6]{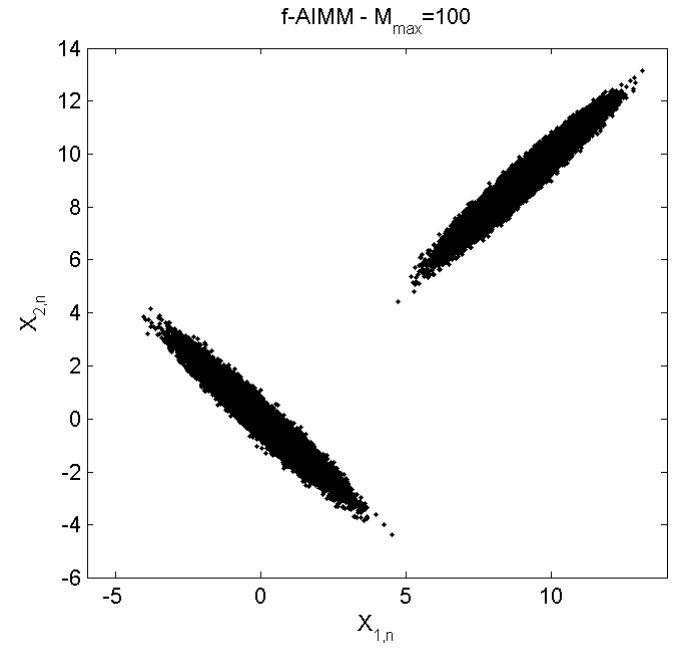}
\includegraphics[scale=.6]{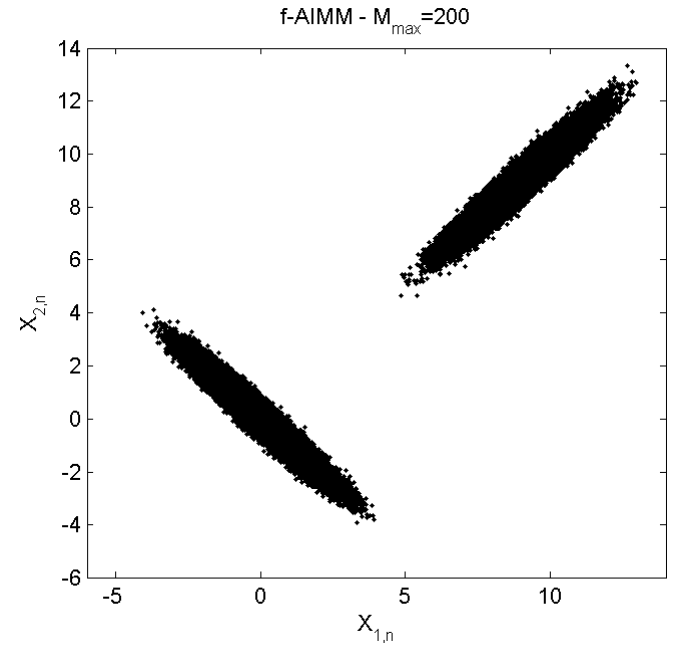}

\includegraphics[scale=.6]{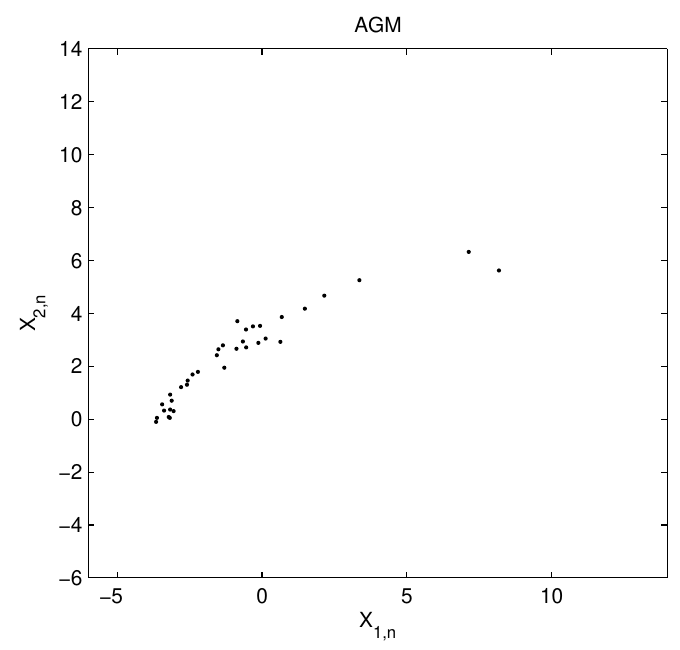}
\includegraphics[scale=.6]{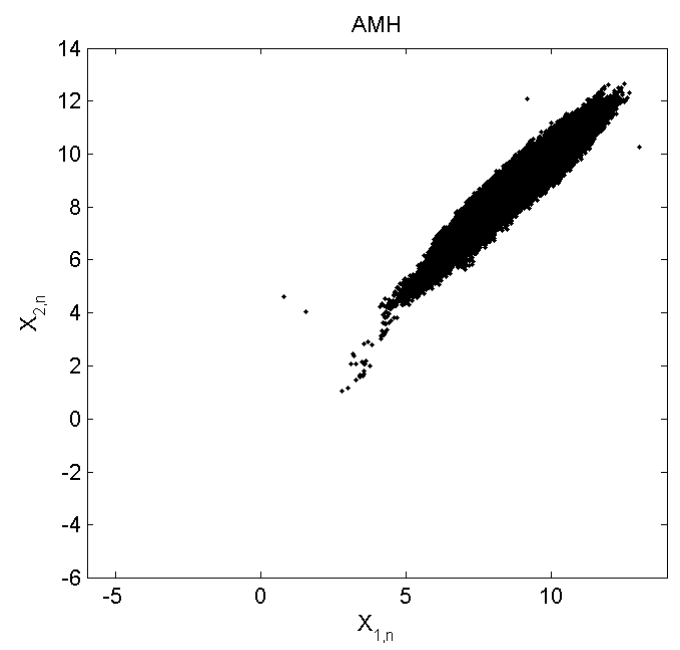}

\caption{Example \ref{example_pi4}: $\pi_4$ target, $d=10$ - Samples $\{(X_{1,n},X_{2,n}),\,n\leq 200,000\}$ from from four different algorithms (from left to right, top to bottom): f--AIMM in two settings $\Mmax=100$ and $\Mmax=200$, AGM and AMH. \label{fig:bimodal10_samples}}

\end{figure}

\begin{figure}
\centering
\includegraphics[scale=.6]{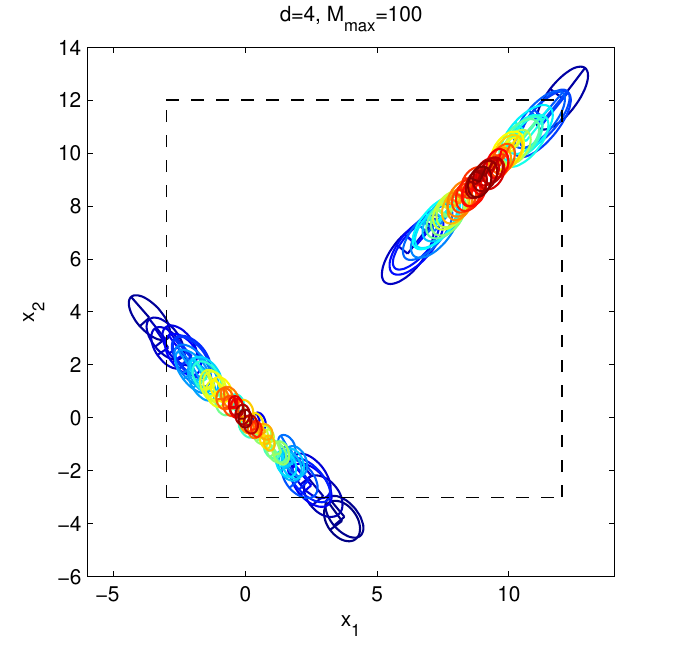}
\includegraphics[scale=.6]{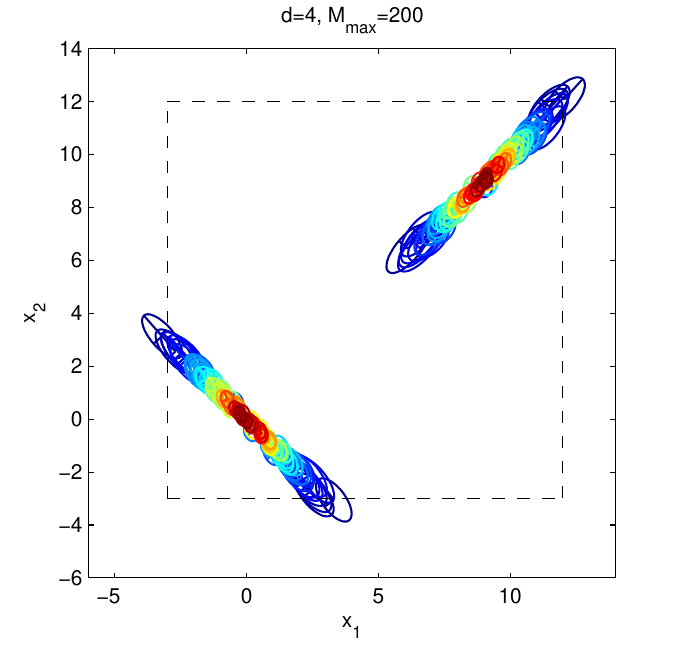}

\includegraphics[scale=.6]{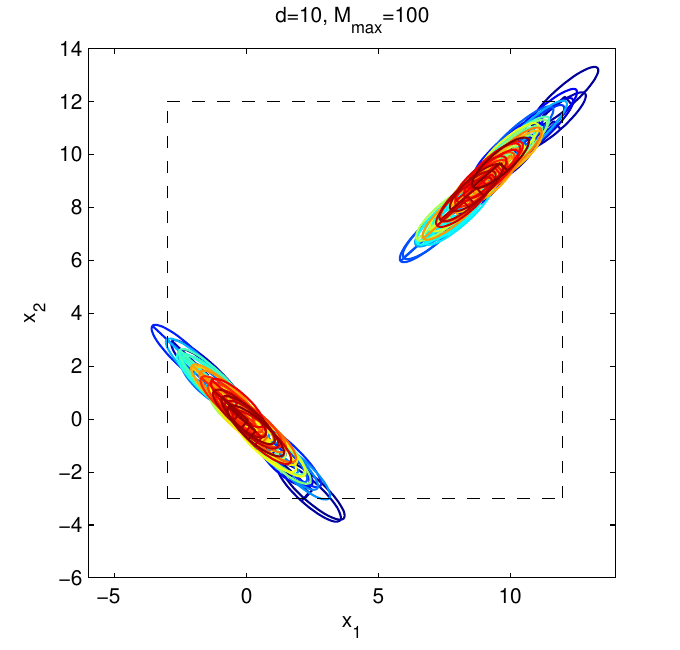}
\includegraphics[scale=.6]{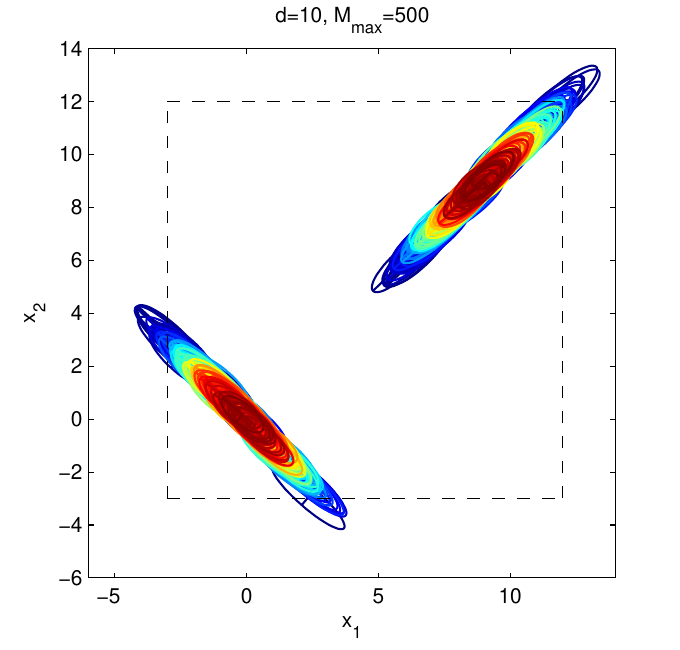}

\caption{(example \ref{example_pi4}) $\pi_4$ target, $d\in\{4,10\}$ - Incremental mixture (colored ellipses) and uniform initial proposal (black dashed rectangle) designed by f--AIMM in the bimodal target $\pi_4$, in dimensions $d=4$ (top row) and $d=10$ (bottom row). Distributions are projected onto the joint marginal space $(\Xset_1,\Xset_2)$. The colors stand for the weights $\{\beta_\ell\}_{\ell=1}^{M_n}$ attached to each component. The region inside the ellipses contains $75\%$ of each Gaussian mass. \label{fig:bimodal_inc}}
\end{figure}

Figures \ref{fig:bimodal4_samples} and \ref{fig:bimodal10_samples} illustrate the experiment in dimensions $d=4$ and $d=10$. On the one hand, in both setups $\Mmax=100$ and $\Mmax=200$, f--AIMM is more efficient at discovering and jumping from one mode to the other. Allowing more kernels in the incremental mixture proposal will result in faster convergence; compare the settings with $\Mmax=100$ and $\Mmax=200$ in Figure \ref{fig:bimodal_inc}.  On the other hand, because of the distance between the two modes, RWMH visits only one while AMH visits both but in an unbalanced fashion. Figure \ref{fig:bimodal10_samples} displays the samples from the joint marginal $(X_1,X_2)$ obtained through f--AIMM
(with two different values of $\Mmax$), AMH and RWMH. Increasing the dimension from $d=4$ to $d=10$ makes AMH unable to visit the two modes after $n=200,000$ iterations. As for AGM, the isolated samples reflect a failure to explore the state space. These facts are confirmed in Table~\ref{tab:bim_comp} which highlights the better mixing efficiency of f--AIMM relative to the four other algorithms and in Table~\ref{tab:bim_mixt} which shows that f--AIMM is the only method which, after $n=200,000$ iterations, visits the two modes with the correct proportion.

\begin{table}
\centering

\caption{Example \ref{example_pi4}: $\pi_4$ target, $d\in\{4,10\}$ - Comparison of f--AIMM with RWMH, AMH, AGM and IM after $n=200,000$ iterations (replicated $100$ times).
\label{tab:bim_comp}}
\begin{subtable}{\textwidth}
\centering
\vspace{.5cm}
\caption{$\pi_4$ in dimension $d=4$}
\begin{tabular}{c|c|c||c|c|c|c|c}
&$\Wst$ & $\Mmax$ & $\ACC$ & $\ESS$ & $\CPU$ & $\EFF$ & $\JMP$\\
\hline
f--AIMM & 5 & 100 & $.69$ & .30 & 408 & $7.3\,10^{-4}$ & 68\\
\hline
\hline
\multicolumn{2}{c|}{RWMH} & -- & .23 & $3.1\,10^{-3}$ & 93 & $3.4\,10^{-5}$ & .09\\
\multicolumn{2}{c|}{AMH} & -- & .26 & $2.7\,10^{-2}$ & 269 & $1.0\,10^{-4}$ & .64\\
\multicolumn{2}{c|}{AGM--MH} & 100 & $3.0\,10^{-3}$ & $5.0\,10^{-4}$ & 3,517 & $1.4\,10^{-7}$ & .29 \\
\multicolumn{2}{c|}{IM} & -- & $3.3\,10^{-4}$ & $5.9\,10^{-4}$ & 81 &  $7.3\,10^{-6}$ & .05
\end{tabular}

\end{subtable}

\vspace{1cm}

\begin{subtable}{\textwidth}
\centering
\vspace{.5cm}
\caption{$\pi_4$ in dimension $d=10$}
\begin{tabular}{c|c|c||c|c|c|c|c}
&$\Wst$ & $\Mmax$ & $\ACC$ & $\ESS$ & $\CPU$ & $\EFF$ & $\JMP$\\
\hline
f--AIMM & 20 & 100 & $.45$ & .18 & 1,232 & $1.4\,10^{-4}$ & 116.4\\
f--AIMM & 10 & 200 & $.64$ & .25 & 1,550 & $1.6\,10^{-4}$ & 200.1\\
\hline
\hline
\multicolumn{2}{c|}{RWMH} & -- & $.38$ & $7.2\,10^{-4}$ & 476 &  $1.7\,10^{-6}$ & .07 \\
\multicolumn{2}{c|}{AMH} & -- & $.18$ & $1.3\,10^{-3}$ & 2,601 & $5.0\,10^{-7}$ & .57 \\
\multicolumn{2}{c|}{AGM--MH} & 100 & $2.6\,10^{-4}$ & $5.0\,10^{-4}$ & 4,459 &  $1.1\,10^{-7}$ &  $7.1\,10^{-3}$\\
\multicolumn{2}{c|}{IM} & -- & $6.4\,10^{-5}$ & $6.1\,10^{-4}$ & 473 & $1.2\,10^{-6}$ & $4.1\,10^{-3}$
\end{tabular}

\end{subtable}
\end{table}

\begin{table}
\centering
\caption{Example \ref{example_pi4}: $\pi_4$ target, $d\in\{4,10\}$ - Mean Square Error (MSE) of the mixture parameter $\lambda$, for the different algorithms
(replicated $100$ times). \label{tab:bim_mixt}}
\begin{tabular}{c|c|c|| c | c}
&$\Wst$ & $\Mmax$ & MSE(${\lambda}$), $d=4$ & MSE(${\lambda}$), $d=10$\\
\hline
f--AIMM & 5 & 100 & .0001 & .06\\
f--AIMM & 20 & 100 & .0006 & $.09$\\
f--AIMM & 10 & 200 & .0024 & $.01$\\
\hline
\hline
\multicolumn{2}{c|}{RWMH} & -- & .25 & .25\\
\multicolumn{2}{c|}{AMH} & -- & .15 & .25\\
\multicolumn{2}{c|}{AGM--MH} & 100 & .22 &.25\\
\multicolumn{2}{c|}{IM} & -- & .03 & .20
\end{tabular}

\end{table}

\newpage
\section{Discussion}
\label{sec:disc}

Although implicitly evaluated in an (non adaptive) independence Metropolis--Hastings transition, the information conveyed by the ratio of importance weights is lost because of the threshold set to one in the M--H
acceptance probability. Indeed, while at $X_n$ and given two realizations of the proposal $Q$, say $\tX_{n+1}$ and $\tY_{n+1}$, the two events $\{W(\tX_{n+1})>W(X_n)\}$ and  $\{W(\tY_{n+1})\gg W(X_n)\}$ result in the
same transition, \ie the proposed move is accepted with probability one, regardless of the magnitude difference between $W(\tX_{n+1})/W(X_n)$ and $W(\tY_{n+1})/W(X_n)$. This paper aims to design an adaptive
MCMC algorithm that makes use of this information by incrementing the independence M--H proposal distribution in the latter case and not in the former.

The general methodology, referred to as Adaptive Incremental Mixture MCMC (AIMM), presented and studied in this paper is a novel adaptive Markov chain Monte Carlo method to sample from challenging distributions. Theoretically, we establish under very mild assumptions, that if it only adapts on a compact set and has bounded jumps, this algorithm generates an ergodic process for any distribution of interest. We show that these conditions can always be enforced by using a specific implementation. In simpler implementations where those conditions may not be verified, the algorithm is nevertheless shown to work well in practice. We provide an even more efficient algorithm, referred to as f-AIMM, that guarantees that the incremental proposal evaluation does not slow down the algorithm. This algorithm can be seen as a series of AIMM algorithms where Gaussian components are progressively dropped. As a consequence, f-AIMM is compatible with the theoretical framework developed at Section \ref{sec:conv}. In particular, provided that it only adapts on a compact set and has bounded jumps, this algorithm is invariant for any target distribution. We illustrate its performance in a variety of challenging sampling scenarios.

Compared to other existing adaptive MCMC methods, AIMM needs less prior knowledge of the target. Its strategy of incrementing an initial naive proposal distribution with Gaussian kernels leads to a fully adaptive exploration of the state space. Conversely, we have shown that in some examples the adaptiveness of some other MCMC samplers may be compromised when an unwise choice of parametric family for the proposal kernel is made. The performance of AIMM depends strongly on the threshold $\Wst$ which controls the adaptation rate. This parameter should be set according to the computational budget available. Based on our simulations, AIMM consistently yielded the best tradeoff between fast convergence and low variance.

The adaptive design of AIMM was inspired by Incremental Mixture Importance Sampling (IMIS) \citep{raftery2010}. IMIS iteratively samples and weights particles according to a sequence of importance distributions that adapt over time. The adaptation strategy is similar to that in AIMM: given a population of weighted particles, the next batch of particles is simulated by a Gaussian kernel centered at the particle having the largest importance weight. However IMIS and AIMM are structurally different since the former is an adaptive importance sampling method while the latter is an adaptive MCMC algorithm. Comparing them on a fair basis is difficult. In particular, the Effective Sample Size estimators for the two methods stem from different approximations. The computational efficiency of IMIS suffers from the fact that, at each iteration, the whole population of particles must be reweighted in order to maintain the consistency of the importance sampling estimator. By contrast, at each transition, AIMM evaluates only the importance weight of the new proposed state. However, since IMIS calculates the importance weight of large batches of particles, it acquires a knowledge of the state space more quickly than AIMM which accepts/rejects one particle at a time.

\begin{figure}
\centering
\includegraphics[scale=.8]{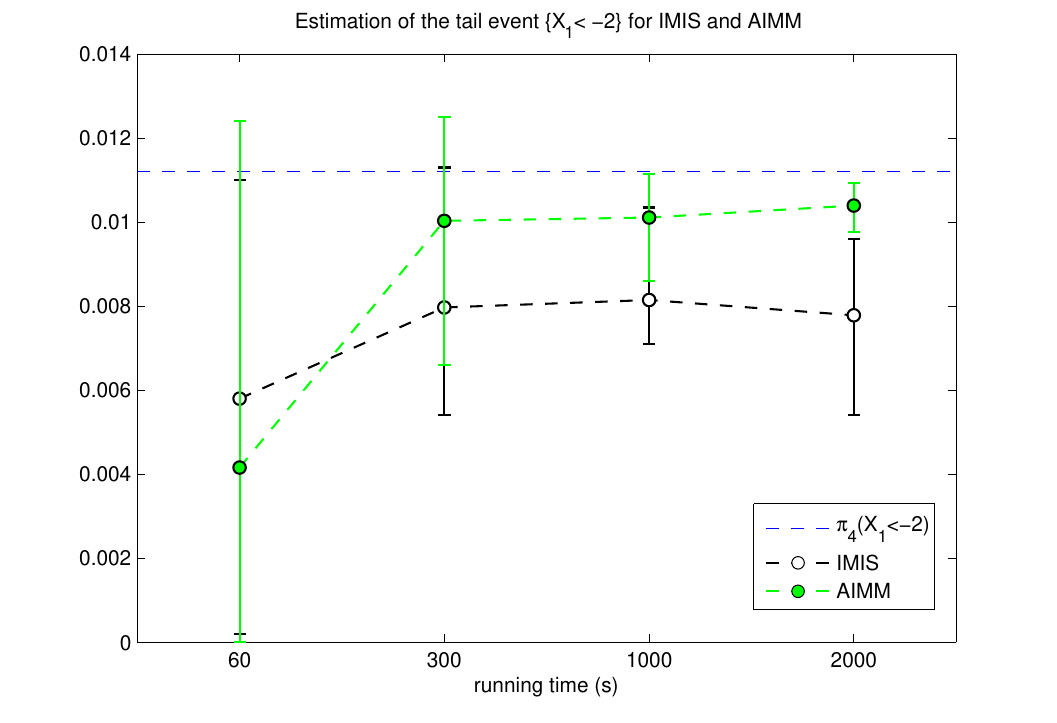}
\caption{Example \ref{example_pi4}: $\targ_4$ target, $d=4$ - Estimation of the probability $\pi_4(X_1<-2)$ (true value, 0.0112) obtained through AIMM and IMIS for different runtimes ($60 ; 300 ; 1,000 ; 2,000$, in seconds). \label{fig:imis_1}}
\end{figure}

\begin{figure}
\centering
\includegraphics[scale=.4]{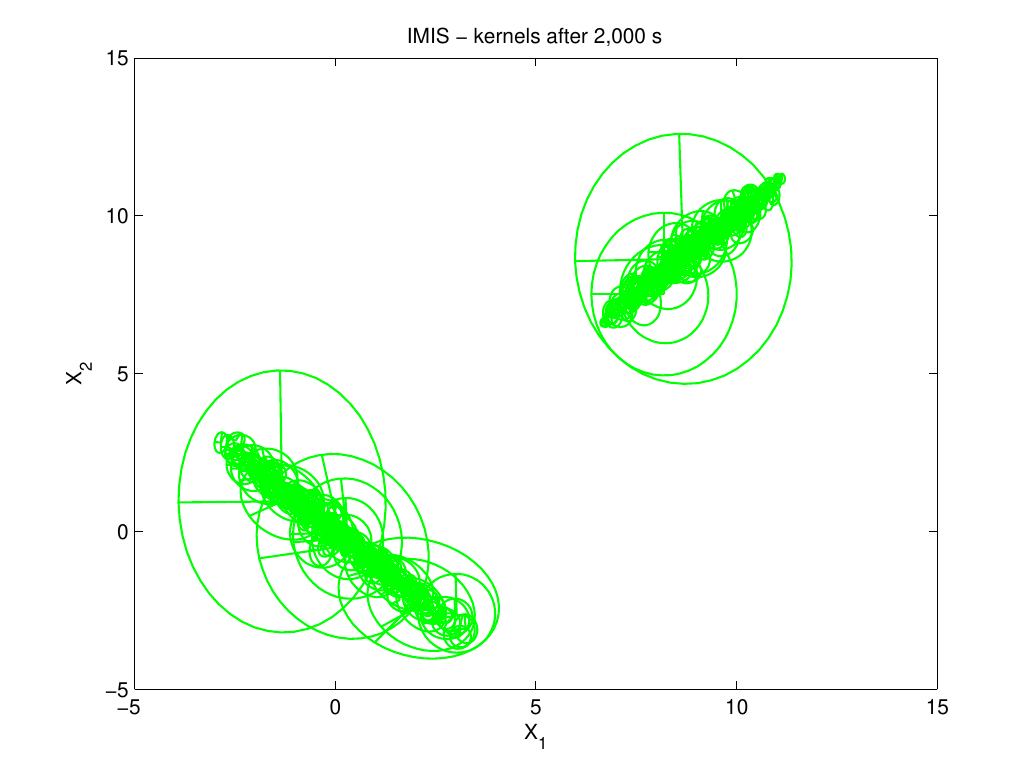}
\includegraphics[scale=.4]{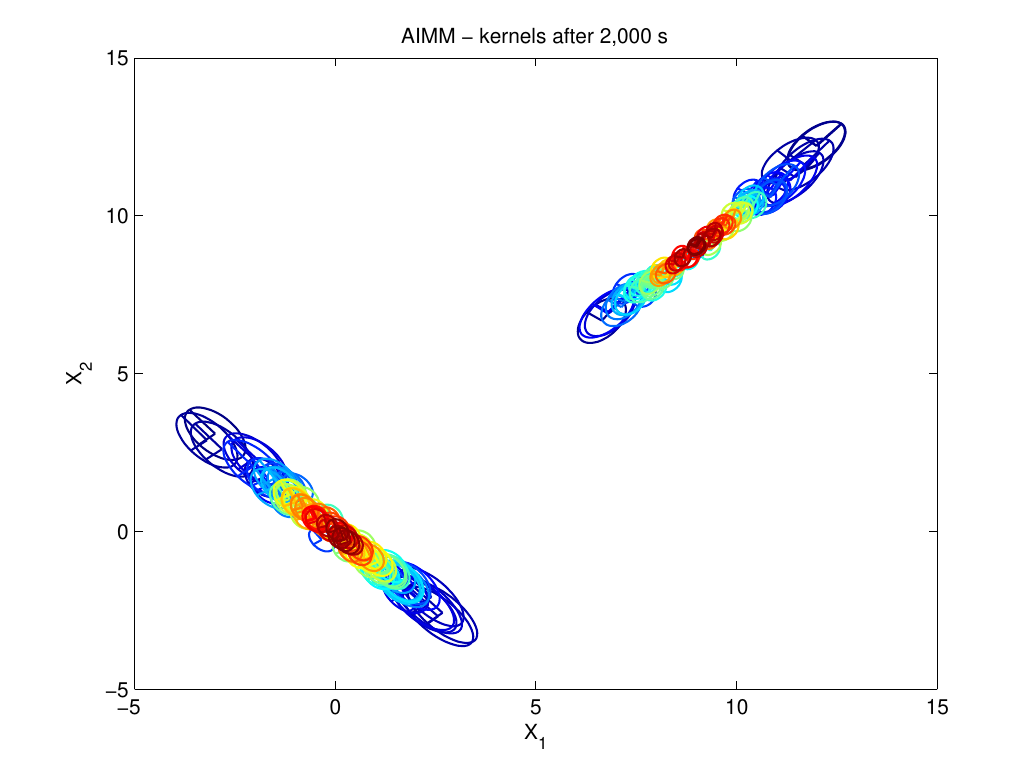}

\includegraphics[scale=.145]{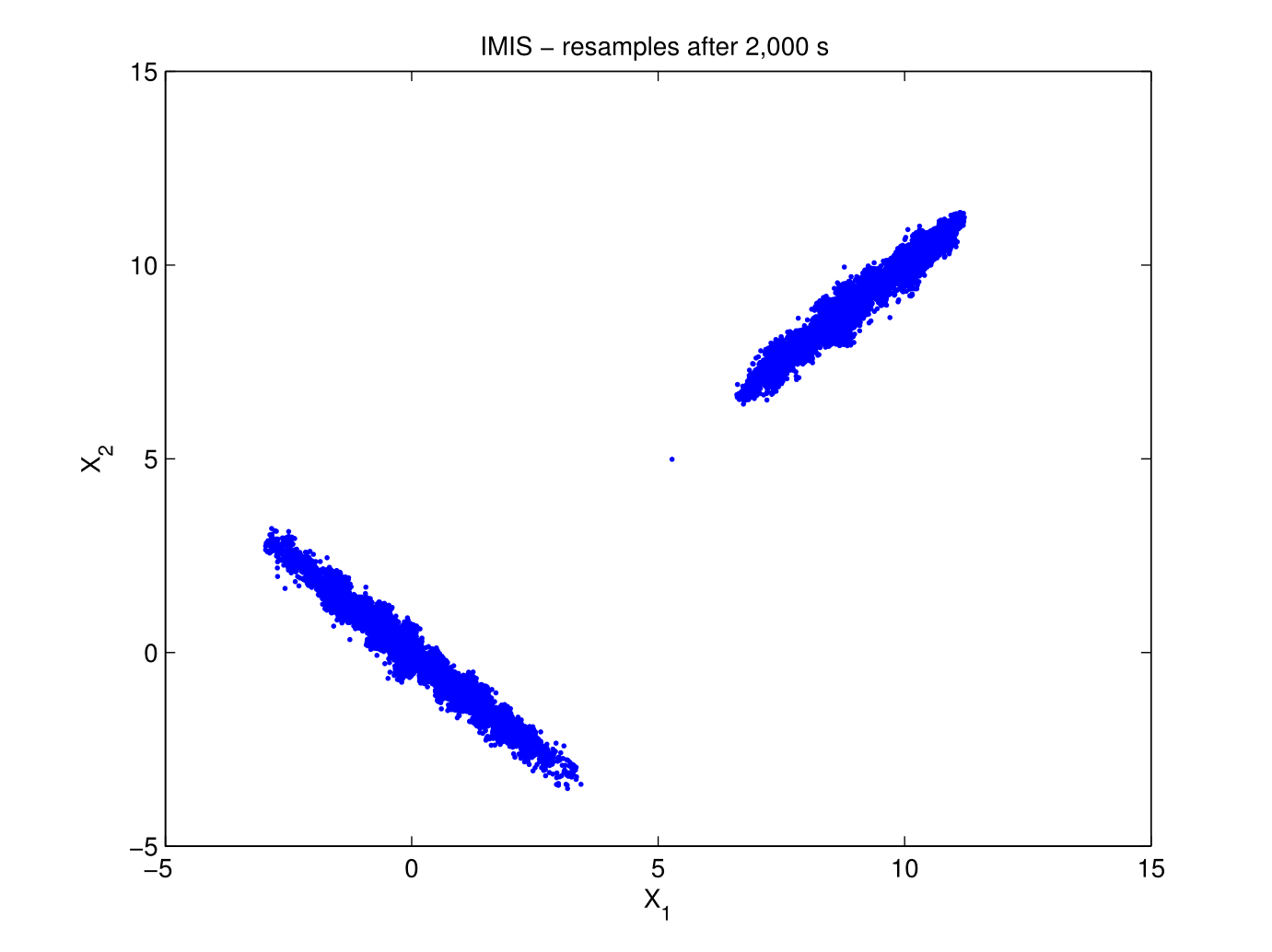}
\includegraphics[scale=.145]{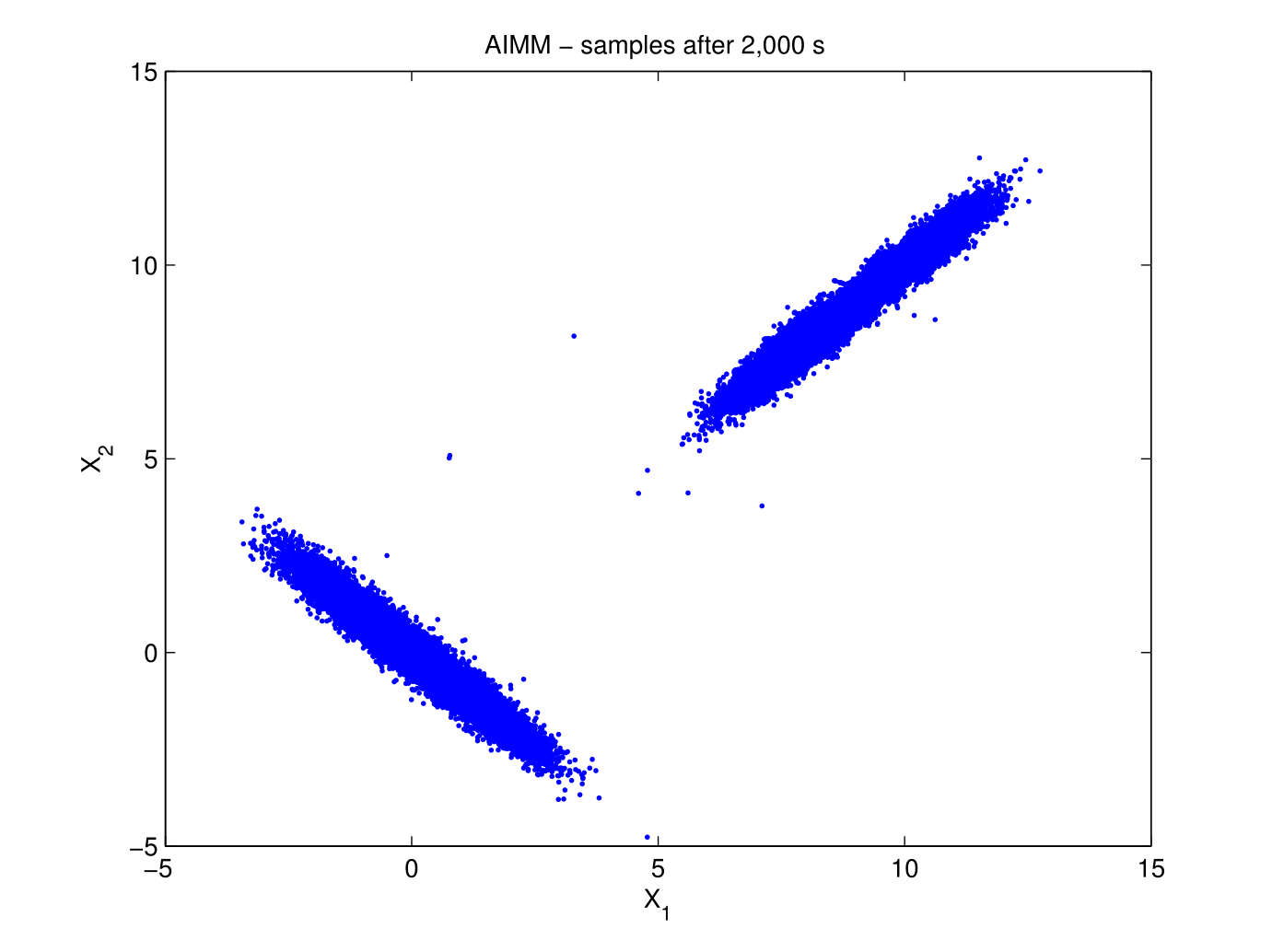}
\caption{Example \ref{example_pi4}: $\targ_4$ target, $d=4$, runtime of $2,000$ seconds - Top: Gaussian kernels created by IMIS and AIMM (colours stand for weights assigned to the kernels and the region inside the ellipses contains $75\%$ of each Gaussian mass). Bottom: resamples from IMIS (using the incremental proposal obtained after $2,000$ seconds) and samples of the AIMM chain.
\label{fig:imis_2}}

\end{figure}

We therefore expect AIMM to be more efficient in situations where the exploration of the state space requires a large number of increments of the proposal and IMIS to be more efficient for short run times. To substantiate this expectation, we have compared the performance of AIMM and IMIS on the last example of Section~\ref{sec:sim} in dimension 4. Figure~\ref{fig:imis_1} reports the estimation of the probability $\pi_4(X_1<-2)$ obtained through both methods for different run times. For short run times IMIS benefits from using batches of particles and gains a lot of information on $\pi_4$ in a few iterations. On the other hand, AIMM provides a more accurate estimate of $\pi_4(X_1<-2)$ after about $150$ seconds. Figure~\ref{fig:imis_2} illustrates the outcome of AIMM and IMIS after running them for $2,000$ seconds. The mixture of incremental kernels obtained by AIMM is visually more appealing than the sequence of proposals derived by IMIS, reinforcing the results from Figure~\ref{fig:imis_1}.

AIMM can be regarded as a transformation of IMIS, a particle-based inference method, into an adaptive Markov chain. This transformation could be applied to other adaptive importance sampling methods, thus designing Markov chains that might be more efficient than their importance sampling counterparts. In a Bayesian context, AIMM could, in addition to sampling the posterior distribution, be used to estimate intractable normalizing constants and marginal likelihoods via importance sampling. Indeed, the incremental mixture proposal produced by AIMM is an appealing importance distribution since it approximates the posterior and is straightforward to sample from.

\section*{Acknowledgements}
The Insight Centre for Data Analytics is supported by Science Foundation Ireland under grant SFI/12/RC/2289. Nial Friel's research was also supported by a Science Foundation Ireland grant: 12/IP/1424. Adrian E. Raftery's research was supported by NIH grants R01 HD054511 and R01 HD070936, Science Foundation Ireland grant 11/W.1/I2079, and the Center for Advanced Study in the Behavioral Sciences at Stanford University. Antonietta Mira's research was supported by a Swiss National Science Foundation grant.

\bibliographystyle{chicago}
\bibliography{biblio}

\begin{thebibliography}{}

\bibitem[\protect\citeauthoryear{Andrieu and Moulines}{Andrieu and
  Moulines}{2006}]{andrieu2006ergodicity}
Andrieu, C. and {\'E}.~Moulines (2006).
\newblock On the ergodicity properties of some adaptive {MCMC} algorithms.
\newblock {\em Annals of Applied Probability\/}~{\em 16}, 1462--1505.

\bibitem[\protect\citeauthoryear{Bai, Roberts, and Rosenthal}{Bai
  et~al.}{2009}]{bai2009containment}
Bai, Y., G.~O. Roberts, and J.~S. Rosenthal (2009).
\newblock On the containment condition for adaptive {Markov chain Monte Carlo}
  algorithms.

\bibitem[\protect\citeauthoryear{Bates, Cullen, and Raftery}{Bates
  et~al.}{2003}]{bates2003bayesian}
Bates, S.~C., A.~Cullen, and A.~E. Raftery (2003).
\newblock Bayesian uncertainty assessment in multicompartment deterministic
  simulation models for environmental risk assessment.
\newblock {\em Environmetrics\/}~{\em 14\/}(4), 355--371.

\bibitem[\protect\citeauthoryear{Capp{\'e}, Douc, Guillin, Marin, and
  Robert}{Capp{\'e} et~al.}{2008}]{cappe2008adaptive}
Capp{\'e}, O., R.~Douc, A.~Guillin, J.-M. Marin, and C.~P. Robert (2008).
\newblock Adaptive importance sampling in general mixture classes.
\newblock {\em Statistics and Computing\/}~{\em 18\/}(4), 447--459.

\bibitem[\protect\citeauthoryear{Craiu, Gray, {\L}atuszy{\'n}ski, Madras,
  Roberts, Rosenthal, et~al.}{Craiu et~al.}{2015}]{craiu2015stability}
Craiu, R.~V., L.~Gray, K.~{\L}atuszy{\'n}ski, N.~Madras, G.~O. Roberts, J.~S.
  Rosenthal, et~al. (2015).
\newblock Stability of adversarial {M}arkov chains, with an application to
  adaptive {MCMC} algorithms.
\newblock {\em The Annals of Applied Probability\/}~{\em 25\/}(6), 3592--3623.

\bibitem[\protect\citeauthoryear{Craiu, Rosenthal, and Yang}{Craiu
  et~al.}{2009}]{craiu2009learn}
Craiu, R.~V., J.~Rosenthal, and C.~Yang (2009).
\newblock Learn from thy neighbor: Parallel-chain and regional adaptive {MCMC}.
\newblock {\em Journal of the American Statistical Association\/}~{\em
  104\/}(488), 1454--1466.

\bibitem[\protect\citeauthoryear{Fort, Moulines, and Priouret}{Fort
  et~al.}{2011}]{fort2011convergence}
Fort, G., E.~Moulines, and P.~Priouret (2011).
\newblock Convergence of adaptive and interacting {Markov chain Monte Carlo}
  algorithms.
\newblock {\em The Annals of Statistics\/}~{\em 39\/}(6), 3262--3289.

\bibitem[\protect\citeauthoryear{G{\aa}semyr}{G{\aa}semyr}{2003}]{gasemyr2003}
G{\aa}semyr, J. (2003).
\newblock On an adaptive version of the {Metropolis–-Hastings} algorithm with
  independent proposal distribution.
\newblock {\em Scandinavian Journal of Statistics\/}~{\em 30\/}(1), 159--173.

\bibitem[\protect\citeauthoryear{Gilks, Roberts, and Sahu}{Gilks
  et~al.}{1998}]{gilks1998adaptive}
Gilks, W.~R., G.~O. Roberts, and S.~K. Sahu (1998).
\newblock Adaptive {M}arkov chain {M}onte {C}arlo through regeneration.
\newblock {\em Journal of the {A}merican {S}tatistical {A}ssociation\/}~{\em
  93}, 1045--1054.

\bibitem[\protect\citeauthoryear{Gilks and Wild}{Gilks and
  Wild}{1992}]{gilks1992adaptive}
Gilks, W.~R. and P.~Wild (1992).
\newblock Adaptive rejection sampling for {G}ibbs sampling.
\newblock {\em Applied Statistics\/}~{\em 41}, 337--348.

\bibitem[\protect\citeauthoryear{Giordani and Kohn}{Giordani and
  Kohn}{2010}]{giordani2010adaptive}
Giordani, P. and R.~Kohn (2010).
\newblock Adaptive independent {M}etropolis--{H}astings by fast estimation of
  mixtures of normals.
\newblock {\em Journal of Computational and Graphical Statistics\/}~{\em 19},
  243--259.

\bibitem[\protect\citeauthoryear{Haario, Saksman, and Tamminen}{Haario
  et~al.}{1999}]{haario1999adaptive}
Haario, H., E.~Saksman, and J.~Tamminen (1999).
\newblock Adaptive proposal distribution for random walk {M}etropolis
  algorithm.
\newblock {\em Computational Statistics\/}~{\em 14}, 375--396.

\bibitem[\protect\citeauthoryear{Haario, Saksman, and Tamminen}{Haario
  et~al.}{2001}]{haario2001adaptive}
Haario, H., E.~Saksman, and J.~Tamminen (2001).
\newblock An adaptive {M}etropolis algorithm.
\newblock {\em Bernoulli\/}~{\em 7}, 223--242.

\bibitem[\protect\citeauthoryear{Hastings}{Hastings}{1970}]{hastings1970monte}
Hastings, W.~K. (1970).
\newblock Monte {C}arlo sampling methods using {M}arkov chains and their
  applications.
\newblock {\em Biometrika\/}~{\em 57}, 97--109.

\bibitem[\protect\citeauthoryear{Hesterberg}{Hesterberg}{1995}]{hesterberg1995weighted}
Hesterberg, T. (1995).
\newblock Weighted average importance sampling and defensive mixture
  distributions.
\newblock {\em Technometrics\/}~{\em 37}, 185--194.

\bibitem[\protect\citeauthoryear{Holden, Hauge, and Holden}{Holden
  et~al.}{2009}]{holden2009adaptive}
Holden, L., R.~Hauge, and M.~Holden (2009).
\newblock Adaptive independent {M}etropolis-{H}astings.
\newblock {\em Annals of Applied Probability\/}~{\em 19}, 395--413.

\bibitem[\protect\citeauthoryear{Li and Barron}{Li and
  Barron}{2000}]{li2000mixture}
Li, J.~Q. and A.~R. Barron (2000).
\newblock Mixture density estimation.
\newblock In {\em Advances in neural information processing systems}, pp.\
  279--285.

\bibitem[\protect\citeauthoryear{Liu}{Liu}{1996}]{liu1996metropolized}
Liu, J.~S. (1996).
\newblock Metropolized independent sampling with comparisons to rejection
  sampling and importance sampling.
\newblock {\em Statistics and Computing\/}~{\em 6}, 113--119.

\bibitem[\protect\citeauthoryear{Luengo and Martino}{Luengo and
  Martino}{2013}]{luengo2013fully}
Luengo, D. and L.~Martino (2013).
\newblock Fully adaptive {G}aussian mixture {M}etropolis-{H}astings algorithm.
\newblock In {\em 2013 IEEE International Conference on Acoustics, Speech and
  Signal Processing (ICASSP)}, pp.\  6148--6152.

\bibitem[\protect\citeauthoryear{Metropolis, Rosenbluth, Rosenbluth, Teller,
  and Teller}{Metropolis et~al.}{1953}]{metropolis1953equation}
Metropolis, N., A.~W. Rosenbluth, M.~N. Rosenbluth, A.~H. Teller, and E.~Teller
  (1953).
\newblock Equation of state calculations by fast computing machines.
\newblock {\em Journal of Chemical Physics\/}~{\em 21}, 1087--1092.

\bibitem[\protect\citeauthoryear{Mira}{Mira}{2001}]{mira2001ordering}
Mira, A. (2001).
\newblock Ordering and improving the performance of {Monte Carlo Markov
  chains}.
\newblock {\em Statistical Science\/}, 340--350.

\bibitem[\protect\citeauthoryear{Pasarica and Gelman}{Pasarica and
  Gelman}{2010}]{pasarica2010adaptively}
Pasarica, C. and A.~Gelman (2010).
\newblock Adaptively scaling the {M}etropolis algorithm using expected squared
  jumped distance.
\newblock {\em Statistica Sinica\/}~{\em 20}, 343--364.

\bibitem[\protect\citeauthoryear{Poole and Raftery}{Poole and
  Raftery}{2000}]{poole2000inference}
Poole, D. and A.~E. Raftery (2000).
\newblock Inference for deterministic simulation models: the {B}ayesian melding
  approach.
\newblock {\em Journal of the American Statistical Association\/}~{\em
  95\/}(452), 1244--1255.

\bibitem[\protect\citeauthoryear{Raftery and Bao}{Raftery and
  Bao}{2010}]{raftery2010}
Raftery, A.~E. and L.~Bao (2010).
\newblock Estimating and projecting trends in {HIV/AIDS} generalized epidemics
  using incremental mixture importance sampling.
\newblock {\em Biometrics\/}~{\em 66}, 1162--1173.

\bibitem[\protect\citeauthoryear{Roberts and Rosenthal}{Roberts and
  Rosenthal}{2001}]{roberts2001optimal}
Roberts, G.~O. and J.~S. Rosenthal (2001).
\newblock Optimal scaling for various {M}etropolis-{H}astings algorithms.
\newblock {\em Statistical {S}cience\/}~{\em 16}, 351--367.

\bibitem[\protect\citeauthoryear{Roberts and Rosenthal}{Roberts and
  Rosenthal}{2007}]{roberts2007}
Roberts, G.~O. and J.~S. Rosenthal (2007).
\newblock Coupling and ergodicity of adaptive {M}arkov chain {M}onte {C}arlo
  algorithms.
\newblock {\em Journal of Applied Probability\/}~{\em 44}, 458--475.

\bibitem[\protect\citeauthoryear{Roberts and Rosenthal}{Roberts and
  Rosenthal}{2009}]{roberts2009examples}
Roberts, G.~O. and J.~S. Rosenthal (2009).
\newblock Examples of adaptive {MCMC}.
\newblock {\em Journal of Computational and Graphical Statistics\/}~{\em 18},
  349--367.

\bibitem[\protect\citeauthoryear{Rosenthal}{Rosenthal}{2003}]{rosenthal2003asymptotic}
Rosenthal, J.~S. (2003).
\newblock Asymptotic variance and convergence rates of nearly-periodic {Markov
  chain Monte Carlo} algorithms.
\newblock {\em Journal of the American Statistical Association\/}~{\em 98},
  169--177.

\bibitem[\protect\citeauthoryear{Rosenthal et~al.}{Rosenthal
  et~al.}{2011}]{rosenthal2011optimal}
Rosenthal, J.~S. et~al. (2011).
\newblock Optimal proposal distributions and adaptive {MCMC}.
\newblock {\em Handbook of Markov Chain Monte Carlo\/}~{\em 4}.

\bibitem[\protect\citeauthoryear{Rosenthal and Yang}{Rosenthal and
  Yang}{2017}]{rosenthal2017ergodicity}
Rosenthal, J.~S. and J.~Yang (2017).
\newblock Ergodicity of combocontinuous adaptive {MCMC} algorithms.
\newblock {\em Methodology and Computing in Applied Probability\/}, 1--17.

\bibitem[\protect\citeauthoryear{Tran, Pitt, and Kohn}{Tran
  et~al.}{2014}]{tran2014adaptive}
Tran, M.-N., M.~K. Pitt, and R.~Kohn (2014).
\newblock Adaptive {M}etropolis--{H}astings sampling using reversible dependent
  mixture proposals.
\newblock {\em Statistics and Computing\/}~{\em 26}, 1--21.

\end{thebibliography}

\begin{appendices}

\section{Proof of Proposition \ref{prop1}}
\label{app1}

\setcounter{condition}{0}
\setcounter{corollary}{0}

In this appendix, we use the following notations
\begin{itemize}
\item $\tQ_n$ is the incremental part of the proposal $Q_n$ at iteration $n$, \ie $\tQ_n\propto\sum_{k=1}^{n-1}\beta_k\phi_k$
\item at iteration $n$, only $M_n\leq n-1$ components among $\phi_1,\ldots,\phi_{n-1}$ have a non-zero weight
\item $\event_n$ is the event that the proposal increments at iteration $n$ of AIMM, \ie $\beta_n>0$ almost surely
\item $\event$ is the event that the proposal increments infinitely often \ie $\event=\cap_n\cup_{k>n}\event_n$
\end{itemize}

The proof of Proposition \ref{prop1} is achieved by first showing that the KL divergence between $\tQ_n$ and $\tQ_{n+1}$ goes to zero. This allows to prove that, in the total variation norm, $\{Q_n\}_n$ adapts less and less and finally that the convergence to zero of the difference between two consecutive transition kernels $K_n-K_{n+1}$ occurs uniformly, provided that some implementation details are satisfied. We first state the following lemma and a corollary that will be needed to establish the diminishing adaptation for the AIMM process.

\begin{lemma}
\label{lemma:1}
The sum of the weights $\bbeta_n$ is not bounded in $\proba_\event$-probability, \ie
$\bbeta_n\overset{\proba_\event}{\to}\infty$.
\end{lemma}

\begin{proof}
Let $M>0$, and note that
\begin{multline}
\label{eq:app:lem1_1}
\proba_\event\{\bbeta_{n+1}<M\}=\proba_\event\{\bbeta_{n+1}<M\,\cap\, \bbeta_{n}<M\}+\proba_\event\{\bbeta_{n+1}<M\,\cap\, \bbeta_{n}\geq M\}\\
=\proba_\event\{\bbeta_{n+1}<M\,|\, \bbeta_{n}<M\}\proba_\event\{\bbeta_n<M\}\leq \proba_\event\{\bbeta_n<M\}
\end{multline}
since $\proba_\event\{\bbeta_{n+1}<M\,\cap\, \bbeta_{n}\geq M\}\leq \proba_\event\{\bbeta_{n+1}<M\,|\, \bbeta_{n}\geq M\}\leq \proba_\event\{\bbeta_{n+1}<\bbeta_n\,|\,\bbeta_n\geq M\}=0$. Note that the inequality in \eqref{eq:app:lem1_1} is strict on $\event_n$. Indeed,
\begin{equation}
\label{eq:app:lem1_2}
\proba_\event\{\beta_{n+1}\geq M-\bbeta_n\,|\, \bbeta_{n}<M,\event_n\}>0\Rightarrow
\proba_\event\{\bbeta_{n+1}< M\,|\, \bbeta_{n}<M,\event_n\}<1\,,
\end{equation}
while
\begin{equation}
\label{eq:app:lem1_3}
%\proba\{\beta_{n+1}\geq M-\bbeta_n\,|\, \bbeta_{n}<M,\overline{\event_n}\}=0\Rightarrow
\proba_\event\{\bbeta_{n+1}< M\,|\, \bbeta_{n}<M,\overline{\event_n}\}=1\,.
\end{equation}

Combining \eqref{eq:app:lem1_1}, \eqref{eq:app:lem1_2} and \eqref{eq:app:lem1_3} shows that $\{\proba_\event\{\bbeta_n<M\}\}_n$ is a decreasing sequence. In particular, it is strictly decreasing an infinite number of times and therefore $\proba_\event\{\bbeta_{n}<M\}\to 0$ which yields $\bbeta_n\overset{\proba_\event}{\to}\infty$.
\end{proof}

\begin{corollary}
\label{coro:1}
On $\event$, the incremental component weight \ie $\beta_n/\bbeta_n$ goes to zero in probability.
\end{corollary}
\begin{proof}
Let $\eps>0$ and note that since $\pi$ is absolutely continuous we have $\sup_{x\in\Xset}\pi(x)=:\bar{\pi}<\infty$ and
\begin{equation}
\label{eq:app:lem1_0}
\proba_\event\left\{\frac{\pi(\tX_n)^\gamma}{\bbeta_n}>\eps\right\}\leq  \proba_\event \left\{\frac{\bar{\pi}^\gamma}{\bbeta_n}>\eps\right\}\,.
\end{equation}
Using Lemma \ref{lemma:1}, the RHS of \eqref{eq:app:lem1_0}  goes to zero as $n\to\infty$ which concludes the proof.

\end{proof}

\subsection{Diminishing adaptation in the KL metric and total variation norm}
\begin{proposition}
\label{prop:2}
Assume that almost surely
\begin{enumerate}[\textbf{A}1.]
\item the covariance matrix of any component of the mixture satisfies $\det\Sigma_k>\delta>0$
\end{enumerate}
Consider the sequence $\{\tD_n\}_{n>0}$, defined as the KL divergence between the incremental components of two consecutive proposals, i.e.~$\tD_n=\KL\{\tQ_n,\tQ_{n+1}\}$. Then $\tD_n$ converges to zero in probability, \ie $\tD_n=o_\proba(1)$.
\end{proposition}

\begin{proof}
\label{proof2}
For all $\eps>0$, we have
\begin{equation}
\label{eq:app:1}
\proba\{\tD_n>\eps\}=\proba\{\tD_n>\eps\cap\bar{\event}\}+\proba\{\tD_n>\eps\cap\event\}\,.
\end{equation}
We will show that both terms in the RHS of Eq. \eqref{eq:app:1} converge to 0.

For the first term, note that $\bar{\event}$ is the event that the mixture will eventually stop incrementing.  Therefore for all $\omega\in\bar{\event}$, there exists $\nu\in\nset$ such that for all $n>n_0$, $\tQ_n=\tQ_{n+1}$. Hence, for any $n>n_0$, $\omega\not\in\{\tD_n>\eps\}$. This implies that $\{\tD_n>\eps\}$ and $\bar{\event}$ are eventually mutually exclusive events, leading to the first term in the RHS of Eq. \eqref{eq:app:1} converging to 0.

We now show that the second term in the RHS of Eq. \eqref{eq:app:1} goes to zero. The case where $\proba(\event)=0$ is trivial. Indeed, $\proba(\event)=0\Rightarrow \proba\{\tD_n>\eps\cap\event\}=0$ and the proof is completed. In the following, we consider the case where $0<\proba(\event)\leq 1$ and show that $\proba\{\tD_n>\eps\cap\event\}$ converge to 0 as well. First, note that we have
\begin{equation}
\label{eq:app:1_2}
\proba\{\tD_n>\eps\cap\event\}\leq \proba\{\tD_n>\eps\,|\,\event\}\,.
\end{equation}
Defining $\bbeta_n=\sum_{k\leq n}\beta_k$, we now show that the RHS of Eq. \eqref{eq:app:1_2} goes to zero in probability.
\begin{multline*}
\tD_n=\int\rmd\tQ_n \log \frac{\tQ_n}{\tQ_{n+1}}\leq \int\rmd\tQ_n \log \frac{\tQ_n}{\tQ_{n+1}}=
\int\rmd\tQ_n \log \frac{\sum_{k=1}^{n-1}\beta_k\phi_k}{\sum_{k=1}^{n}\beta_k\phi_k}\times \frac{\bar{\beta}_{n}}{\bar{\beta}_{n-1}}\\
\leq \int\rmd\tQ_n \left|\log \frac{\sum_{k=1}^{n-1}\beta_k\phi_k}{\sum_{k=1}^{n}\beta_k\phi_k}\right|+\int\rmd\tQ_n\left| \log\frac{\bar{\beta}_{n}}{\bar{\beta}_{n-1}}\right|=
\int\rmd\tQ_n \log \frac{\sum_{k=1}^{n}\beta_k\phi_k}{\sum_{k=1}^{n-1}\beta_k\phi_k}+ \log\frac{\bar{\beta}_{n}}{\bar{\beta}_{n-1}}\,,
\end{multline*}
where the last equality holds since for all $x\in\Xset$, $\sum_{k=1}^{n-1}\beta_k\phi_k(x)\leq \sum_{k=1}^{n}\beta_k\phi_k(x)$. Since we assume that $\pi$ is absolutely continuous, there exists a finite constant $\overline{\pi}>0$ such that for all $x\in\Xset$, the pdf of $\pi$ satisfies $\pi(x)<\overline{\pi}$, which implies that $\beta_{n}<\overline{\pi}^\gamma$. Using the log inequality, this yields:
$$
\tD_n\leq \overline{\pi}^\gamma\left\{\int\rmd\tQ_n \frac{\phi_{n}}{\sum_{k=1}^{n-1}\beta_k\phi_k}+ \frac{1}{\bar{\beta}_{n-1}}\right\}=
\frac{\overline{\pi}^\gamma}{\bbeta_{n-1}} \left\{1+\int\rmd\phi_n\right\}\,,%\\=
%\frac{1}{\sum_{k=1}^{M_n} b_k} \left\{1+\int\rmd\phi_{M_n+1}W_n\frac{Q_n}{\tQ_n}\right\}\,,
$$
 We now write that by inclusion of events,
$$
\proba_\event\{\tD_n>\eps\}\leq
\proba_\event\left\{\frac{2\overline{\pi}^\gamma}{\bbeta_{n-1}}>\eps\right\}=\proba_\event\left\{\bbeta_{n+1}\leq \frac{2\overline{\pi}^\gamma}{\eps}\right\}\,.
$$
The proof is completed by applying Lemma \ref{lemma:1}.
\end{proof}

\begin{proposition}
\label{prop:3}
Assume that almost surely
\begin{enumerate}[\textbf{A}1.]
\item the covariance matrix of any component of the mixture satisfies $\det\Sigma_k>\delta>0$
\end{enumerate}
The sequence of random variables $\{D_n:=\KL\{Q_n,Q_{n+1}\}\}_{n>0}$ goes to zero in probability.
\end{proposition}

\begin{proof}
By straightforward algebra, we have:
\begin{multline*}
\label{eq:Dn}
D_n=\int\rmd Q_n\log\frac{Q_n}{Q_{n+1}}=\omega_n\int\rmd Q_0\log \frac{Q_n}{Q_{n+1}}+(1-\omega_n)\int\rmd \tQ_n\log \frac{Q_n}{Q_{n+1}}\\
\leq
\omega_n\int\rmd Q_0\log \frac{Q_0+(1-\omega_n)\tQ_n}{\omega_{n+1}Q_0}+
(1-\omega_n)\int\rmd \tQ_n\log \frac{\omega_nQ_0+(1-\omega_n)\tQ_n}{(1-\omega_{n+1})\tQ_{n+1}}\\
=
\omega_n\int\rmd Q_0\log \frac{ Q_0}{\omega_{n+1}Q_0}\left\{1+\frac{(1-\omega_n)\tQ_n}{Q_0}\right\}+
(1-\omega_n)\int\rmd \tQ_n\log \frac{(1-\omega_n)\tQ_n}{(1-\omega_{n+1})\tQ_{n+1}}\left\{1+\frac{\omega_nQ_0}{(1-\omega_n)\tQ_n}\right\}\\
\leq
\omega_n\log \frac{1}{\omega_{n+1}}+\omega_n\int\rmd Q_0\log \left\{1+\frac{(1-\omega_n)\tQ_n}{Q_0}\right\}
+(1-\omega_n)\tD_n+(1-\omega_n)\int\rmd \tQ_n\log \left\{1+\frac{\omega_nQ_0}{(1-\omega_n)\tQ_n}\right\}\\
\leq
\omega_n\log \frac{1}{\omega_{n+1}}+(1-\omega_n)\tD_n
+\omega_n\int\rmd Q_0\frac{(1-\omega_n)\tQ_n}{Q_0}
+(1-\omega_n)\int\rmd \tQ_n\frac{\omega_nQ_0}{(1-\omega_n)\tQ_n}\\
=
\omega_n\log \frac{1}{\omega_{n+1}}+(1-\omega_n)\tD_n
+\omega_n{(1-\omega_n)}
+(1-\omega_n)\frac{\omega_n}{(1-\omega_n)}
\leq
\tD_n+v_n\,,
\end{multline*}
where we have used the log inequality $\log(1+x)< x$ for all $x>0$ and defined $v_n:=-\omega_n\log {\omega_{n+1}}+2\omega_n$.
The proof is completed by noting that
\begin{equation}
\label{eq:Dn4}
\proba\{D_n>\eps\}\leq \proba\{\tD_n+v_n>\eps\}
\end{equation}
and applying Proposition \ref{prop:2} to get that $\tD_n=o_{\proba}(1)$ which, combined with the fact that $v_n=o(1)$, shows that the RHS of Eq. \eqref{eq:Dn4} converges to zero.
\end{proof}

\begin{corollary}
Assume that almost surely
\begin{enumerate}[\textbf{A}1.]
\item the covariance matrix of any component of the mixture satisfies $\det\Sigma_k>\delta>0$
\end{enumerate}
Then the difference between two consecutive proposals converges in total variation to zero in probability, \ie $\left\|Q_n-Q_{n+1}\right\|=o_\proba(1)$.
\end{corollary}

\begin{proof}
This result follows from combining Proposition \ref{prop:2} together with Pinsker inequality,
\begin{equation}
\label{eq:pinsker}
\|Q_n-Q_{n+1}\|\leq \sqrt{\frac{1}{2}\KL\{Q_n,Q_{n+1}\}}\,.
\end{equation}
Indeed, for all $\eps>0$, we have using Eq. \eqref{eq:pinsker} that
$$
\proba\{\|Q_{n}-Q_{n+1}\|^2>\eps\}\leq \proba\{\KL\{Q_n,Q_{n+1}\}>2\eps\}\,,
$$
which shows that $\|Q_n-Q_{n+1}\|^2=o_\proba(1)$. The proof is completed by using the continuous mapping theorem since $x\mapsto\sqrt{x}$ is continuous on $[0,1]$ and therefore preserves the limit.
\end{proof}

\subsection{Diminishing adaptation in the sup norm}
We want to prove that
$$
\sup_{x\in\Xset}\|K_{n+1}(x,\cdot)-K_{n}(x,\cdot)\| ,
$$
converges to zero in probability, which corresponds to the diminishing assumption of \cite{roberts2007}. We first establish the following result.

\begin{proposition}
\label{prop:unif}
Assume that almost surely
\begin{enumerate}[\textbf{A}1]
\setcounter{enumi}{0}
\item the covariance matrix of any component of the mixture satisfies $\det\Sigma_k>\delta>0$
\end{enumerate}
Then the difference between two consecutive proposal probability density functions converge uniformly to zero in probability \ie $\sup_{x\in\Xset}|Q_{n+1}(x)-Q_n(x)|=o_{\proba}(1)$.
\end{proposition}
\begin{proof}
Let $\eps>0$, we will show that $\lim_n\proba\{\sup_{x\in\Xset}|Q_{n+1}(x)-Q_n(x)|>\eps\}= 0$. Using similar arguments to those in the proof of Proposition \ref{prop:2}, we have that
\begin{multline*}
\proba\left\{\sup_{x\in\Xset}|Q_{n+1}(x)-Q_n(x)|>\eps\right\}=\proba\left\{\sup_{x\in\Xset}|Q_{n+1}(x)-Q_n(x)|>\eps\,\cap\, \event\right\}\\
+\underbrace{\proba\left\{\sup_{x\in\Xset}|Q_{n+1}(x)-Q_n(x)|>\eps\,\cap\, \bar{\event}\right\}}_{\searrow 0 }\,,
\end{multline*}
and
\begin{equation*}
\proba\left\{\sup_{x\in\Xset}|Q_{n+1}(x)-Q_n(x)|>\eps\,\cap\, \event\right\}\leq
\proba_\event\left\{\sup_{x\in\Xset}|Q_{n+1}(x)-Q_n(x)|>\eps\right\}\,.%\\
%=\proba\{\sup_{x\in\Xset}|Q_{n+1}(x)-Q_n(x)|>\eps\,\cap\,\event_n \,|\, \event\}\leq
%\proba\{\sup_{x\in\Xset}|Q_{n+1}(x)-Q_n(x)|>\eps \,|\, \event,\event_n\}\,.
\end{equation*}
Now by straightforward algebra we have that
\begin{multline}
\label{eq:prop:unif1}
|Q_{n+1}(x)-Q_n(x)|=\left|(\omega_{n+1}-\omega_n)Q_0(x)+\frac{1-\omega_{n+1}}{\bbeta_{n+1}}\sum_{k=1}^{{M_n}+1}\beta_k\phi_k(x)-
\frac{1-\omega_{n}}{\bbeta_{n}}\sum_{k=1}^{M_n}\beta_k\phi_k(x)\right|\\
\leq \frac{\kappa}{(1+\kappa n)(1+\kappa(n+1))}Q_0(x)+\sum_{k=1}^{M_n} \beta_k\phi_k(x)\left|\frac{1-\omega_{n+1}}{\bbeta_{n+1}}-\frac{1-\omega_{n}}{\bbeta_{n}}\right|+\frac{\beta_{M_n+1}}{\bbeta_{n+1}}\phi_{M_n+1}(x)\\
\leq \frac{\kappa}{(1+\kappa n)^2}{Q_0}(x)+(1-\omega_{n})\left|1-\frac{(1-\omega_{n+1})\bbeta_n}{(1-\omega_n)\bbeta_{n+1}}\right|\tQ_n(x)+\frac{\beta_{M_n+1}}{\bbeta_{n+1}}{\phi_{M_n+1}}(x)\\
\leq\frac{\kappa}{(1+\kappa n)^2}\overline{Q_0}+\left|1-\frac{1-\omega_{n+1}}{1-\omega_n}\left(1-\frac{\beta_{M_n+1}}{\bbeta_{n+1}}\right)\right|\overline{\tQ_n}+\frac{\beta_{M_n+1}}{\bbeta_{n+1}}\overline{\phi_{M_n+1}}=:R_n\,,
\end{multline}
where for any real-valued function $f:\Xset\to\rset$, we have defined $\overline{f}:=\sup_{x\in\Xset}{f}(x)$. Since $\sup_{x\in\Xset}|Q_{n+1}(x)-Q_n(x)|\leq R_n$, we have that
$$
\proba_\event\left\{\sup_{x\in\Xset}|Q_{n+1}(x)-Q_n(x)|>\eps \right\}\leq
\proba_\event\left\{R_n>\eps \right\}\,.
$$
The proof is concluded by noting that $R_n$ is the sum of three random variables, each of which converges to zero in $\proba_\event$-probability:
\begin{itemize}
\item ${\kappa\overline{Q_0}}/{(1+\kappa n)^2}=o(1)$,
\item the middle term in \eqref{eq:prop:unif1} is upper bounded by
$$
\left|1-\frac{1-\omega_{n+1}}{1-\omega_n}\left(1-\frac{\beta_{M_n+1}}{\bbeta_{n+1}}\right)\right|\leq \left|1-\frac{1-\omega_{n+1}}{1-\omega_n}\right|+
\frac{\beta_{M_n+1}}{\bbeta_{n+1}}\frac{1-\omega_{n+1}}{1-\omega_n}
$$
and noting that ${1-\omega_{n+1}}\slash {1-\omega_n}\to 1$, the first terms converges to zero surely and, using Corollary \ref{coro:1}, the second one converges to zeros in $\proba_\event$-probability,
\item the last term converges to zero in $\proba_\event$-probability, again using Corollary \ref{coro:1}.
\end{itemize}
\end{proof}

\begin{proposition}
\label{prop:cauchy}
Assume that
\begin{enumerate}[\textbf{A}1.]
\setcounter{enumi}{0}
\item the covariance matrix of any component of the mixture satisfies $\det\Sigma_k>\delta>0$
\item the incremental mixture (unnormalized) weights are defined as:
\begin{equation}
\label{eq:new_weights}
\beta_n=\frac{\eta+\pi(\tX_n)^\gamma}{(1+\eta)^n}\1_{\event_n}
\qquad\text{and}\qquad\omega_n=\frac{1}{1+\sum_{k=1}^n\beta_k}\,,
\end{equation}
where $\eta>0$ is a constant, typically $\eta\ll 1$.
\end{enumerate}
Then, $\{Q_n\}_n$ converges uniformly in probability to a distribution $Q$, \ie $\sup_{x\in\Xset}|Q_n(x)-Q(x)|=o_\proba(1)$
\end{proposition}
\begin{proof}
Uniform convergence being equivalent to uniformly Cauchy, we show that $\{Q_n\}_n$ is uniformly Cauchy, in probability, \ie
\begin{equation}
\label{unifCauchy}
\sup_{x\in\Xset}\sup_{\ell\in\nset}|Q_{n+\ell}(x)-Q_n(x)|=o_\proba(1)\,.
\end{equation}
Using the same reasoning that in the proof of Proposition \ref{prop:2}, it is sufficient to prove that the convergence occurs in $\proba_\event$-probability only. Similarly to Proposition \ref{prop:unif}, we write
$$
|Q_{n+\ell}-Q_n|\leq |\omega_{n+\ell}-\omega_n|\overline{Q_0}+\left|\frac{1-\omega_{n+\ell}}{\bbeta_{n+\ell}}-\frac{1-\omega_n}{\bbeta_n}\right|
\sum_{k=1}^n\beta_k\overline{\phi_k}+\frac{1}{\bbeta_{n+\ell}}\sum_{k=n+1}^{n+\ell}\beta_k\overline{\phi_k}\,,
$$
where we have used the same notations than in the proof of Proposition \ref{prop:unif}.

First, we note that
\begin{equation}
\label{eq:step1}
|\omega_{n+\ell}-\omega_n|=\frac{\omega_0(\bbeta_{n+\ell}-\bbeta_n)}{(\omega_0+\bbeta_n)(\omega_0+\bbeta_{n+\ell})}\leq \frac{\bbeta_{n+\ell}-\bbeta_n}{\omega_0}
\end{equation}
and since
\begin{equation}
\label{eq:bbeta}
\bbeta_{n+\ell}-\bbeta_n=\sum_{k=n+1}^{n+\ell}\frac{\eta+\pi(\tX_k)^\gamma}{(1+\eta)^k}\leq \frac{\eta+\overline{\pi}^\gamma}{(1+\eta)^n}\sum_{k=1}^\ell\frac{1}{(1+\eta)^k}=
\frac{\eta+\overline{\pi}^\gamma}{(1+\eta)^n}\frac{1-\{1/(1+\eta)\}^\ell}{\eta}\,,
\end{equation}
clearly $|\omega_{n+\ell}-\omega_n|\leq{(\eta+\overline{\pi}^\gamma)}(1+\eta)^{-n}\slash{\eta\omega_0}$. Second, we have
\begin{equation*}
\left|\frac{1-\omega_{n+\ell}}{\bbeta_{n+\ell}}-\frac{1-\omega_n}{\bbeta_n}\right|=\left|\frac{1}{\omega_0+\bbeta_{n+\ell}}-\frac{1}{\omega_0+\bbeta_n}\right|\leq \frac{\bbeta_{n+\ell}-\bbeta_n}{\omega_0^2}\leq\frac{{(\eta+\overline{\pi}^\gamma)}}{\eta \omega_0^2(1+\eta)^n}
\end{equation*}
Moreover, by assumption \textbf{A}2, we have that $\overline{\phi_k}\leq 1/(2\pi)^{d/2}\sqrt{\delta}$, so
$$
\sum_{k=1}^n\beta_k\overline{\phi_k}\leq \frac{1}{(2\pi)^{d/2}\sqrt{\delta}}\bbeta_n\,.
$$
Since
$$
\bbeta_n\leq {\eta+\overline{\pi}^\gamma}\sum_{k=1}^n\frac{1}{(1+\eta)^k}=\frac{\eta+\overline{\pi}^\gamma}{\eta}\{1-(1/(1+\eta))^n\}<
\frac{\eta+\overline{\pi}^\gamma}{\eta}\,,
$$
we have that
\begin{equation}
\label{eq:step2}
\left|\frac{1-\omega_{n+\ell}}{\bbeta_{n+\ell}}-\frac{1-\omega_n}{\bbeta_n}\right|\sum_{k=1}^n\beta_k\overline{\phi_k}\leq
\frac{{(\eta+\overline{\pi}^\gamma)}^2}{\eta^2 \omega_0^2(2\pi)^{d/2}\sqrt{\delta}}\frac{1}{(1+\eta)^n}\,.
\end{equation}
Finally, using Eq. \eqref{eq:bbeta}, we have
\begin{equation}
\label{eq:step3}
\frac{1}{\bbeta_{n+\ell}}\sum_{k=n+1}^{n+\ell}\beta_k\overline{\phi_k}\leq \frac{\bbeta_{n+\ell}-\bbeta_n}{\omega_0}\leq
\frac{\eta+\overline{\pi}^\gamma}{\eta\omega_0(1+\eta)^n}\,.
\end{equation}
Combining Eqs. \eqref{eq:step1}, \eqref{eq:bbeta}, \eqref{eq:step2} and \eqref{eq:step3}, we have that
\begin{equation}
\label{eq:step4}
|Q_{n+\ell}(x)-Q_n(x)|\leq Z_n:=\frac{1}{(1+\eta)^n}\frac{\eta+\overline{\pi}^\gamma}{\eta\omega_0}\left\{1+\overline{Q_0}+\frac{\eta+\overline{\pi}^\gamma}{\eta\omega_0}\right\}\,,
\end{equation}
so that for all $\eps>0$,
$$
\proba_\event\left\{\sup_{x,\ell}|Q_{n+\ell}(x)-Q_n(x)|>\eps\right\}\leq \proba_\event\{Z_n>\eps\}
$$
which converges to zero as $\{Z_n\}_{n}$ is a deterministic sequence that converges to zero. The proof is completed.
\end{proof}

\begin{proposition}
\label{prop:cauchy_reciprocate}
Assume that
\begin{enumerate}[\textbf{A}1.]
\setcounter{enumi}{0}
\item the covariance matrix of any component of the mixture satisfies $\det\Sigma_k>\delta>0$
\item the incremental mixture (unnormalized) weights are defined as:
\begin{equation}
\label{eq:new_weights}
\beta_n=\frac{\eta+\pi(\tX_n)^\gamma}{(1+\eta)^n}\1_{\event_n}
\qquad\text{and}\qquad\omega_n=\frac{\omega_0}{\omega_0+\sum_{k=1}^n\beta_k}\vee \lambda\,,
\end{equation}
where $\eta>0$ and $\lambda\in(0,1)$ are some constants, typically $(\eta,\lambda)\ll 1$.
\item the initial kernel $Q_0$ is subexponential or satisfies $Q_0(x)\propto \exp\{\psi(x)\}$ where $\psi=o(x^2)$ $(x\to\infty)$, \ie $Q_0$ has heavier tail than a Gaussian, \eg multivariate Laplace or t-distribution.
\end{enumerate}

Then, the function $\{H_n:=1/Q_n\}_n$ converges uniformly in probability to a function $H$, \ie $\sup_{x\in\Xset}|H_n(x)-H(x)|=o_\proba(1)$
\end{proposition}

\begin{proof}
As in the proof of Proposition \ref{prop:cauchy}, we show that $\{H_n\}_n$ is uniformly Cauchy convergent in probability. Under Assumption \textbf{A}2, we have that for all $x\in\Xset$ and $\ell\in\nset$, for $n$ sufficiently large, $\omega_n=\lambda$ and
$$
Q_n=\lambda Q_0+(1-\lambda)\frac{\sum_{k=1}^{n}\beta_k\phi_k}{\bbeta_n}\,.
$$
\begin{multline*}
|H_{n+\ell}-H_n|=\frac{|Q_{n+\ell}-Q_n|}{Q_{n+\ell}Q_n}\\
=\frac{(1-\lambda)\left|\sum_{k=1}^{n+\ell}\beta_k\phi_k/\bbeta_{n+\ell}-\sum_{k=1}^{n}\beta_k\phi_k/\bbeta_{n}\right|}{\left(\lambda Q_0+(1-\lambda){\sum_{k=1}^{n}\beta_k\phi_k}\slash{\bbeta_n}\right)\left(\lambda Q_0+(1-\lambda){\sum_{k=1}^{n+\ell}\beta_k\phi_k}\slash{\bbeta_{n+\ell}}\right)}\\
\leq
\frac{(1-\lambda)}{\lambda^2Q_0^2}\left|\sum_{k=1}^{n}\beta_k\phi_k\left(\frac{1}{\bbeta_{n+\ell}}-\frac{1}{\bbeta_n}\right)+
\frac{1}{\bbeta_{n+\ell}}\sum_{k=n+1}^{n+\ell}\beta_k\phi_k\right|\\
\leq \frac{(1-\lambda)}{\lambda^2}\left\{\left|\frac{1}{\bbeta_{n+\ell}}-\frac{1}{\bbeta_n}\right|\sum_{k=1}^{n}\beta_k\frac{\phi_k}{Q_0^2}
+\frac{1}{\bbeta_{n+\ell}}\sum_{k=n+1}^{n+\ell}\beta_k\frac{\phi_k}{Q_0^2}\right\}\\
\leq \frac{(1-\lambda)}{\lambda^2\bbeta_n^2}\left\{{\sum_{k=n+1}^{n+\ell}\beta_k}\sum_{k=1}^{n}\beta_k\frac{\phi_k}{Q_0^2}
+\bbeta_n\sum_{k=n+1}^{n+\ell}\beta_k\frac{\phi_k}{Q_0^2}\right\}\,.
\end{multline*}
Now by assumption \textbf{A}3. the function ${\phi_k}/{Q_0^2}$ is bounded for all $x\in\Xset$ and there exists $M<\infty$ such that for any Gaussian pdf $\phi$, $\phi/Q_0^2<M$. This yields:
$$
\sup_{x\in\Xset}|H_{n+\ell}(x)-H_n(x)|\leq
2\frac{(1-\lambda)M}{\lambda^2\bbeta_n}{\sum_{k=n+1}^{n+\ell}\beta_k}\leq Y_n:=2\frac{(1-\lambda)M}{\lambda^2\bbeta_n}\frac{\eta+\overline{\pi}^\gamma}{\eta(1+\eta)^n}\,.
$$
Upon $\event$, for $n$ sufficiently large $\bbeta_n>0$ almost surely and $Y_n\to 0$. The proof is completed by noting that
$$
\proba_\event\left\{\sup_{x,\ell}|H_{n+\ell}(x)-H_n(x)|>\eps\right\}\leq \proba_\event\{Y_n>\eps\}\to 0\,.
$$
\end{proof}

\begin{corollary}
\label{coro:2}
Assume that assumptions \textbf{A}1, \textbf{A}2 and \textbf{A}3 of Proposition \ref{prop:cauchy_reciprocate} hold, then
$$
\sup_{x\in\Xset}\left|1-\frac{Q_n(x)}{Q(x)}\right|=o_\proba(1)\,.
$$
\end{corollary}
\begin{proof}
We have
$$
\sup_{x\in\Xset}\left|1-\frac{Q_n(x)}{Q(x)}\right|=\sup_{x\in\Xset}Q_n(x)\left|\frac{1}{Q_n(x)}-\frac{1}{Q(x)}\right|\leq \sup_{x\in\Xset}Q_n(x)\sup_{x\in\Xset}\left|\frac{1}{Q_n(x)}-\frac{1}{Q(x)}\right|\,,
$$
and the proof is completed by noting that, for a sufficiently large $n$, we have
$$
Q_n\leq \lambda \overline{Q_0}+(1-\lambda)\frac{1}{(2\pi)^{d/2}\sqrt{\delta}}
$$
and applying Proposition \ref{prop:cauchy_reciprocate}.
\end{proof}

We can now prove Proposition \ref{prop1}
\setcounter{proposition}{0}
\begin{proposition}
\label{prop:diminish_cauchy}
Assume that \textbf{A}1--\textbf{A}3 of Proposition \ref{prop:cauchy_reciprocate} hold and that in addition, we have:
\begin{enumerate}[\textbf{A}1.]
\setcounter{enumi}{3}
\item the event $\event_n$ upon which the mixture increments is defined with $\uW>0$
\end{enumerate}
Then, AIMM satisfies diminishing adaptation, \ie
$$
\sup_{x\in\Xset}\|K_{n+1}(x,\cdot)-K_n(x,\cdot)\|=o_{\proba}(1)\,.
$$
\end{proposition}

\begin{proof}
From Proposition \ref{prop:cauchy}, we know that there exists a probability density function $Q$ such that $\{Q_n\}_n$ converges uniformly to $Q$. Denoting by $K$ the MH transition kernel with proposal $Q$, we have that
$$
\|K_{n+1}(x,\cdot)-K_n(x,\cdot)\|\leq \|K_{n}(x,\cdot)-K(x,\cdot)\|+\|K_{n+1}(x,\cdot)-K(x,\cdot)\|\,.
$$
To establish diminishing adaptation, it is thus sufficient to prove that $\sup_{x\in\Xset}\|K_{n}(x,\cdot)-K(x,\cdot)\|=o_{\proba_\event}(1)$. In particular, we can write
\begin{multline}
\label{eq:dimish3}
\|K_{n}(x,\cdot)-K(x,\cdot)\|=\int\left|(Q(y)-Q_n(y))\alpha(x,y)-Q_n(y)(\alpha_n(x,y)-\alpha(x,y))\right|\rmd y+\\
\left|\int\alpha(x,y)(Q(y)-Q_n(y))\rmd y-\int Q_n(\rmd y)(\alpha_n(x,y)-\alpha(x,y))\right|\\
\leq 2\sup_{x\in\Xset}\left|Q(x)-Q_n(x)\right|+2\int Q_n(\rmd y)\left|\alpha_n(x,y)-\alpha(x,y)\right|\,.
\end{multline}
Now note that for all $(x,y)\in\Xset^2$,
$$
\left|\alpha_n(x,y)-\alpha(x,y)\right|\leq \left|\frac{\pi(y) Q_n(x)}{\pi(x)Q_n(y)}-\frac{\pi(y) Q(x)}{\pi(x)Q(y)}\right|
$$
we have
\begin{multline}
\label{eq:dimish2}
\int Q_n(\rmd y)\left|\alpha_n(x,y)-\alpha(x,y)\right|\leq \int \frac{\pi(y)}{\pi(x)}\left|Q_n(x)-\frac{Q(x)Q_n(y)}{Q(y)}\right|\rmd y\\
\leq
\int \frac{\pi(y)}{\pi(x)}\left|Q_n(x)-Q(x)+Q(x)\left(\frac{Q_n(y)}{Q(y)}-1\right)\right|\rmd y\\
\leq
\int \frac{\pi(y)}{\pi(x)}\left|Q_n(x)-Q(x)\right|\rmd y+\int \frac{W(y)}{W(x)}\left|{Q_n(y)}-Q(y)\right|\rmd y\\
\leq \int \frac{\pi(y)}{W(x)}\left|1-\frac{Q_n(x)}{Q(x)}\right|\rmd y+\int \frac{W(y)}{W(x)}\left|{Q_n(y)}-Q(y)\right|\rmd y
\leq \frac{2}{\underline{W}}\sup_{x\in\Xset}\left|1-\frac{Q_n(x)}{Q(x)}\right|\,.%\\
%\leq \left(\frac{1}{\inf_{x\in\Xset} \pi(x)}+\frac{\Wst}{\underline{W}}\right)\sup_{x\in\Xset}\left|{Q_n(x)}-Q(x)\right|\,.
\end{multline}
Indeed, the proposal limit $Q$ must satisfy $\underline{W}<W:=\pi/Q<\Wst$. Combining Eqs. \eqref{eq:dimish3} and \eqref{eq:dimish2} yields
$$
\sup_{x\in\Xset}\|K_{n}(x,\cdot)-K(x,\cdot)\|\leq 2\left(\sup_{x\in\Xset}\left|{Q_n(x)}-Q(x)\right|+\frac{1}{\underline{W}}\sup_{x\in\Xset}\left|1-\frac{Q_n(x)}{Q(x)}\right|\right)\,,
$$
and applying Proposition \ref{prop:cauchy} and Corollary \ref{coro:2} completes the proof.
\end{proof}

\section{Proof of Proposition \ref{prop2}}
\label{app2}
We show that the AIMM version presented in Section \ref{sec:conv:aimm2} satisfies diminishing adaptation and containment and is thus ergodic \cite[Theorem 2]{roberts2007}.

\subsection{Diminishing Adaptation}
First we note that diminishing adaptation is implied by Assumption \textbf{B}3. Indeed, using the same approach that in Appendix \ref{app1} and introducing the quantity $\overline{\Delta}_n:=\sup_{x\in\Xset}\|\overline{P}_{n+1}(x,\,\cdot\,)-\overline{P}_{n}(x,\,\cdot\,)\|$, we have for all $\eps>0$
\begin{equation}
\label{eq:app2:1}
\proba\{\overline{\Delta}_n>\eps\}=\proba\{\overline{\Delta}_n>\eps\,|\,\event\}\proba\{\event\}+\proba\{\overline{\Delta}_n>\eps\,|\,\event^c\}
\proba\{\event^c\}\,,
\end{equation}
where for readability we have denoted by ${\event}^c$ the complementary event of ${\event}$ ($\event=\cap_n\cup_{k>n}\event_k$). Now on $\event^c$, there exists a $n_0<\infty$ such that for all $n>n_0$ the mixture does not increment anymore. In such a case, $\proba\{\overline{\Delta}_n>\eps\,|\,\event^c\}=0$ as soon as $n>n_0$ since for all $n>n_0$, $\proba\{M_n=M_{n_0}\}=1$. This implies that $\proba\{\overline{P}_{n+1}(x,\,\cdot\,)=\overline{P}_{n}(x,\,\cdot\,)\}=1$ which is straightforward if $\omega_n$ is a function of $n$ through $M_n$. Now by definition of $\overline{\event}=\cap_n\cup_{k>n}\overline{\event}_k$ ($\overline{\event}_n=\event_n\cap \{M_n<M\}$), upon $\event$, there exists an $n_1<\infty$ such that $M_{n_1}=M$ and for all $n>n_1$, $\proba\{M_n=M\}=1$. Similarly to the reasoning with $\event^c$, we conclude that for all $n>n_1$, $\proba\{\overline{\Delta}_n>\eps\,|\,\event\}=0$. Diminishing adaptation follows from Eq. \ref{eq:app2:1} by writing that for all $\eps>0$ and $n>\max(n_0,n_1)$, $\proba\{\overline{\Delta}_n>\eps\}=0$.

\subsection{Containment}
We now show that under the Assumptions of Proposition \ref{prop2}, the AIMM process satisfies the containment condition \ie $C_n(\eps)$ defined at Eq. \eqref{eq:cont} is bounded in probability. We use Theorem 1 from \cite{rosenthal2017ergodicity}, which extends the seminal development made in \cite{craiu2015stability} and in particular Theorem 21 therein. In \cite{rosenthal2017ergodicity}, adaptive MCMC are analysed in a parametric adaptation framework, \ie under the assumption that there exists a set $\Ycal\subset \rset^p$ such that any adaptive proposal belongs to a family of distributions parameterized by $\Ycal$. Theorem 1 \citep{rosenthal2017ergodicity} identifies six assumptions, that we now enumerate for completeness, that are sufficient to guarantee the containment condition. We define for all $r>0$, $\comp_r:=\{x\in\Xset,\,\min_{y\in\comp}\|x-y\|\leq r\}$.
\begin{enumerate}[\textbf{RY}-1]
\item $\Ycal$ is compact.
\item The process evolves through a $\pi$-Harris ergodic transition kernel that does not allow moves larger than some fixed distance $D>0$.
\item The process only adapts within a compact set $\comp$.
\item $Q_0$ is bounded above by $Q_0(\rmd y)\leq \zeta \nu_\ast(\rmd y)$ where $\nu_\ast$ is a probability measure concentrated on $\comp_{2D}\backslash \comp_D$.
\item There exists two positive constants $(\varepsilon,\delta)$ such that the fixed kernel $P$, \ie the M-H kernel with proposal $Q_0$, is bounded below by $P(x,\rmd y)\geq \varepsilon \text{Leb}(\rmd y)$ whenever $(x,y)\in J$ with $\|x-y\|\leq \delta$ and $J\supset \comp_{2D}\backslash \comp_D$.
\item The function $(x,y,\gamma)\mapsto Q_\gamma(x,y)$ is combocontinuous, \ie it is a finite combination of continuous functions (which generalizes the concept of piecewise continuous functions where some pieces are not necessarily intervals).
\end{enumerate}
We now show that all those assumptions are satisfied by the AIMM process described at Section \ref{sec:conv:aimm2}.

In the context of Proposition \ref{prop2}, we define $\Ycal$ as the following union of sets
\begin{equation}
\label{eq:app2:2}
\Ycal:=\cup_{\ell=0}^M\left\{\mathsf{A}^\ell,\mathsf{B}^\ell,\mathsf{C}^\ell\right\}\,,
\end{equation}
such that $\mathsf{A}:=\{\beta\in\rset,\;0\leq \beta\leq \overline{\pi}^\gamma\}$, $\mathsf{B}:=\{\mu\in\rset^d,\;-L\leq \mu_i\leq L\}$ and
$\mathsf{C}:=\{\Sigma\in\mathcal{M}_d(\rset),\;-L\leq \Sigma_{i,j}\leq L\}$. Indeed for all $n\in\nset$, there exists $\gamma=\{\beta_1,\ldots,\beta_\ell,\mu_1,\ldots,\mu_\ell,\Sigma_1,\ldots,\Sigma_\ell\}$ with $\ell\leq M$ such that  $\oQ_n^\comp\equiv\oQ_{\gamma}^\comp:=\overline{\omega}_n^\comp Q_0+(1-\overline{\omega}_n^\comp)\sum_{l=1}^\ell\beta_l\Phi(\,\cdot\,|\,\mu_l,\Sigma_l)\slash \sum_{l=1:\ell}\beta_l$ and $\gamma\in\Ycal$. Clearly, $\Ycal$ is a finite union of compact sets and is thus compact and \textbf{RY}-1 holds. By construction, the second part of \textbf{RY}-2 and \textbf{RY}-3 hold. Moreover, it can be noted that for any $D>0$, the Metropolis algorithm that consists in proposing $\tX_{n+1}\sim \overline{Q}_n^\comp$ and accepting with probability $\1_{\|\tX_{n+1}-X_n\|\leq D}\overline{\alpha}^\comp(X_n,\tX_{n+1})$, with $\overline{\alpha}^\comp$ is defined as $\alpha$ \eqref{eq:alpha} up to replacing $Q_n$ by $\overline{Q}_n^\comp$, is still $\pi$-invariant. This follows from noting that this algorithm would generate a $\pi$-reversible Markov chain. Since it is also $\phi$-irreducible and aperiodic, $\oP_\gamma^\comp$ is Harris ergodic for $\pi$, for any $\gamma\in\Ycal$. This checks \textbf{RY}-2. \textbf{RY}-4 is straightforward by taking $Q_0$ as any distribution dominated by the Lebesgue measure, \eg multivariate Gaussian, multivariate T-student, etc. Assumption \textbf{RY}-5 follows from $Q_0$ and $\pi$ being absolutely continuous distributions on $(\Xset,\Xalg)$. Finally, we note that for a fixed $\gamma\in\Ycal$, $(x,y)\mapsto \oQ^\comp_\gamma(x,y)$ is piecewise continuous. In fact this function is continuous on the two complementary sets $\comp$ and $\overline{\kappa}$. Moreover, we note that for a fixed $y\in\Xset$, $(\gamma,x)\mapsto \oQ_\gamma^\comp(x,y)=\sum_{l=1}^M\1_{\ell(\gamma)=l}\oQ_{\gamma}^\comp(x,y)$, where $\ell(\gamma)$ is the number of components in the mixture $\oQ_\gamma^\comp$. Since for all $\ell\in\{0,\ldots,M\}$, $x\in\Xset$, the mapping $(x,\gamma)\mapsto \oQ_{\gamma}^\comp(x,y)$ is piecewise continuous, $(\gamma,x)\mapsto \oQ_\gamma^\comp(x,y)$ is combocontinuous and \textbf{RY}-6 holds. The proof is completed.

\section{Simulation details}
\label{app5}
\subsection{Competing algorithms}
\label{app:alg}
\begin{itemize}
\item Adaptive Metropolis--Hastings (AMH) \citep{haario2001adaptive}: a benchmark adaptive MCMC algorithm with non-independence proposal kernel
\begin{equation*}
Q_n(x_n,\cdot)=\left\{
\begin{array}{lc}
\Psi_d(\,\cdot\,;\,x_n,\Sigma_0) & \text{if}\; n\leq N_0\\
p\Psi_d(\,\cdot\,;\,x_n,\Sigma_0) +(1-p)\Psi_d(\,\cdot\,;\,x_n,\Sigma_n) & \text{otherwise}
\end{array}
\right.
\end{equation*}
where $\Psi_d(\,\cdot\,;\,m,S)$ is the $d$-dimensional Gaussian distribution with mean $m$, covariance matrix $S$, $\Sigma_0$ is an initial covariance matrix and $N_0$
the number of preliminary iterations where no adaptation is made. We consider the version of AMH proposed in \cite{roberts2009examples}, in which there is a strictly positive
probability ($p=.05$) of proposing through the initial Gaussian random walk. The covariance matrix of the adaptive part of the proposal is written $\Sigma_n=s_d\Gamma_n$,
where $\Gamma_n$ is defined as the empirical covariance matrix from all the previous states of the $X_1,\ldots,X_n$, and $s_d=2.4^2/d$ is a scaling factor.
The parameters $N_0$ and $\Sigma_0$ were set to be equal to the corresponding AIMM parameters.

\item Adaptive Gaussian Mixture Metropolis--Hastings (AGM) \citep{luengo2013fully}: an adaptive MCMC with an independence proposal kernel defined as a mixture of
Gaussian distributions
    \begin{equation}
    Q_n=\sum_{\ell=1}^{M}\omega_{\ell,n}\Psi_d(\,\cdot\,;\,\mu_{\ell,n},\Lambda_{\ell,n})\,.
    \end{equation}
    In AGM, the number of components $M$ is fixed. The proposal distribution $Q_n$ is parameterised by the vector
    $\Omega_n=\{\omega_{\ell,n},\mu_{\ell,n},\Lambda_{\ell,n}\}_{\ell=1}^M$. Given the new state $X_{n+1}$ of the chain, the next parameter $\Omega_{n+1}$ will
    follow from a deterministic update ; see \cite[section 4]{luengo2013fully} for more details. In the implementation, we set the number of components $M$ identical to the corresponding AIMM parameter $\Mmax$.

\item For the sake of comparing with non--adaptive methods, we also include the Independence Sampler (IM) (see \eg \cite{liu1996metropolized}) and the random
walk Metropolis--Hastings (RWMH) (see \eg \cite{roberts2001optimal}) using the Matlab build--in function \textit{mhsample} with default settings.
\end{itemize}

\subsection{Performance indicators}
\label{app:perf}
The quality of a sampler is based on its ability to explore the state space fully and quickly without getting trapped in specific states (mixing). The KL divergence between $\targ$ and the chain distribution (if available) and the \textit{Effective Sample Size} are indicators that allow us to assess those two properties. Related statistics such as the chain \textit{Jumping distance} and the \textit{Acceptance rate} of the sampler are also reported. We also provide the chain \textit{Efficiency}, which penalizes the Effective Sample Size by the computational burden generated by the sampler.

\begin{itemize}
\item For a stochastic process $\{X_k,\,k\leq n\}$, we recall that when sampling from $\targ$ is feasible, the KL divergence between the target and the chain distribution can be approximated by
    \begin{equation}
    \KL=\frac{1}{L}\sum_{\ell=1}^L\log\left\{\frac{\targ(Z_\ell)}{\Law(Z_\ell\,|\,X_{1:n})}\right\}\,, \quad Z_\ell\sim \targ\,,
    \end{equation}
    where $\Law(\,\cdot\,|\,X_{1:n})$ is the kernel density estimation of the stochastic process (obtained using the routine \texttt{ksdensity} provided in Matlab in the default setting). We stress that using approximated values for $\KL$ are not an issue here as we are first and foremost interested in comparing different samplers, with lower KL being the better.

\item The jumping distance measures how fast the sampler explores the state space, see \cite{pasarica2010adaptively}. For a stochastic process $\{X_k,\,k\leq n\}$
it is estimated by:
\begin{equation}
\JMP=\frac{1}{n-1}\sum_{k=1}^{n-1}\|X_{k+1}-X_k\|^2 ,
\end{equation}
and the larger the squared distance the better.

\item The mixing rate of a of a one dimensional chain $(d=1)$ is classically evaluated with the Effective Sample Size (ESS), which is approximated by
\begin{equation}
\label{eq:ess}
\ESS=1\bigg\slash \left\{1+2\sum_{t=1}^T\hat{\rho}_{t}\right\}\, ,
\end{equation}
where $\hat{\rho}_{t}$ denotes the empirical lag $t$ covariance estimated using the sample path of $X_{1:n}$, $T=\min(1000,t_0)$ and $t_0$ is the smaller lag such that $\hat{\rho}_{t_0+\ell}<.01$, for all $\ell>0$. $\text{ESS}\in(0,1)$ (ESS=1 corresponds to \iid samples) and the higher $\ESS$, the better the algorithm. When $d>1$, $\ESS$ is set as the minimum ESS among the $d$ marginal ESS's.

\item The tradeoff between the computational efficiency and the precision is estimated by $\EFF$
\begin{equation}
\EFF=\ESS\bigg\slash\tau\,,
\end{equation}
where $n$ is the total number of iterations performed by the sampler and $\tau$ the CPU time (in second) required to achieve the $N$ transitions
\citep{tran2014adaptive}.
\end{itemize}

\subsection{AGM targeting $\pi_2$}
\label{app:agm}
\begin{figure}
\centering
\includegraphics[scale=.42]{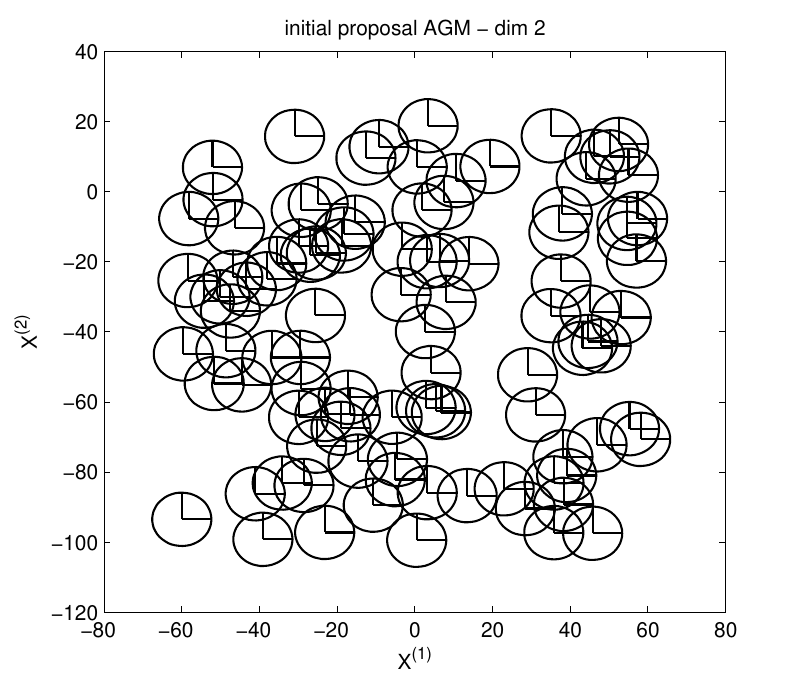}
\includegraphics[scale=.42]{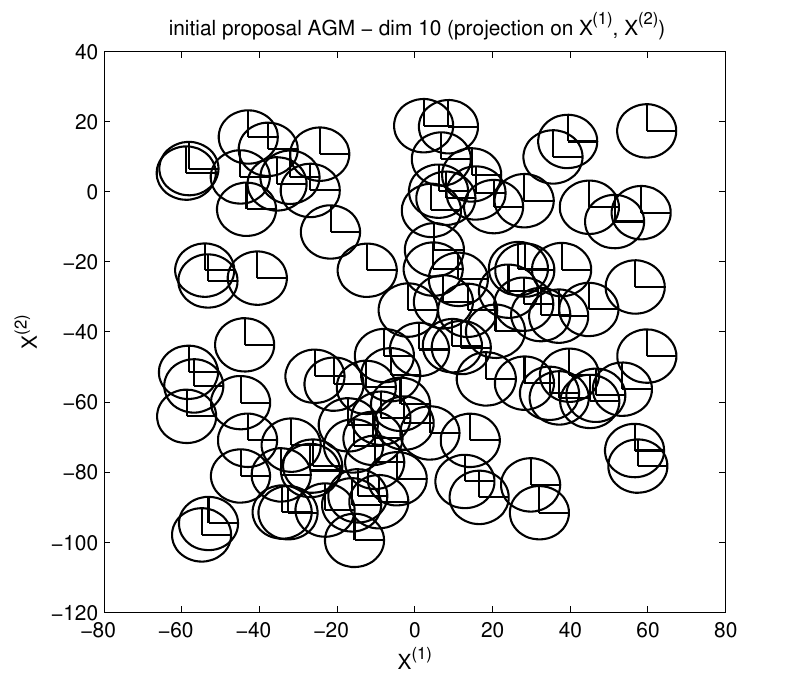}

\includegraphics[scale=.42]{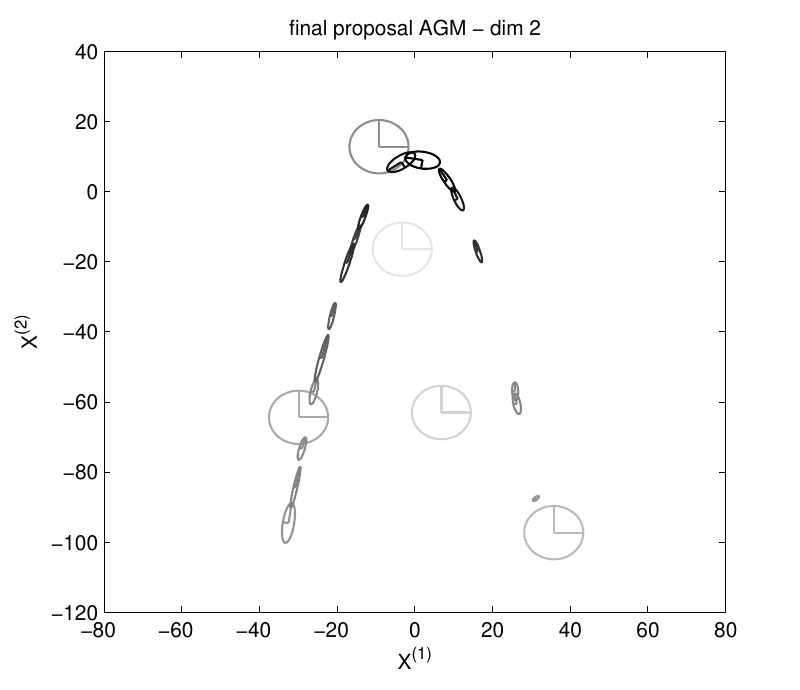}
\includegraphics[scale=.42]{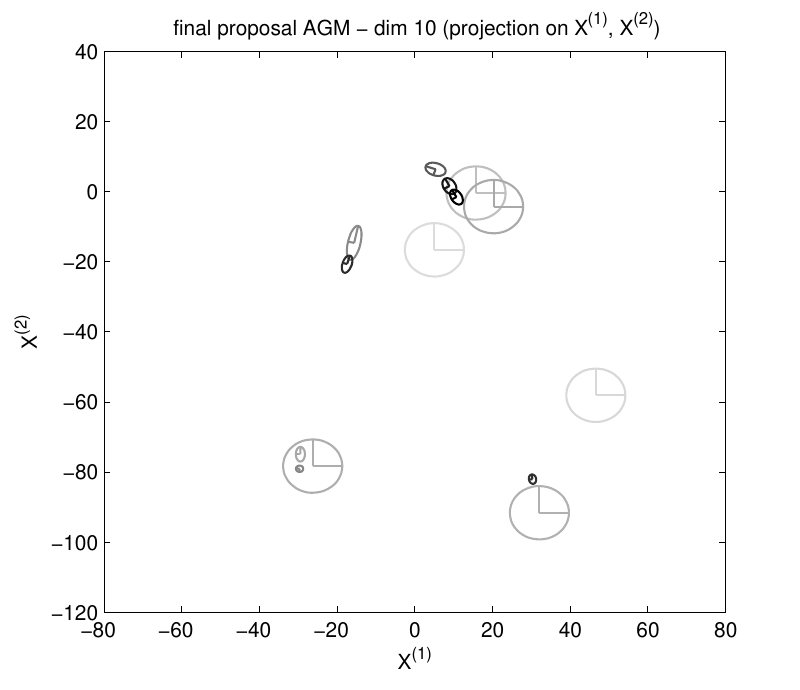}

\includegraphics[scale=.152]{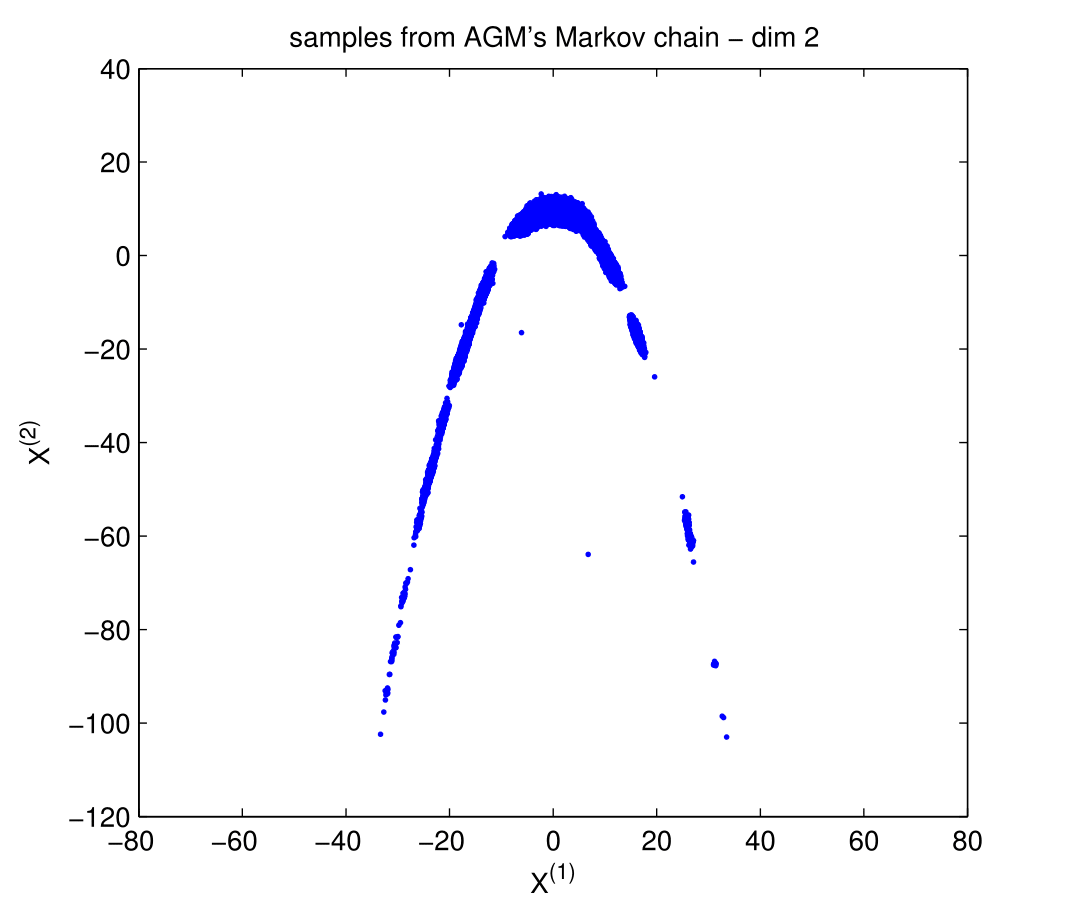}
\includegraphics[scale=.152]{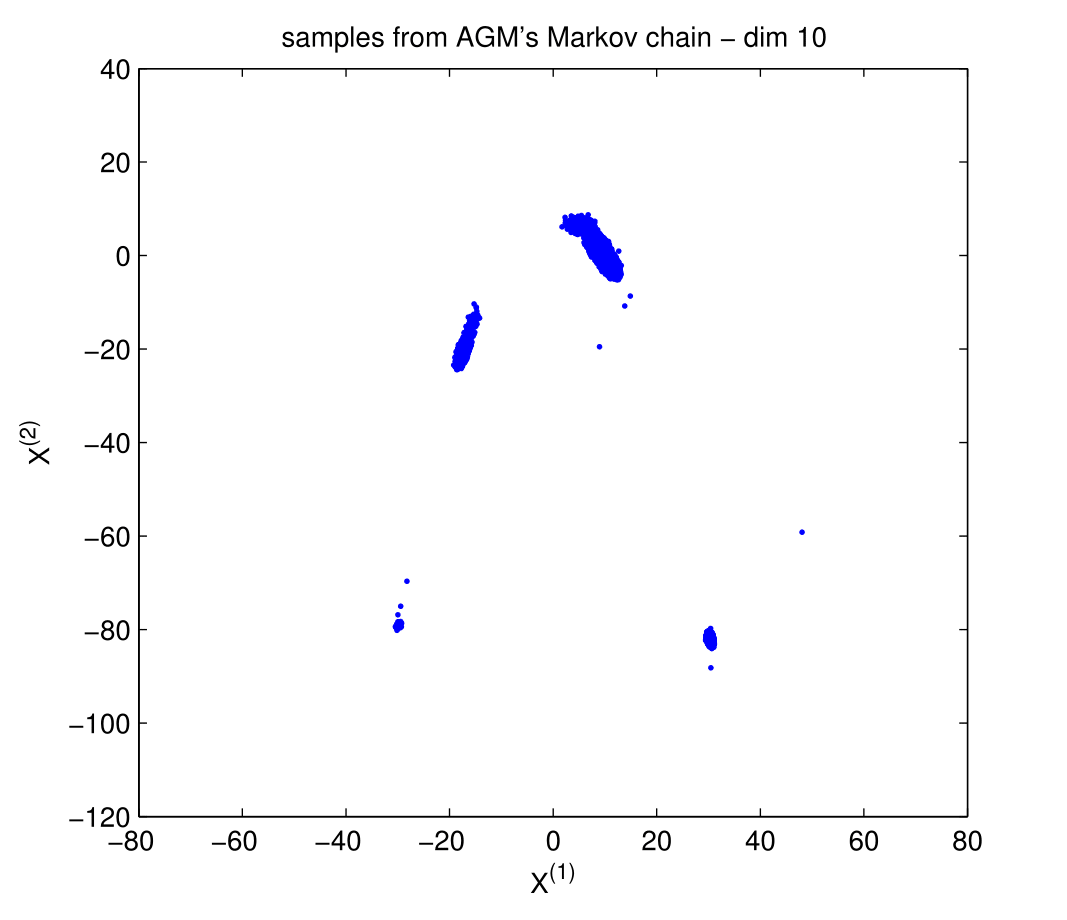}
\caption{(example \ref{example_pi2}) $\pi_2$ target, $d\in\{2,10\}$ - AGM sampling from $\pi_2$ in dimensions $d=2$ (left column) and $d=10$ (right column, projection onto the first two dimensions). First and second rows: initial mixtures with centers drawn uniformly and adapted mixtures after $n=200,000$ iterations. The grey levels on the ellipses stand for the weights of the components (from white to black for larger weights) and the region inside the ellipses contains $75\%$ of each Gaussian mass. Third row: samples $\{X_{1,2},X_{2,n}\,,n\leq 200,000\}$ from AGM. \label{fig:banana_agm}}
\end{figure}

Although similar to AIMM, AGM adapts a Gaussian mixture proposal to the target, the distribution of the resulting chain remains far from $\pi_2$ after $200,000$
iterations. Figure \ref{fig:banana_agm} gives a hint to understand why AGM is not efficient in sampling from $\pi_2$. By construction, AGM adapts locally a component of the proposal, provided that some samples are available in its vicinity. As a result, components initially located where the probability mass is non-negligible will adapt well to the target but the others will see their weight shrinking to zero (in dimension two, out of one hundred initial kernels having the same weight, the ten kernels with highest weight hold $0.99$ of the total mixture weight after $200,000$ iterations). This gives those kernels with initial poor location a high inertia, which prevents them moving away from low density regions to reach regions that are yet to be supported by the proposal. AIMM's opportunistic increment process explores the state space more efficiently.

\end{appendices} 
\end{document}